\documentclass[11pt]{article}
\usepackage{tikz}
\usepackage{geometry}
\usepackage[english]{babel}
\usepackage[utf8]{inputenc}
\usepackage[T1]{fontenc}
\usepackage{indentfirst}
\usepackage{amsmath}
\usepackage{amssymb}
\usepackage{amsthm}
\usepackage{proof}
\usepackage{eufrak}
\usepackage{graphicx}
\usepackage{psfrag}
\usepackage{mathtools}
\usepackage{xfrac}
\usepackage{float}

\usepackage{graphicx}
\usepackage{tikz}

\usetikzlibrary{decorations.markings}

\tikzset{line/.style={line width=0.25mm},
curve/.style={line,smooth,tension=1},
->-/.style={decoration={
  markings,
  mark=at position #1 with {\arrow[>=stealth]{>}}},postaction={decorate}},
-<-/.style={decoration={
  markings,
  mark=at position #1 with {\arrow[>=stealth]{<}}},postaction={decorate}},
}

\allowhyphens

\newcommand{\bs}[1]{{\boldsymbol{#1}}}
\newcommand{\ket}[1]{{\left|#1\right\rangle}}
\newcommand{\bra}[1]{{\left\langle #1\right|}}

\newcommand{\Piu}{P}
\newcommand{\Pid}{N}

\pgfdeclarelayer{edgelayer}
\pgfdeclarelayer{nodelayer}
\pgfsetlayers{edgelayer,nodelayer,main}
\tikzstyle{none}=[inner sep=0pt]
\tikzstyle{rn}=[circle,fill=Red,draw=Black,line width=0.8 pt]
\tikzstyle{gn}=[circle,fill=Lime,draw=Black,line width=0.8 pt]
\tikzstyle{yn}=[circle,fill=Yellow,draw=Black,line width=0.8 pt]
\tikzstyle{simple}=[-,draw=Black,line width=2.000]
\tikzstyle{arrow}=[-,draw=Black,postaction={decorate},decoration={markings,mark=at position .5 with {\arrow{>}}},line width=2.000]
\tikzstyle{tick}=[-,draw=Black,postaction={decorate},decoration={markings,mark=at position .5 with {\draw (0,-0.1) -- (0,0.1);}},line width=2.000]
\usetikzlibrary{arrows}
\tikzstyle{black}=[fill=black, shape=circle]
\tikzstyle{arrow}=[->]
\tikzstyle{red arrow}=[draw=red, ->]


\newcommand{\be}{\begin{equation}}
\newcommand{\ee}{\end{equation}}
\newcommand{\bea}{\begin{eqnarray}}
\newcommand{\eea}{\end{eqnarray}}

\usepackage{ifpdf}

\ifpdf
\usepackage{epstopdf}
\usepackage[pdftex,colorlinks,urlcolor=blue,citecolor=blue,linkcolor=black]{hyperref}
\else

\usepackage[hypertex,colorlinks,urlcolor=blue,citecolor=blue,linkcolor=black]{hyperref}
\fi
\begin{document}

\thispagestyle{empty}

\begin{center}%
{\LARGE\textbf{\mathversion{bold}%
Constrained integrability and anyonic chains}\par}

\vspace{1cm}
{\textsc{Matthew Blakeney$^{a, b}$, Luke Corcoran$^{ c}$, Marius de Leeuw$^{ c}$}}
\vspace{8mm} \\
\textit{
$^a$  Perimeter Institute for Theoretical Physics, Waterloo, Canada \\
[5pt]
}
\vspace{.2cm}
\textit{
$^b$  Department of Physics and Astronomy,\\
University of Waterloo, Waterloo, ON, Canada\\
[5pt]
}
\vspace{.2cm}
\textit{
$^c$ School of Mathematics \& Hamilton Mathematics Institute, \\
Trinity College Dublin, Ireland\\
[5pt]
}

\vspace{1cm}
\texttt{matt.blakeney20@gmail.com}\\ 
\texttt{\{corcorl2, deleeuwm\}@tcd.ie} \\
%

\par\vspace{15mm}

\textbf{Abstract} \vspace{5mm}

\begin{minipage}{13cm}

We review the notion of Yang-Baxter integrability for spin chains that have Hilbert spaces with constraints, such as a Rydberg blockade. We focus
on anyonic chains, whose constraints arise from the fusion rules of the fusion categories on which they are based. We discuss the emergence of Temperley-Lieb algebras and present a new result on which types of 
anyonic chains exhibit them. We then give an overview of known results for integrable anyonic chains and extend them to several fusion categories up to rank $7$. Using a modification of the boost operator formalism, we find several new integrable anyonic chains and discuss some of their properties. These include spin-$\frac32$ models for $\mathfrak{su}(2)_k$ fusion categories, anyonic chains based on the Tambara-Yamagami fusion categories TY$(\mathbb{Z}_n)$, and product fusion categories Fib$\times$Fib and Fib$\times$Ising. We review recent results for spin chains based on the Haagerup-Izumi fusion category HI$(\mathbb{Z}_3)$, and present preliminary numerics for a HI$(\mathbb{Z}_5)$ model.
 
\end{minipage}
\end{center}


\newpage 

\tableofcontents
\bigskip
\hrule

\section{Introduction}

In recent years there has been a surge of interest in quantum mechanical spin chains on \textit{constrained} Hilbert spaces. These are models where the usual tensor product Hilbert space structure $(\mathbb{C}^n)^{\otimes L}$ is broken due to physical constraints. On the one hand, there are anyonic chains which realize non-invertible symmetries \cite{McGreevy:2022oyu, Shao:2023gho, Bhardwaj:2023kri, Schafer-Nameki:2023jdn, Kaidi:2026urc}. Anyonic chains are one-dimensional models which describe the fusion of objects in a fusion category $\mathcal{C}$.  On the other hand, one can consider spin models where certain states are energetically suppressed and effectively forbidden. A prototypical example in this case are so-called \textit{Rydberg-blockaded chains}. These are spin-$\frac12$ models where states with neighboring down spins $\ket{\downarrow\downarrow}$ are excluded, which notably have been experimentally realized \cite{Urban2008ObservationOR, Bernien:2017ubn}. These two descriptions can coincide, for example in the case of the \textit{golden chain}. This model arises naturally as an anyonic chain for the Fibonacci fusion category \cite{Feiguin:2006ydp}, and also as a critical point in a two-parameter Rydberg-blockaded chain \cite{Fendley_2004}.

Constrained models can exhibit interesting and exotic features. Phase diagrams have been mapped out for several anyonic chains, for example those based on $\mathfrak{su}(2)_k$ \cite{Gils_2013, Vernier_2017} and $\mathfrak{so}(5)_2$ \cite{Finch:2014ina}, and several critical phases/integrable points have been identified. The corresponding conformal field theories (CFTs) to which these phases correspond range from Virasoro minimal models to $\mathcal{N}=1$ superconformal models and $\mathbb{Z}_k$ parafermions. Recent investigations into more exotic Haagerup-Izumi fusion categories has led to the discovery of a critical lattice model with Haagerup symmetry \cite{Huang:2021nvb, Vanhove:2021zop}. Anyonic chains also come equipped with non-local topological symmetry operators inherited from the underlying fusion category. At criticality these symmetries are expected to descend to topological defects of the continuum CFT, where they constrain the operator content and the symmetry-allowed relevant deformations \cite{Pfeifer:2010xi, Buican:2017rxc}. In the Rydberg-blockaded case, one of the simplest systems to consider is the PXP model, which has been shown to contain quantum many-body scars \cite{Turner_2018, Turner:2018yco}. This behavior has been thought to be related to a nearby integrable model \cite{Khemani:2019huc} and an approximate $\mathfrak{su}(2)$ structure \cite{Choi:2018cfo}. Integrable Rydberg-blockaded models up to range 5 have been explored, and include the constrained XXZ model \cite{constrained1}, off-critical golden chain \cite{Bianchini:2014bfa}, and double golden chain \cite{Corcoran:2024ofo}.

Integrable spin chains are special one-dimensional models with distinctive features. They possess a large amount of symmetry, which often leads to exact solutions in various dynamical scenarios. Integrability is generally associated with models on factorized Hilbert spaces, due to $R$-matrix formulations which necessarily rely on this \cite{Faddeev:1996iy, Gombor:2021nhn}. However, different definitions of integrability such as the existence of higher conserved charges or Poissonian level statistics do not. There are several examples of integrable spin chains in both the anyonic chain and Rydberg-blockaded case, including the examples mentioned above. The golden chain is integrable, and furthermore all fusion category projectors onto the identity fusion channel correspond to a Temperley-Lieb (TL) algebra and are therefore integrable \cite{Blakeney:2025ext}. Explorations into spin-$1$ anyonic chains have revealed integrable structures based on the Birman-Murakami-Wenzl (BMW) algebra, a generalization of Temperley-Lieb \cite{Vernier_2017,Finch:2014ina}. Due to the local constraints, interesting models on constrained Hilbert spaces are typically of range-3 and higher. Therefore any $R$-matrix formulation of integrability in the constrained case would build on the medium-range formulation of \cite{Gombor:2021nhn}. This was completed in \cite{Corcoran:2024ofo}, which allows for a systematic classification of integrable constrained models in both the Rydberg-blockaded and anyonic chain cases. This approach is based on the boost operator \cite{1982JETP...55..306T, links2001ladder} method of generating higher charges. 

One bottleneck in the study of anyonic chains is that they are constructed from \textit{$F$-symbols} of the underlying fusion category $\mathcal{C}$. These are a set of numbers $(F^{a}_{bcd})_{ef}\in \mathbb{C}$, which are obtained as the solution to a large set of coupled cubic equations known as the pentagon equations. There is no known method to solve these equations in general, and typically solutions for higher-rank categories are not available. However, much progress has been made recently for fusion categories of low-rank \cite{gert_thesis}. For example, recent calculations of the $F$-symbols of the Haagerup $\mathcal{H}_3$ fusion category \cite{osborne2019fsymbols} and higher Haagerup-Izumi categories \cite{Huang:2020lox} led to the identification of the Haagerup CFT mentioned above. Recently, fusion categories up to rank 7, complete with exhaustive sets of $F$-symbols, have been tabulated in AnyonWiki \cite{gert_vercleyen_anyonwiki}. This is accompanied by a Mathematica package Anyonica \cite{vercleyen_anyonica_2025} which allows for efficient symbolic implementations of anyonic chains, and integration into numerical workflows.

In this paper we discuss several aspects of integrability in constrained systems, with a focus on anyonic chains. We review known results in the anyonic chain literature systematically, and compare everything in a uniform setting. Using the constrained boost operator formalism, we classify integrable anyonic chains for several fusion categories up to rank 7. In some cases we reproduce the known models, and in lesser-studied cases we construct new integrable spin chains. In these cases we perform numerical analyses to check for critical behavior. The paper is organized as follows. In section \ref{sec:fusionandanyon} we review the construction of anyonic chains from fusion categories, introducing only the structure needed in later sections. In section \ref{sec:properties} we review two general properties of anyonic chains; firstly that they commute with a set of topological operators $Y_b$ which furnish a representation of the underlying fusion category, and secondly that in most cases one can construct anyonic chains which are built from TL operators. In this case we generalize a result that identity projectors $a \otimes a \rightarrow 1$, for $a\in\mathcal{C}$, are TL \cite{Blakeney:2025ext} to channels $a\otimes a \rightarrow b$, where $b\in \mathcal{C}$ is any invertible object. In section \ref{sec:fib} we review the simplest anyonic chain, the golden chain, which is based on the Fibonacci fusion category. We discuss the properties of this model in detail, which are useful to compare against when we consider more complicated anyonic chains. We also give details of numerical methods which we apply throughout the paper. In section \ref{sec:rydberg} we emphasize that there are interesting constrained models beyond anyonic chains, by considering the Rydberg-blockade constraint. We review the Yang--Baxter formulation of integrability in this case and introduce the constrained boost operator formalism, which allows for a systematic search of integrable models on constrained Hilbert spaces.

In section \ref{sec:su2k} we review $\mathfrak{su}(2)_k$ anyonic chains, a large family of models which generalize the golden chain, from the point of view of integrability. By extrapolating from $k = 2$ to $k=7$ we highlight the main features of these models, and verify old results for integrable chains up to spin-1 \cite{Vernier_2017}. For $k \geq 6$ it is possible to form spin-$\frac32$ anyonic chains. The case $k=6$ has an extra symmetry and so we find a large number of integrable spin-$\frac32$ chains in this case. $k\geq 7$ is generic and we find three isolated integrable points, which we also verify numerically for $k = 8, 9$. Finally, in section \ref{sec:other} we consider several further fusion categories, where we review and extend results for integrable anyonic chains. In particular we consider the fusion categories $\mathfrak{so}(5)_2$, the Haagerup-Izumi fusion categories HI$(\mathbb{Z}_n)$, the Tambara-Yamagami categories TY($\mathbb{Z}_n$), and products Fib$\times$Fib and Fib$\times$Ising. In the latter three cases we find new integrable models. In the Fib$\times$Fib case one of our models matches closely with a critical point recently studied in \cite{Antunes:2025huk}. For HI$(\mathbb{Z}_3)$ there are no integrable anyonic chains beyond the TL case. However, in light of the $c\sim2$ critical model $\mathrm{P}^{\rho}_\rho$ studied in \cite{Huang:2021nvb}, we briefly study the analogous model for HI$(\mathbb{Z}_5)$. We find some evidence of criticality with $c\sim 3$, but further numerics are needed to make any precise conjectures.

\section{Anyonic chains and fusion categories}\label{sec:fusionandanyon}
In this section, we give an introduction to anyonic chains from a physically motivated standpoint. We introduce only the structure required for this paper and avoid the burden of complete mathematical definitions. Physically-motivated introductions to the additional structures which can be added to fusion categories and we omit here (such as braiding and modularity) can be found in \cite{Trebst_2008,Wolf:2020qdo,beer2018categoriesanyonstravelogue}, and mathematical introductions to fusion categories can be found in \cite{Kirillov,etingof2017fusioncategories}.

This section is laid out as follows: first, the notion of a fusion ring is introduced to describe the fusion and splitting of lines in fusion diagrams. Next, we physically motivate the structure of a fusion category by discussing changes of basis and describe the calculus on fusion diagrams that can be employed. Finally, we discuss the constrained Hilbert space of anyonic chains and the local operators which act on it, and summarize the procedure for defining an anyonic chain model.

\subsection{Fusion rings}
Anyonic chains generalize the notion of a spin chain, modeling particle interactions on a one-dimensional ordered set of sites. In generalizing from spin chains to anyonic chains, the local degrees of freedom are upgraded from spins to topological charges/line species, and the chain is constructed by repeatedly joining (i.e. `fusing') together lines of some chosen species.

Concretely, we start with a set of lines of species $a$, and interpreting diagrams such that causality flows from top to bottom, the leftmost line recurrently fuses with its neighbor. The intermediate line species, $\{x_1,x_2,...,x_L\}$, along the spine of this `fusion diagram' are the degrees of freedom on the anyonic chain.

\begin{figure}[H]\centering
\includegraphics{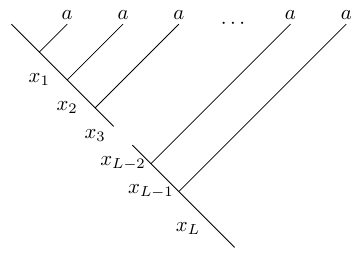}
\caption{An anyonic chain fusion diagram with the degrees of freedom, $\{x_1,x_2,...,x_L\}$, along the spine of the fusion diagram.}
\label{fig:ChainDiagram}
\end{figure}

In order to define an anyonic chain, the set of allowed line species in the fusion diagrams of the theory must be defined. Furthermore, a set of self-consistent rules which tell us what species of line may be produced when any two lines are fused is required. This information defines a fusion ring.

We will work with the formal definition of fusion rings given in \cite{vercleyen_slingerland_Rings}. In summary, this definition states that the species of lines in our theory are elements $\{a,b,\dots\}$ of a basis $B$ with a product called `fusion' defined on them by `fusion rules'
\begin{equation}\label{eq:fusionring}
    a\otimes b=\bigoplus_{c\in B}N_{ab}^cc,
\end{equation}
where $B$ and $N_{ab}^c$ must satisfy the following conditions:

\begin{itemize}
    \item $B$ must contain an element $1$, physically corresponding to the possibility that two lines may annihilate to produce the vacuum/trivial line.
    \item  The fusion ring's fusion multiplicities, $N_{ab}^c$,  must be non-negative integers. This is required in order to interpret the fusion multiplicities as the number of ways in which a line of species $c$ can result from the fusion of lines of species $a$ and $b$. 
    \item There exists a linear involution $\bar{\cdot}:a\mapsto \bar{a}$ such that $N_{a\bar{b}}^1=\delta_{ab}$. This ensures that for every line, there is a dual species of line with which it can be pairwise created/annihilated.
    \item $N_{ab}^c=N_{\bar{a}c}^b\text{ }\forall a,b,c \in B$, enforcing that a line of species $c$ can split into two lines of species $a$ and $b$ in the same number of ways as $a$ and $b$ can fuse to produce a $c$. This essentially ensures that fusion and splitting processes are related schematically by `time reversal'.
\end{itemize}

Additionally,  because the multiplication rule defined by \eqref{eq:fusionring} is the multiplication rule of a ring, it must be associative. This ensures that the order in which we pairwise fuse a collection of lines doesn't affect the species obtained from their cumulative fusion.

The multiplicity of a fusion ring is defined as $\text{max}(N_{ab}^c)$, and its rank is the size of $B$. In this paper and much of the literature, only multiplicity-1, rank-$\leq7$ fusion rings are considered. A systematic approach to defining fusion rings of a given rank and multiplicity was developed in \cite{vercleyen_slingerland_Rings} and an online database of the classification of low rank fusion rings can be found in \cite{gert_vercleyen_anyonwiki}.

It is often helpful to consider the lines in fusion diagrams such as figure \ref{fig:ChainDiagram} to be worldlines of particles. In this picture, the species/charge associated with different line segments in fusion diagrams can be thought of as the species of the particle or effective species of a bound state at that point in the worldline's history.

While this worldline picture will be invoked to provide intuition surrounding the mathematical machinery of fusion categories, one must keep in mind that the `time' parametrizing the flow of causation on these worldlines is not physical. That is, given that a fusion diagram describes the state of an anyonic chain at some given point in time, the direction in which causation flows on fusion diagrams is necessarily spacelike. It is in this sense that the `time reversal' in the final bullet point above is only schematic.

Having defined a fusion ring, it may seem as if we have ample mathematical structure in hand to define systems of lines which can fuse and split. However, if we examine the case of three lines fusing more closely, it becomes apparent that we require more machinery to discuss quantum mechanics on such systems in full. The machinery required for this purpose is that of a fusion category.

\subsection{Fusion categories}
To understand why additional mathematical structure is required, let us consider the following setup. We have three particles of species $a$, $b$, and $c$, with one measurement device measuring the species of the bound state produced by fusing $a$ and $b$, while the other device measures the bound state produced by fusing all three particles together. Suppose we make these measurements and obtain species types $d$ and $e$ for the product of the total and intermediate fusions respectively, collapsing the wavefunction in the measurement basis. This procedure can be interpreted as worldlines in a fusion diagram as demonstrated below.

\begin{figure}[H]
\centering
\includegraphics{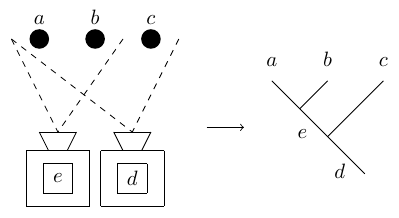}
\caption{A pair of measurements on three particles, and the corresponding eigenstate in the diagrammatic description.}
\end{figure}

Now, if we remove the measurement device detecting the product of the fusion of $a$ and $b$, and instead point it at particles $b$ and $c$, we may now ask what the state of the system is in the new measurement basis.

\begin{figure}[H]
\centering
\includegraphics{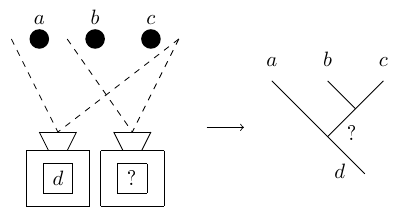}
\caption{A different measurement basis for the fusion of the same three particles.}
\end{figure}

We know that the system is in an eigenstate of the previous measurement basis, so the overall product of fusion must still be $d$. This is consistent with the associativity of the fusion ring describing the fusion of these particles into bound states. However, we do not know anything about the quantum state in the local $b\otimes c$ bound state measurement basis. In general, the state in this basis will be some superposition of the species which can be produced by fusing $b$ and $c$. Therefore, we parametrize the superposition by a set of complex coefficients $(F_{d}^{a,b,c})_{e,f}$ containing indices indicating the species of all particles involved:

\begin{equation}\label{eq:Fmove}
\includegraphics{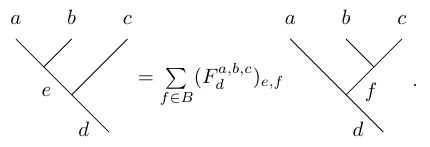}
\end{equation}

This change of basis is referred to as an $F$-move, and $(F_d^{a,b,c})_{e,f}$ is referred to as the $e,f$ entry of an $F$-matrix or as an $F$-symbol. Naturally, if $f$ can not be produced as a bound state of $b$ and $c$, then the coefficient $(F_d^{a,b,c})_{e,f}$ is zero. We now look for additional constraints which are natural to impose on these change of basis matrices.

In the interest of leaving non-unitary systems available for analysis, we avoid enforcing that the $F$-matrices be unitary.\footnote{With that said, note that non-unitarity of $F$-symbols does not imply non-unitarity of all anyonic chain models built from that fusion category - it doesn't even imply non-unitarity of the fusion category, because unitarity of $F$-symbols is not invariant under the gauge transformations outlined below.} Instead, let us consider the fusion of four particles. In this case, there are five bases in which the intermediate products of fusion can be measured, so we can expect to derive self-consistency conditions for $(F_d^{a,b,c})_{e,f}$. Suppose we start in a basis where the four particles fuse from left to right consecutively and we want to change to a basis where they fuse from right to left consecutively. One route between these bases involves one intermediate step, while another route involves two intermediate steps:

\begin{equation}
\includegraphics[width=\linewidth]{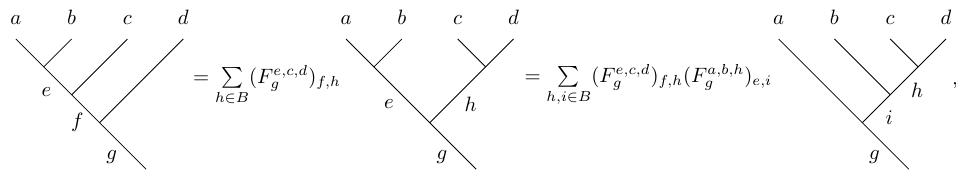}
\end{equation}

\begin{equation}
    \includegraphics[width=\linewidth]{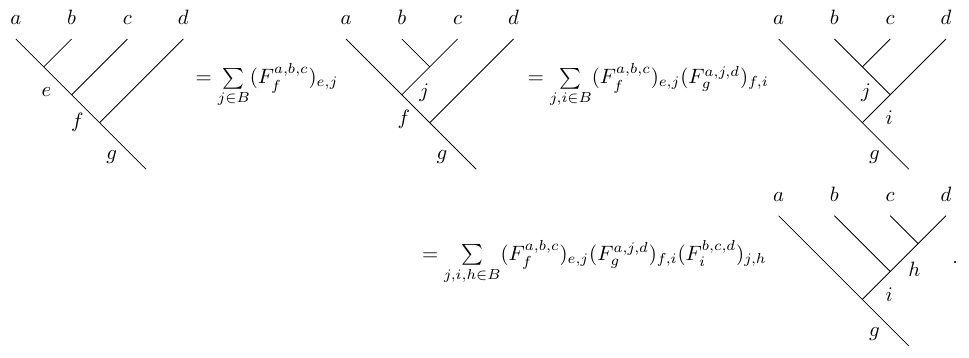}
\end{equation}

Because these diagrammatic expressions all represent the same state, we have a consistency condition - namely, for any given set of particles $a$ through $i$, the following set of equations must hold:
\begin{equation}\label{eq:pentagon}
    (F_g^{e,c,d})_{f,h}(F_g^{a,b,h})_{e,i}=\sum\limits_{j\in B}(F_f^{a,b,c})_{e,j}(F_g^{a,j,d})_{f,i}(F_i^{b,c,d})_{j,h}.
\end{equation}
These equations are known as the `pentagon equations' because the commutative diagram representation of the consistency condition is a closed cycle with five edges. The pentagon equations are invariant under the following gauge transformations:
\begin{equation}\label{eq:gauge}
    (F_{d}^{a,b,c})_{e,f}\rightarrow\frac{g_{a,b,e}g_{e,c,d}}{g_{b,c,f}g_{a,f,d}}(F_{d}^{a,b,c})_{e,f}.
\end{equation}
These gauge transformations can be thought of as arising from an ambiguity in the definition of our diagrams. Because the fusion diagrams represent basis vectors for states of our quantum system, we may choose to rescale all vertices where $a$ and $b$ fuse to produce $c$ by $g_{a,b,c}$. In order to respect this rescaling, the transformation law inherited by the $F$-symbols is precisely the gauge transformation \eqref{eq:gauge}.

We emphasize that the fusion diagram alone is not a vector in the Hilbert space, it is a basis vector. It must be multiplied by a basis-dependent component which transforms inversely to the fusion diagram under gauge transformations in order to define a vector in the Hilbert space.

A theorem due to MacLane \cite{maclane_1978} states that the pentagon equations, as consistency conditions for the associativity of the fusion of four lines, ensure that the fusion of $n>4$ lines is also consistent with associativity. Therefore, a self-consistent description of the fusion of lines is defined by a set of fusion rules which define a fusion ring and a solution to the pentagon equations defined by that fusion ring. This effectively defines a fusion category - supplementing the data of a fusion ring with a solution to the pentagon equations is said to `categorify' the fusion ring.

As categories $\mathcal{C}$ are defined by a set of objects and morphisms between them, the elements of $B$ in the definition of the fusion ring (i.e. the species of lines in the theory) are identified as the simple objects of the fusion category, $\text{Obj}(\mathcal{C})$. In practice, we use the notation $a\in\mathcal{C}$ to denote the species of lines in our theory because it is clear in context that we are referring to the objects in the fusion category. In the same way, we define $|\mathcal{C}|$ as the number of simple objects, i.e. the rank of the fusion ring.

The pentagon equations are in general a set of $\mathcal{O}(n^9)$ cubic polynomial equations in $\mathcal{O}(n^6)$ variables, up to $\mathcal{O}(n^3)$ gauge degrees of freedom \cite{Huang:2020lox}. As such, solving the pentagon equations for any given set of fusion rules is a non-trivial task. In fact, a solution is not guaranteed to exist for every fusion ring. The $F$-symbols for $\mathfrak{su}(2)_k$ fusion categories are known for all $k$ \cite{KirillovReshetikhin1988}, and the $F$-symbols for multiplicity free fusion rules up to rank-7 have been systematically solved \cite{gert_thesis}. Other systematic approaches to the problem of categorifying low-rank fusion rings are presented in \cite{aboumrad2022quantumcomputinganyonsfmatrix,Ardonne_2010,maeurer_2026_18760250,hagge2015geometricinvariantsfusioncategories,Liu_2022}, although the approach of \cite{gert_thesis} has produced the most complete dataset of low-rank fusion categories, compiled in an online database \cite{gert_vercleyen_anyonwiki}. 

Beyond this, apart from some special cases such as $\mathbb{Z}_{2n+1}$ Haagerup-Izumi fusion rules \cite{Huang:2020lox} or integral fusion categories \cite{alekseyev2026classificationintegralmodulardata}, the pentagon equations for fusion rules of rank greater than 8 have thus far generally withstood modern methods for solving polynomial systems of equations, simply due to the computational complexity involved. However, this complexity can be avoided in some cases as some higher rank fusion categories can be constructed from known low-rank fusion categories - this is the approach taken in \cite{maeurer2025computingcenterfusioncategory,maeurer2026fsymbolsrsymbolsdrinfeldcenter}.

Besides unitarity, there are layers of additional structure which may be added to that of a fusion category, starting with a braiding structure. This is necessary in the case of $2+1$d systems, where anyonic excitations are defined by unitary modular tensor categories \cite{KITAEV20062,Barkeshli_2019,Kitaev_2012}. However, in the present work, we consider $1+1$d systems. This is important because if we had chosen to consider an ambient two dimensional space, `exchange' would be a well-defined notion, and as such we would have to develop exchange statistics for our particles. By consequence of not requiring self-consistent exchange statistics, working with worldlines in $1+1$d allows one to consider a larger set of models than one could in $2+1$d. Specifically, we are free to consider non-commutative fusion rings. On the flip side, we are precluded from considering exchanges when building up a set of non-trivial local operators on our system.

With this in mind, we note that referring to our systems as an `anyonic chains' is generally a misnomer because anyons are defined by their exchange statistics. However, the title can be justified by considering anyonic chains as models of the boundary dynamics of a $2+1$d theory containing anyons - boundary excitations of such theories can contain species types which do not allow for self-consistent exchange statistics.

This may be stated more precisely as follows: it was shown in \cite{Kitaev_2012} that while the excitations in the bulk of a Levin-Wen model \cite{Levin_2005} (which are realizations of a Turaev-Viro TQFT \cite{kirillov2011stringnetmodelturaevviroinvariants}) are associated with objects in the monoidal center of a unitary tensor category $\mathcal{C}$ (which necessarily admits a notion of `exchange'), the excitations on a gapped boundary are associated with the simple objects of $\mathcal{C}$ itself. Further, whether gapless or not, $1+1$d systems with categorical symmetry can be lifted to the gapped boundaries of a Turaev-Viro TQFT \cite{Lootens2024}, via the SymTFT approach \cite{Bottini:2025hri}. As such, one may choose to think of anyonic chains as models describing dynamics of edge modes of some genuinely anyonic system as justification for the adjective `anyonic'.

\subsection{Fusion diagram calculus}
Having identified fusion categories as the relevant mathematical structure for describing the fusion of lines in $1+1$d theories, the calculus of fusion diagrams can be used to perform calculations using our quantum states.

The $F$-moves \eqref{eq:Fmove} defined by $F$-symbols are the first tool at our dispense for dealing with fusion diagrams, allowing us to perform transformations which reorder the chronology of fusion. 

The trace operation is defined by connecting coinciding lines on the top and bottom of the fusion diagram, and the trace of an individual line is defined as the quantum dimension of the species of that line:
\begin{equation}
    \includegraphics{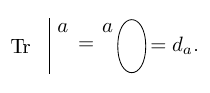}
    \label{eq:traceDef}
\end{equation}

The set of quantum dimensions of the particle types in the theory are defined by a solution to the algebraic equations given by fusion rules:
\begin{equation}\label{eq:qdims}
    d_ad_b=\sum_{c\in\mathcal{C}}N_{ab}^cd_c.
\end{equation}
However, this set of equations generally has multiple solutions for a given set of fusion rules. The relevant solution varies based on the categorification of the fusion ring being considered.  For a given categorification, the quantum dimensions of the objects in the fusion category are given by the $F$-symbols as follows:
\begin{equation}
    d_a=\left|(F_a^{a,\bar{a},a})_{1,1}\right|^{-1}.
\end{equation}

Based on \eqref{eq:qdims}, the quantum dimensions can be viewed as a generalization of the dimensions of the irreducible representations of a Lie algebra. However, they may be non-integer valued, reflecting the fact that they do not arise in a tensor product Hilbert space. For example in the Kitaev chain \cite{Kitaev_2001}, the edge modes have quantum dimension $\sqrt{2}$, reflecting the fact that a single qubit of information is encoded nonlocally across the pair of edge modes. Another way of interpreting the quantum dimensions is that, assuming total ignorance of the fusion channel prior to fusion, $d_c/d_ad_b$ is the probability of $c$ being produced when $a$ and $b$ are fused \cite{Trebst_2008}.

Note that the action of joining the two ends of the line when tracing can be more physically described as introducing an $\bar{a}$ line and enforcing that the pair of lines are created from and annihilated back into the vacuum object. As such, the process of tracing a fusion diagram can be thought of as associating an amplitude with the process whereby a set of worldlines are pairwise created from and annihilated into the vacuum, with half of the worldlines interacting with each other before the pairwise annihilation occurs.

The simplest quantum states have three lines meeting at a vertex - the states in our anyonic chains' Hilbert space are formed by gluing together such states. The bra dual to a ket is represented in diagrammatic language by mirroring the diagram about a horizontal axis. As such, the norm of a state is represented by joining these diagrams and tracing:
\begin{equation}
    \includegraphics{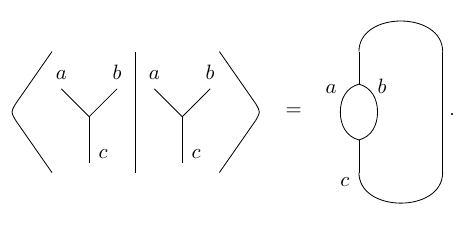}
    \label{eq:innerProduct}
\end{equation}

A common choice of normalization for this diagram is to evaluate it to $\sqrt{d_ad_bd_c}$. With this choice of normalization, the fusion calculus' resolution of identity and `bigon' relations are given as follows:
\begin{equation}
    \includegraphics{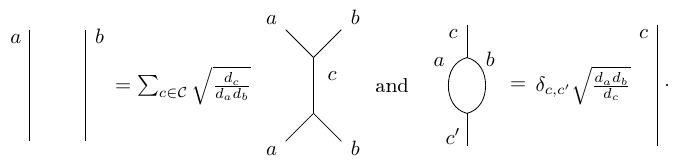}
    \label{eq:BigonAndIdentity}
\end{equation}
Without the additional structure of braiding, these relations and the $F$-moves provide enough machinery to evaluate any closed diagram. 

Note that all of these relations are invariant under gauge transformations, which transform basis covectors inversely to basis vectors. Each of these diagrammatic relations can be viewed as coming from joining the edges of basis vectors and covectors, so they pick up no prefactor under gauge transformations.

\subsection{States and operators on anyonic chains}
Recall that when we discuss anyonic chains, we are considering fusion diagrams with external lines of species $a\in\mathcal{C}$ which successively join together - we are free to choose $\mathcal{C}$ and $a$ when defining our model. The species of the resultant lines (which will also be elements of $\mathcal{C}$) are the degrees of freedom of the anyonic chain. In the anyonic chain literature, it is conventional to rotate fusion diagrams $45^\circ$ anti-clockwise, such that the causal flow of the fusion diagram is from top left to bottom right. Rotating the anyonic chain fusion diagram \ref{fig:ChainDiagram} in this way leaves us with the products of fusion (the dynamical variables) in a horizontal line:
\begin{figure}[H]
    \centering
    \includegraphics{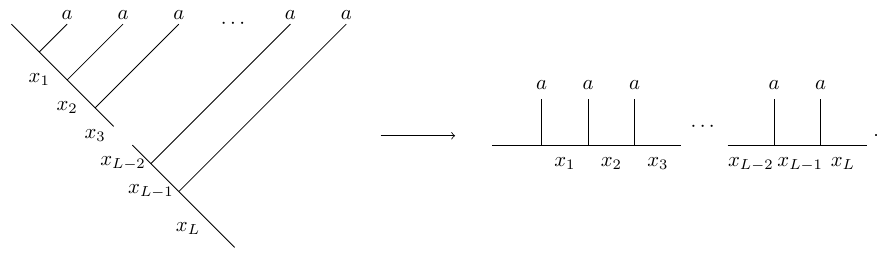}
    \label{fig:Rotation}
\end{figure}
This basis will be referred to as the canonical anyonic chain basis, and these basis vectors will be denoted $\ket{x_1,x_2,...,x_L}$. The fusion rules constrain the a priori $\left(\mathbb{C}^{|\mathcal{C}|}\right)^{\otimes L}$ Hilbert space by requiring physical states to have $x_i$ be an element of the fusion product of $x_{i-1}$ and $a$. Throughout this review, this constraint is represented by an adjacency graph where connected nodes are valid neighbors on the anyonic chain.

We also impose an additional constraint on the Hilbert space by taking the anyonic chain to have periodic boundary conditions. Periodic boundary conditions impose not only that the leftmost and rightmost external $a$ lines are neighbors, but also that $x_L$ is the species on the edge before $x_1$ on the chain. The fact that an anyonic chain's periodic boundary conditions includes the second of these two identifications is crucial to the existence of the topological symmetry discussed in section \ref{sec:topsym}.

Because the Hilbert space is constrained by fusion rules, the dimension of the Hilbert space grows as $d_a^L$ rather than $|\mathcal{C}|^L$. This can be understood from the schematic definition of quantum dimensions of objects as the generalization of the dimension of an irreducible representation. The analogous statement in a spin-1/2 chain setting can be obtained by considering the chain in terms of eigenvectors of $(\sum_{i=0}^j\vec{\sigma}_i)^2$ rather than the conventional $\sigma_{z,i}$-eigenvector basis. Although there is an infinite set of highest weight representations of $\mathfrak{su}(2)$, the Hilbert space grows as $d_{1/2}^L=2^L$ because of the `fusion rule' $1/2\otimes1/2=0\oplus1$.

The Hamiltonians discussed in this review are built from projection operators $\mathrm{P}^{a}_{b,i}$ acting on the anyonic chain. These projection operators project the fusion of two $a$ lines either side of site $i$ on the chain into the $b$ channel. As such, their projective effect is most clearly observed in a basis related by an $F$-move to the canonical basis depicted above
\begin{equation}
    \includegraphics{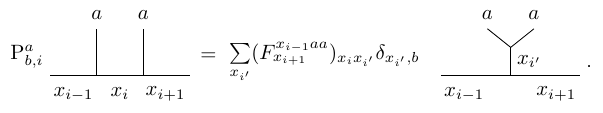}
    \label{eq:ProjectorDef}
\end{equation}
Changing basis back to the canonical anyonic chain basis, the local action of $\mathrm{P}^{a}_{b,i}$ is
\begin{equation}
    \includegraphics{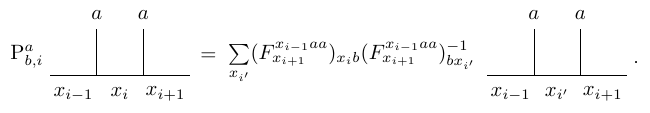}
    \label{eq:ProjectorComps}
\end{equation}
While these local operators are nearest-neighbor operators in terms of external lines, the dependence of $F$-symbols on the cumulative product of a fusion process makes them range-3 in the canonical anyonic chain basis.

These projection operators can be represented graphically by the following diagram:
\begin{equation}
    \includegraphics{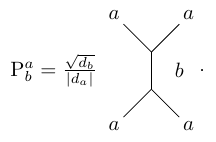}
    \label{eq:Intertwiner}
\end{equation}
 The local action of the operator on the $i$th site of the anyonic chain is defined by joining its lower edges with the pair of external edges either side of $x_i$. Then, using $F$-moves and \eqref{eq:BigonAndIdentity}, this action reduces to \eqref{eq:ProjectorComps}. This diagrammatic representation is natural in light of the operator being a projector, as it is simply the normalized outer product of the basis ket where two $a$ lines fuse to produce $b$ with its dual bra.
 
 Another sense in which this diagrammatic representation is natural is that the gauge-invariance of the projectors follows immediately from the gauge-invariance of the relations \eqref{eq:BigonAndIdentity}. This may also be observed in the $F$-symbol based definition \eqref{eq:ProjectorComps} by performing a gauge transformation on both the $F$-symbols and the fusion diagrams. Because our Hamiltonians will be built from these projectors, they will also be invariant under these gauge transformations.

 From this diagrammatic representation, it is clear that our projection operators are a basis for a subset of the intertwiners discussed in \cite{Lootens_2023}. Specifically, we consider the subset of intertwiners which don't change the species of our anyonic chain's external lines. It has been shown that fusion category symmetric nearest neighbor local Hamiltonians must be composed of intertwiners \cite{Lootens_2021}. The fusion categorical symmetry of the anyonic chains discussed in this paper is discussed in section \ref{sec:topsym}.

 Note that the diagrammatic representation \eqref{eq:Intertwiner} of the projection operators is just a term in the resolution of identity \eqref{eq:BigonAndIdentity}. Therefore we have
  \begin{equation}
    \sum_{b\in\mathcal{C}}\mathrm{P}^a_{b,i}=1.
    \label{eq:sumProjectors}
 \end{equation}

 In fact, because $\mathrm{P}^a_{b,i}=0$ if $b$ is not in the decomposition of $a\otimes a$ into simple objects, the sum in \eqref{eq:sumProjectors} may equivalently be carried out over $b\in a\otimes a$.

\subsection{Defining an anyonic chain}
With all of the above definitions in hand, we can now give an operational procedure for defining an anyonic chain.

First, we choose a fusion category from which to build the anyonic chain - that is, choose a set of line species and fusion rules which define a fusion ring, and find a solution to the pentagon equations \eqref{eq:pentagon} to categorify these fusion rules.

Next, we choose an object in the fusion category to associate with each external line of the anyonic chain. In this review, only anyonic chains with the same object on every external line are considered. As we have discussed, the choice of these objects defines the constrained Hilbert space of internal line species labels through the fusion rules. Associating object $a$ with all external lines and imposing periodic boundary conditions results in a constrained Hilbert space of dimension $\approx(d_a)^n$.

In order to define a non-trivial system via the present procedure, the objects on the external lines should be chosen to be `non-invertible' - that is, they should be chosen to have quantum dimension greater than 1. This is because the fusion of lines in the fusion diagram must be non-deterministic in order for the anyonic chain to have a Hilbert space of dimension greater than 1. Note that this is only true for anyonic chains defined via the procedure outlined here - a non-trivial system with external lines of dimension 1 can be defined by introducing defects, as in \cite{Bridgeman:2019kpu}.

With the physical Hilbert space completely defined, we may now define dynamics on the Hilbert space by specifying a Hamiltonian. In order to generalize the spin-1/2 Heisenberg-XXX and spin-1 Haldane models whose nearest-neighbor Hamiltonians can be written in terms of projectors into different spin channels, the prototypical examples of anyonic spin chains \cite{Feiguin:2006ydp,Gils_2013} have Hamiltonians defined by a linear combination of projection operators $\mathrm{P}^{a}_{b,i}$ (defined by \eqref{eq:ProjectorComps}). In keeping with this philosophy, we define the Hamiltonians of our anyonic chains as a general linear combination of such projectors, summed over all sites on the chain:
\begin{equation}
    \label{eq:Hamiltonian}
        H^{a}_{\mathcal{C}}=\sum_{b\in a \otimes a}\alpha_b\sum_{i=1}^{n}\mathrm{P}^{a}_{b,i},
\end{equation}
where in general $\alpha_b\in\mathbb{C}$. As has been discussed in previous sections, such a Hamiltonian is invariant under gauge transformations of fusion diagrams. One can study this operator as a function of the parameters $\alpha_b$, for example identifying critical phases/integrable points.

The procedure to define an anyonic chain may be summarized as follows:
\begin{itemize}
    \item Choose a fusion category $\mathcal{C}$, i.e.\ a fusion ring of objects $a,b,c,\dots$ and a solution $(F^{a,b,c}_{d})_{ef}$ to the pentagon equations \eqref{eq:pentagon}.
    \item Choose a non-invertible object $a\in \mathcal{C}$ as the external object in the anyonic chain and impose periodic boundary conditions. This leads to a constrained Hilbert space $V^{a}_{\mathcal{C}}$ of dimension $\sim d_a^L$.
    \item Define the general Hamiltonian $H^{a}_{\mathcal{C}}$ for the system as \eqref{eq:Hamiltonian} - i.e. a linear combination of projection operators $\mathrm{P}^{a}_{b,i}$ applied to all sites of the anyonic chain.
\end{itemize}

\section{General properties of anyonic chains}\label{sec:properties}
In this section we first discuss the existence of topological symmetries on periodic anyonic chains. This property holds in general and its existence has been noted since the study of anyonic chains began with the golden chain in \cite{Feiguin:2006ydp}, where this symmetry operator was used to justify the absence of certain relevant terms from the low energy effective field theory. 

Then, we discuss the existence of local operators which generate a representation of the Temperley-Lieb algebra. We show that this property holds when the chosen external object $a\in\mathcal{C}$ is such that $a\otimes a$ contains an object of quantum dimension 1 in its semisimple decomposition, generalizing the statement made in \cite{Blakeney:2025ext}, where the quantum dimension-1 object was required to be the identity object. The existence of this structure implies the existence of integrable models --- it was in the context of exactly solvable models that the algebra was first introduced \cite{TemperleyLieb1971} and is widely applied \cite{Pearce:1990ila,PASQUIER1987162,ikhlef2009temperley,Wang_1996}.

\subsection{Topological symmetry}\label{sec:topsym}
The existence of a topological symmetry is a general property of anyonic chains with periodic boundary conditions which follows directly from the language of fusion categories. For every simple object in the fusion category, there is an associated topological symmetry operator, $Y_c$, which creates a periodic line of species $c$ running parallel to the spine of the anyonic chain.  

Let us now derive the action of such an operator in the conventional anyonic chain basis. Having created a line running parallel to an anyonic chain, one may fuse it into the chain on one site using the resolution of identity, \eqref{eq:BigonAndIdentity}:
\begin{equation}
    \includegraphics{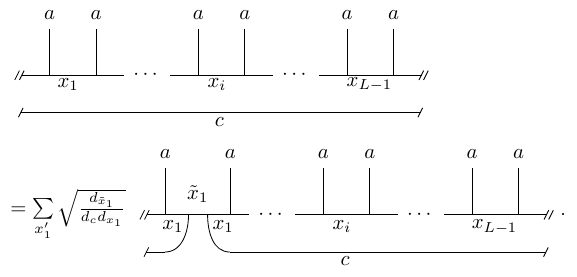}
    \label{eq:TopStep1}
\end{equation}

Recalling that the causal flow of fusion goes from top left to bottom right in the anyonic chain convention, we locally have $x_L$ and $a$ fusing to produce $x_1$, then a $c$ fusing with that $x_1$ to produce a $\tilde x_1$. In the standard fusion diagram notation, we can change to a basis where the $c$ edge fuses into the chain before the $a$ edge using the following $F$-move:
\begin{equation}
    \includegraphics{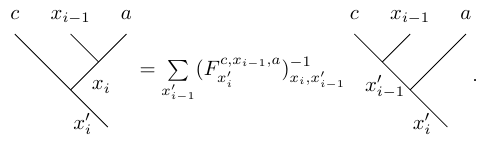}
    \label{eq:TopStep2}
\end{equation}

This change of basis transformation can be repeatedly applied to pass the vertex where $c$ fuses into the spine all the way around the chain, yielding
\begin{equation}
    \includegraphics{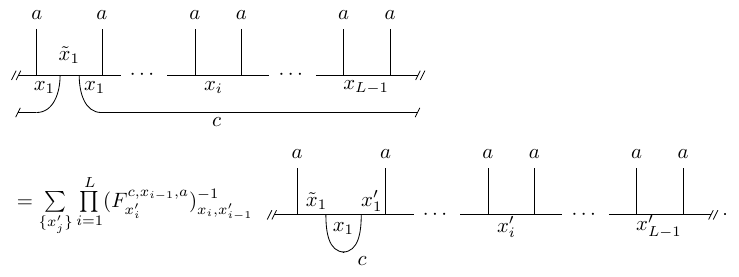}
    \label{eq:TopStep3}
\end{equation}

Now, the bigon identity \eqref{eq:BigonAndIdentity} allows the bubble on the first site of the chain to be rewritten as a $x_1'$ line with a prefactor of $\sqrt{d_cd_{x_1}/d_{\tilde{x}_1}}$, and enforces $x_1'=\tilde{x}_1$. Therefore, in the conventional anyonic chain basis, the topological symmetry operators are defined as:
\begin{equation}
    \includegraphics{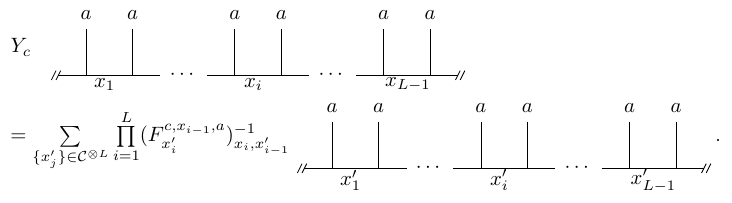}
    \label{eq:NonInvDef}
\end{equation}

In order to show that such an operator commutes with an anyonic chain Hamiltonian of the form \eqref{eq:Hamiltonian}, we will simply show that it commutes with any local projector. Making a choice of basis such that the external edges either side of $x_i$ are fused, such that $\mathrm{P}^{a}_{b,i}=\delta_{x_i',b}$. In this basis, the action of $Y_c$ may be deduced in the same way from its definition as an operator which creates a closed line of species $c$:
\begin{equation}
    \includegraphics{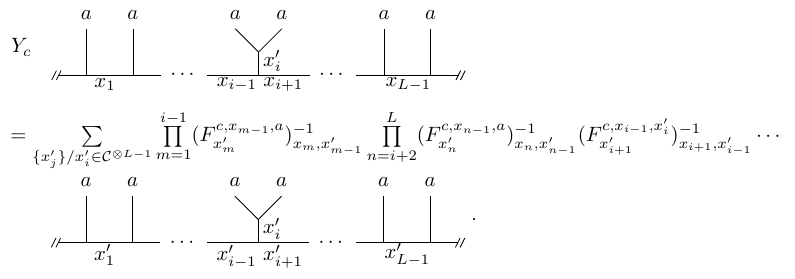}
    \label{eq:TopStep4}
\end{equation}

In this basis, it is clear that whether $\mathrm{P}^{a}_{b,i}=\delta_{x_i',b}$ is applied before or after $Y_c$, the effect is the same. Therefore, we immediately have
\begin{equation}
    [Y_c,\mathrm{P}^{a}_{b,i}]=0,
\end{equation}
so these operators are symmetries of Hamiltonians of the form \eqref{eq:Hamiltonian}.

\subsection{Temperley-Lieb structure}\label{sec:TL}
A set of operators $\{X_i\}$ are generators of a representation of the Temperley-Lieb algebra with scalar parameter $\delta_{TL}$ if they satisfy the following three relations:
\begin{equation}\label{eq:TLrels}
    \begin{aligned}
        &\bullet \hspace{0.25cm} X_i^2=\delta_{TL}X_i\\
        &\bullet \hspace{0.25cm} [X_i,X_j]=0\text{ where }|i-j|\geq2\\
        &\bullet \hspace{0.25cm} X_iX_{i\pm1}X_i=X_i
    \end{aligned}
\end{equation}
In general, the existence of this structure in a model implies integrability because of the close connection between the Temperley-Lieb algebra, the braid group, and the Yang-Baxter equation. In \cite{Blakeney:2025ext}, it was shown that for chains defined using a self-dual external object, the operators $\frac{1}{(F_a^{a,a,a})_{1,1}^{-1}}\mathrm{P}^{a}_{1,i}$ define a Temperley-Lieb algebra. In the present work, we make a more general statement. 

On any anyonic chain with external object $a$ such that $a\otimes a=b\oplus\cdots$, where $d_b=1$, the following operators generate a representation of the Temperley-Lieb algebra with $\delta_{TL}=\pm d_a$:
\begin{equation}\label{eq:TLX}
    X_i=\pm d_a \mathrm{P}^{a}_{b,i}.
\end{equation}

In the unitary $b=1$ case, the statement made in \cite{Blakeney:2025ext} follows from the statement made here. The proof of the present statement follows.

The projectors $\mathrm{P}^a_{b,i}$ square to themselves - this can be seen by using \eqref{eq:BigonAndIdentity}:

\begin{equation}
    \includegraphics{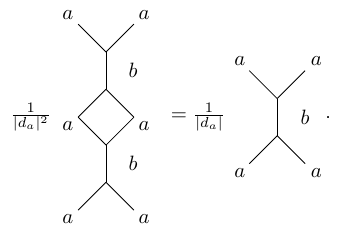}
    \label{eq:TL1}
\end{equation}
From this, it follows that any $X_i=\delta_{TL}\mathrm{P}^a_{b,i}$ satisfies the first Temperley-Lieb condition.

The second Temperley-Lieb condition is trivial in the diagrammatic notation, where order of operations is only apparent for operators acting on the same edges. In short, we immediately have
\begin{equation}
    \includegraphics{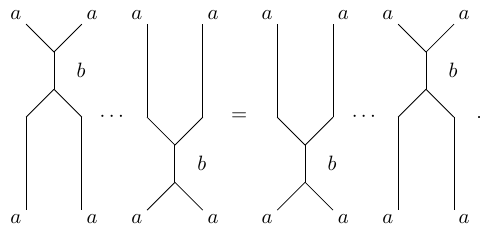}
    \label{eq:TL2}
\end{equation}
It follows from this that any set of operators proportional to the projectors $\mathrm{P}^a_{b,i}$ satisfy the second Temperley-Lieb condition.

The third Temperley-Lieb condition is less straightforward. Graphically, $X_iX_{i+1}X_i=X_i$ is
\begin{equation}
    \includegraphics{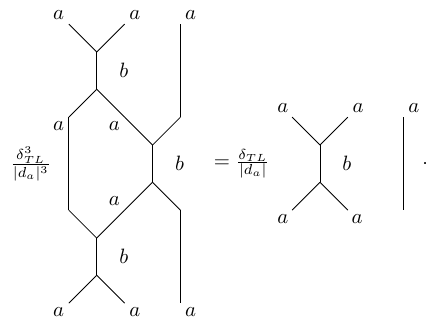}
    \label{eq:TL3New}
\end{equation}

The left hand side can be simplified by considering the following two subdiagrams:
\begin{equation}
    \includegraphics{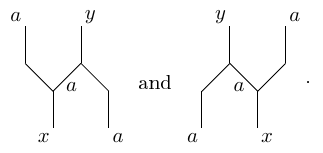}
    \label{eq:FundOps}
\end{equation}
The first of these can be simplified using \eqref{eq:BigonAndIdentity} and an $F$-move:
\begin{equation}
    \includegraphics{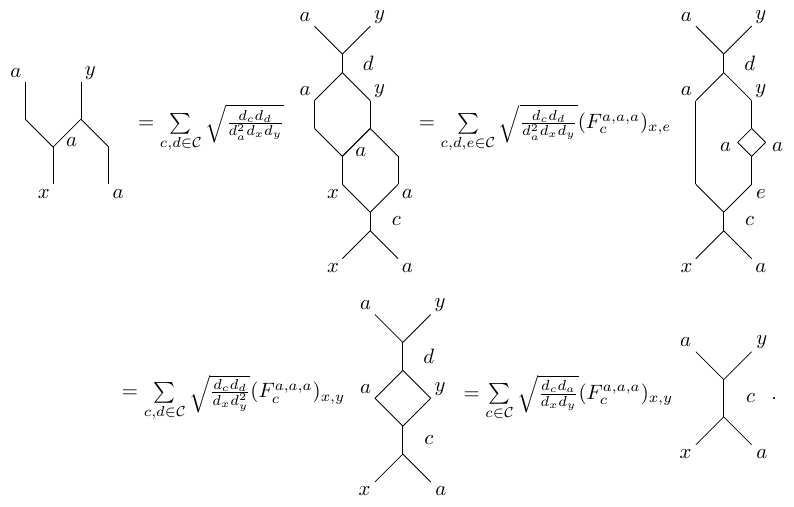}
    \label{eq:FundOp1}
\end{equation}
The second one can be simplified in the same way:
\begin{equation}
    \includegraphics{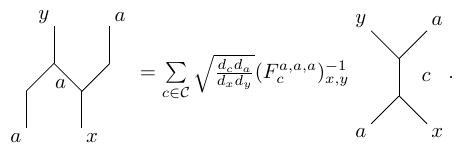}
    \label{eq:FundOp2}
\end{equation}

Taking $x=y=b$, we have that $c=ab$ because $d_b=1$ implies a deterministic outcome upon fusing. Now, stacking these two subdiagrams to produce the middle section of the left hand side of \eqref{eq:TL3New}, we have
\begin{equation}
    \includegraphics{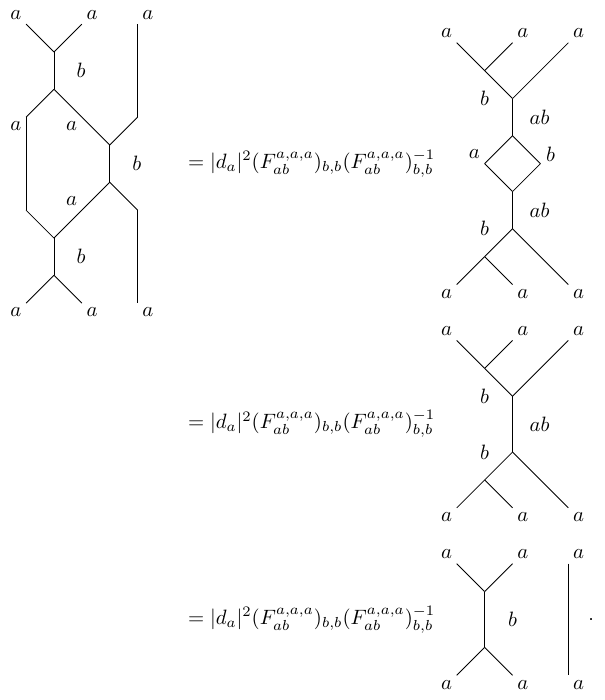}
\end{equation}
The $X_iX_{i-1}X_i=X_i$ condition simplifies similarly. Substituting this simplified diagram back into the left hand side of \eqref{eq:TL3New}, it follows that
\begin{equation}
    \delta_{TL}^2=\frac{1}{(F_{ab}^{a,a,a})_{b,b}(F_{ab}^{a,a,a})_{b,b}^{-1}}
\end{equation}
provides us with a set of Temperley-Lieb operators.

This condition can be simplified to $\delta_{TL}^2= d_a^2$. This can be seen by considering two distinct ways of carrying out the following simplification using \eqref{eq:BigonAndIdentity}:
\begin{equation}
  \includegraphics{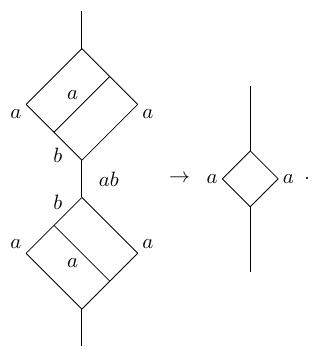}
\end{equation}
One can either use the resolution of identity twice, then the bigon relation, or one can apply an $F$-move to a splitting and fusing process, then two bigon relations and a resolution of identity. The upshot is that
\begin{equation}
    d_a^2(F_{ab}^{a,a,a})_{b,b}(F_{ab}^{a,a,a})_{b,b}^{-1}=1.
\end{equation}
Combining this with our finding from the first relation - that $X_i=\delta_{TL} \mathrm{P}^a_{b,i}$ satisfies the first Temperley-Lieb relation - we are provided with the Temperley-Lieb generators of \eqref{eq:TLX}.

Beyond supplying us with a set of integrable anyonic chain models, the existence of a Temperley-Lieb structure provides information about the criticality of the Temperley-Lieb integrable models. Specifically, because the Temperley-Lieb structure implies a correspondence to the XXZ chain with anisotropy parameter $\Delta = \delta_{TL}/2$, it was argued in \cite{Blakeney:2025ext} that if the external object has quantum dimension $|d_a|\leq2$, the corresponding Temperley-Lieb integrable models are critical, while the Temperley-Lieb models with $|d_a|>2$ are not. The same criterion was applied in the context of string-net ladder models in \cite{Schulz:2014cfa}, where the Temperley-Lieb parameter was the total quantum dimension of the fusion category rather than the quantum dimension of a particular object.

\section{Fibonacci fusion category}\label{sec:fib}
Having described the construction of anyonic chains in general, we focus on the simplest example of a fusion category producing a non-trivial anyonic chain: the Fibonacci fusion category Fib. The corresponding anyonic chain has been dubbed the \textit{golden chain} \cite{Feiguin:2006ydp}, which has several interesting properties. The construction of the model is pedagogically reviewed in \cite{Trebst_2008}, which also considers longer-range models \cite{Trebst2_2008}. Over the years Fib has served as a testbed for various constructions based on fusion categories. This includes the topological symmetry/topological defect correspondence \cite{Buican:2017rxc}, and the study of non-invertible symmetries in two-dimensional quantum field theory \cite{Huang:2021zvu}. In this section we review the construction of the golden chain from Fib and describe some of its properties. We go into a reasonable amount of detail in this section, in particular with regard to numerics, since all of these techniques generalize to the more complicated fusion categories discussed in later sections.

\subsection{The golden chain}
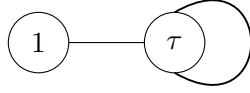
\begin{figure}[H]
\centering

\begin{tikzpicture}[scale=1]
\draw (0,0) circle [radius=0.4] node {$1$};
\draw (1.8,0) circle [radius=0.4] node {$\tau$};

\draw (0.4,0) -- (1.4,0);
\draw[line] (1.8,0.4) to[out=30,in=-30,looseness=5] (1.8,-0.4);

\end{tikzpicture}
\caption{Adjacency rules for the Fibonacci Hilbert space. Nodes represent simple objects. If there is an edge between a pair of objects then this is an allowed pair in $V_{\text{Fib}}$.}
\label{fig:allowedstateFib}
\end{figure}

 Fib contains two simple objects $1$ and $\tau$, which fuse via
\begin{equation*}
\tau \otimes \tau = 1 \oplus \tau.
\end{equation*}
For this fusion rule, there are two independent solutions to the pentagon identity \eqref{eq:pentagon}. For the unitary solution, a gauge can be chosen such that the only non-trivial $F$-symbol is given by
\begin{equation}
F^{\tau\tau\tau}_\tau = \begin{pmatrix} \varphi^{-1} & \varphi^{-1/2} \\ \varphi^{-1/2} & -\varphi^{-1} \end{pmatrix},
\end{equation}
where $\varphi = \frac{1+\sqrt{5}}{2}$ is the golden ratio. The non-unitary solution can be obtained using the Galois conjugation $\varphi \rightarrow -1/\varphi$ \cite{Ardonne_2011,Hsieh:2022hgi}. In both cases there is one non-trivial choice of external object, $a = \tau$, and two projectors $\mathrm{P}^{\tau}_1$ and $\mathrm{P}^{\tau}_\tau$, defined by \eqref{eq:ProjectorComps}. Only one of these operators is independent due to \eqref{eq:sumProjectors}, i.e.\ $\mathrm{P}^{\tau}_{1,i}+\mathrm{P}^{\tau}_{\tau,i}=1$. These operators act on the constrained Hilbert space $V_{\text{Fib}}=(\mathbb{C}^2)^{\otimes L}/\sim$ of dimension $\lceil\varphi^L\rceil$. Each local $\mathbb{C}^2$ is span$(\ket{1}, \ket{\tau})$, and the relation $\sim$ enforces the constraint forbidding adjacent `1' particles. This Hilbert space constraint can be summarized in the fusion graph in figure \ref{fig:allowedstateFib}.
Using the explicit representation in terms of $F$-symbols \eqref{eq:ProjectorComps} the unitary projector $\mathrm{P}^{\tau}_1$ can be calculated. After appropriate normalization it is exactly the golden chain
\begin{align}\label{eq:Hgolden}
 H_{\text{Golden}} &= J\varphi^{-1}\left(\sum_{i=1}^L \varphi^{-1/2} P_i X_{i+1} P_{i+2} + \varphi N_i P_{i+1}N_{i+2}+\varphi^{-1}P_iN_{i+1}P_{i+2} + P_i P_{i+1} P_{i+2}\right)\nonumber \\
 &\coloneqq J\sum_{i=1}^L \mathcal{H}_{i,i+1,i+2},
\end{align}
where we take periodic boundary conditions. $P$ and $N$ are projectors onto the $\tau$ and $1$ states respectively:
\begin{align}\label{eq:PNops}
&P \ket{\tau} = \ket{\tau}, &P \ket{1} = 0, \nonumber \\
&N \ket{\tau} = 0, &N \ket{1} = \ket{1}, 
\end{align}
and $X: \ket{\tau} \leftrightarrow \ket{1}$ is the standard Pauli matrix. For example, on a length-4 chain the allowed states are $\{\ket{1\tau1\tau}, \ket{1\tau \tau \tau}, \ket{\tau 1 \tau 1},  \ket{\tau 1 \tau \tau}, \ket{\tau \tau 1 \tau}, \ket{\tau\tau\tau 1}, \ket{\tau\tau\tau\tau}\}$ and in this basis the Hamiltonian \eqref{eq:Hgolden} reads

\begin{equation}
 H_{\text{Golden}}  =J\varphi^{-1}\left(
\begin{array}{ccccccc}
 2 (\varphi+\varphi^{-1}) &\varphi^{-1/2}& 0 & 0 &\varphi^{-1/2}& 0 & 0 \\
\varphi^{-1/2}& \varphi  & 0 & 0 & 0 & 0 &\varphi^{-1/2}\\
 0 & 0 & 2 (\varphi+\varphi^{-1}) &\varphi^{-1/2}& 0 &\varphi^{-1/2}& 0 \\
 0 & 0 &\varphi^{-1/2}& \varphi  & 0 & 0 &\varphi^{-1/2}\\
\varphi^{-1/2}& 0 & 0 & 0 & \varphi  & 0 &\varphi^{-1/2}\\
 0 & 0 &\varphi^{-1/2}& 0 & 0 & \varphi  &\varphi^{-1/2}\\
 0 &\varphi^{-1/2}& 0 &\varphi^{-1/2}&\varphi^{-1/2}&\varphi^{-1/2}& 4 \\
\end{array}
\right)\end{equation}

\subsection{Integrability, criticality, and topological symmetry}
The golden chain has several interesting properties, which we enumerate in this section.

\par\medskip
\noindent\textbf{Integrability.}\quad
First of all, the golden chain is an integrable model. By one definition, this means that the Hamiltonian \eqref{eq:Hgolden} sits in an infinite family of mutually commuting operators $Q_i$ on the constrained Hilbert space $V_{\text{Fib}}$:

\begin{equation}\label{eq:QiQj}
 [Q_i, Q_j] = 0, \qquad Q_2 \equiv H_{\text{golden}}.
\end{equation}
The first charge generates shifts along the lattice, i.e.\ $U = \exp(i Q_1)$. The second charge is the Hamiltonian and for the golden chain we have \cite{Corcoran:2024ofo}
\begin{align}\label{eq:Q3pauli}
Q_3&=\sum_{j=1}^Li P_j\left(\frac{X_{j+1}Y_{j+2}-Y_{j+1}X_{j+2}}{2}+ \varphi^{5/2}(P_{j+1}Y_{j+2}-Y_{j+1}P_{j+2})\right)P_{j+3},
\end{align}
where $Y$ is the second Pauli matrix. $Q_3$ can be obtained from $Q_2$ using a higher-range analog of the so-called boost operator \cite{Gombor:2021nhn}:
\begin{equation}\label{eq:Q3def}
Q_3 = \sum_{i=1}^L [\mathcal{H}_{i,i+1,i+2},\mathcal{H}_{i+1,i+2,i+3}+\mathcal{H}_{i+2,i+3,i+4}].
\end{equation}
With the definition \eqref{eq:Q3def} it is widely expected that the Reshetikhin condition $[Q_2, Q_3] = 0$ is sufficient to ensure the commutation of a full set of commuting charges \eqref{eq:QiQj}. We note that the density in \eqref{eq:Q3def} is range-5, whereas the explicit expression in terms of Pauli matrices is range-4. This is due to the fact that upon forming the operator \eqref{eq:Q3def} and projecting to allowed states, a range 4 density is sufficient to produce the overall operator $Q_3$. This already hints that some of the usual integrability objects should be modified as the product form of the Hilbert space is destroyed. We discuss this in section \ref{sec:rydberg}.

One other way to prove the integrability of the golden chain is by constructing an explicit map of $e_i \coloneqq \varphi \mathcal{H}_{i,i+1,i+2}$ onto the $\mathfrak{su}(2)_3$ RSOS model \cite{Andrews:1984af}. The operator $e_i$ satisfies the Temperley--Lieb algebra\footnote{Here $e_i e_{i+1}e_i$ acts on an open Fib-constrained Hilbert space of length 4.}
\begin{align}\label{eq:TL}
    &e_i^2=\varphi e_i, \notag\\ &e_ie_{i\pm1}e_i=e_i, \notag\\ &e_ie_j=e_je_i\text{ where }|i-j|\geq2,
\end{align}
which is also indicative of its integrability. Alternatively, one can see integrability by taking the Hamiltonian limit of Baxter's hard square model at the critical point \cite{Fendley_2004}. 
\par\medskip
\noindent\textbf{Criticality.}\quad
The golden chain \eqref{eq:Hgolden} provides a lattice description of two CFTs in the continuum limit. Listing the eigenvalues of the lattice model in increasing order $E_0 < E_1 < \dots$,
the energy gap of critical systems behaves as
\begin{equation}\label{eq:gap}
\Delta E_i \coloneqq E_i - E_0 = v \frac{\Delta_i}{L^{z}} + \dots,
\end{equation}
where $v$ is a non-universal constant, $\Delta_i$ is the dimension of the corresponding field theory operator, and $z=1$ for a CFT. In particular, for a CFT we expect
\begin{equation}\label{eq:loggap}
\log  \Delta E_i \sim - \log L, \qquad L \rightarrow \infty.
\end{equation}
 Furthermore, Let $\ket{\psi_0}$ be the ground state of a spin chain corresponding to a CFT in the continuum limit with central charge $c$. One can split the overall Hilbert space $(\mathbb{C}^n)^{\otimes L} = (\mathbb{C}^n)^{\otimes \ell} \otimes (\mathbb{C}^n)^{\otimes L-\ell} \coloneqq V^\ell \otimes V^{L-\ell}$, and compute the entanglement entropy
\begin{equation}
 S_\ell \coloneqq \text{tr}_{V^\ell} (\ket{\psi_0}\bra{\psi_0}).
\end{equation}
 It can be shown that this quantity takes the asymptotic form \cite{Calabrese:2004eu}
\begin{equation}\label{eq:EEdef}
 S_\ell \sim \frac{c}{3} \log\left(\frac{L}{\pi}\sin\frac{\pi \ell}{L}\right), \qquad L\rightarrow \infty.
\end{equation}
It turns out that the golden chain \eqref{eq:Hgolden} corresponds to a CFT in both the antiferromagnetic ($J<0$) and ferromagnetic ($J>0$) regimes. For $J<0$ it corresponds to the tri-critical Ising model with $c = 7/10$, and for $J > 0$ (so the energy spectrum is read in reverse order) it corresponds to the three-state Potts model with $c=4/5$. We give numerical details and subtleties which arise for anyonic chains in section \ref{sec:goldennumerics}. 

\par\medskip
\noindent\textbf{Topological symmetry.}\quad
The golden chain commutes with the non-local operator $Y_\tau$ \cite{Feiguin:2006ydp} defined in general in section \ref{sec:topsym}, also discussed in detail in \cite{Gils_2013}. This operator satisfies the fusion algebra:
\begin{equation}
Y_\tau^2 = 1 + Y_\tau.
\end{equation}
Therefore the Hilbert space splits into topological sectors labeled by the eigenvalues of this operator 
\begin{equation}
y = \varphi, -\varphi^{-1},
\end{equation}
corresponding respectively to the global fusion channel of the chain $1$ or $\tau$. In the continuum limit, $Y_\tau$ descends to a topological defect of the associated CFT, and this severely restricts the allowed relevant perturbations, underlying the stability of the critical phase \cite{Pfeifer:2010xi, Buican:2017rxc}.

\subsection{Numerics}\label{sec:goldennumerics}

 For a precise mapping of the golden chain to the aforementioned CFTs, one can shift/rescale the spectrum such that the energies $E_i$ correspond to scaling dimensions $\Delta_i$ in the CFT via\footnote{After reducing to appropriate momentum sectors, see for example \cite{Feiguin:2006ydp}.}
\begin{equation}\label{eq:EiCFT}
E_i = E_\infty L + \frac{v}{L} \left(\Delta_i - \frac{c}{12}\right).
\end{equation}
 In practice this requires knowing the $\Delta_i$ beforehand, so a good first check for criticality is:
 \begin{itemize}
  \item Compute the gap $\Delta E_1$ and verify \eqref{eq:loggap}. 
  \item Compute the half-chain entanglement entropy $S_{L/2}$ and verify \eqref{eq:EEdef}, while extracting $c$.
 \end{itemize}
 
 For anyonic chain Hamiltonians it is possible to compute the energy gap $\Delta E_1$ to high values of $L$ using the density matrix renormalization group (DMRG) approach \cite{PhysRevLett.69.2863, RevModPhys.77.259}. This approach bypasses the exponential growth of the spin-chain Hilbert space $(\mathbb{C}^n)^L$, and computes $E_0$ and $\ket{\psi_0}$ by assuming the ground state is well-approximated by matrix product state ansatz with bond dimension $D$. DMRG then performs an iterative local energy minimization, sweeping back and forth $m$ times along the chain until convergence. The first excited state $E_1, \ket{\psi_1}$ can be obtained by running the algorithm on a Hamiltonian orthogonalized with respect to $\ket{\psi_0}$. Due to the nature of this algorithm, it necessarily runs on models with a Hilbert space of product form. To adapt this algorithm to anyonic chains, one approach is to add an energy penalty to states which do not obey the fusion category constraint. For example, for the golden chain we can run the algorithm on 
 \begin{equation}\label{eq:Hdeformed}
 H_{\text{Golden}}' =  H_{\text{Golden}} + U \sum_{i=1}^L N_i N_{i+1}
 \end{equation}
 While this is the method used in this paper, we note that approaches more tailored to anyonic systems have been put forward \cite{Pfeifer:2015pna, 2024zndo..10654900V, Devos:2025yoj}. In practice, one can compute $E_0, E_1$, and hence the gap $\Delta E_1$ by picking appropriate values for the maximum bond dimension $D$, number of sweeps $m$, and repulsion parameter $U$.\footnote{Increasing $D$ and $n$ typically increases the accuracy of the algorithm, but also the runtime. In some cases the low-energy physics is insensitive to the repulsion term and so $U=0$ is possible.} \footnote{In certain cases the ground state is degenerate due to a symmetry, in which case the gap is defined with respect to a higher energy level.}
 
Using the ground state $\ket{\psi_0}$ obtained from DMRG it is also possible to extract the entanglement entropies $S_\ell$. We note that in the case of anyonic chains there is an apparent difficulty in defining \eqref{eq:EEdef} due to the non-factorized nature of the Hilbert space. Because of this it is not possible to define the partition of the Hilbert space or partial trace. Despite this, it is customary to define $S_\ell$ as the entanglement entropies computed using the deformed models such as \eqref{eq:Hdeformed}.\footnote{Equivalently, one can compute the entanglement entropy of the model \eqref{eq:Hgolden} acting on the Hilbert space $(\mathbb{C}^2)^L$, so that rows/columns corresponding to disallowed states are simply zero. In this case it is necessary to discard zero eigenvalues of the reduced density matrix.}

\begin{figure}[t]
  \centering
  \begin{minipage}[t]{0.49\textwidth}
    \centering
    \includegraphics[width=\linewidth]{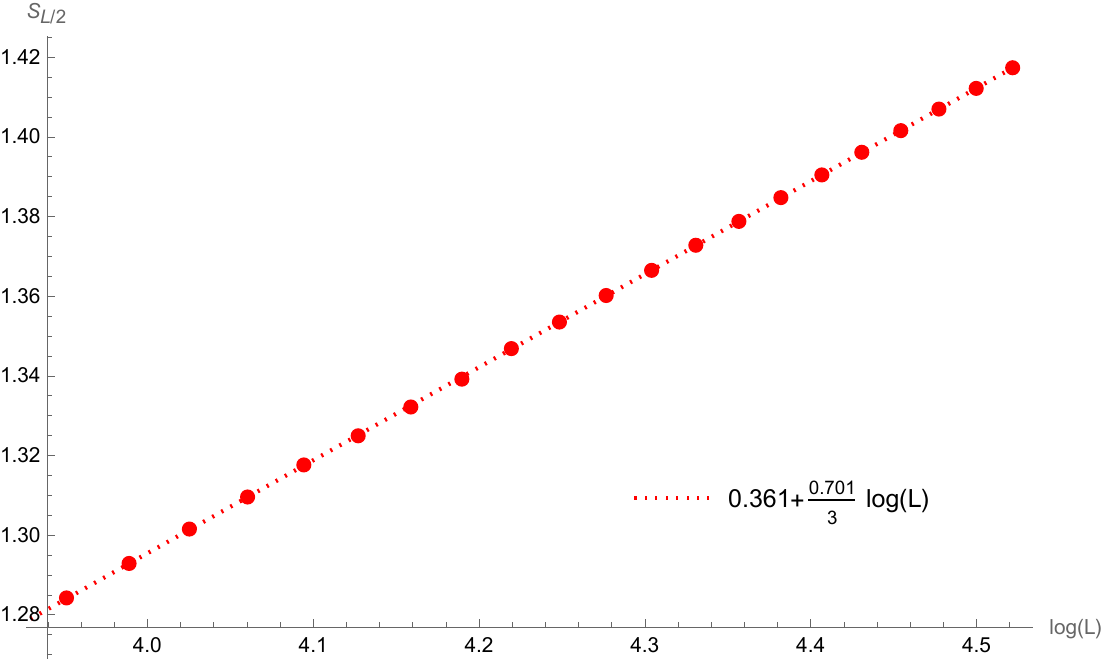}
  \end{minipage}\hfill
  \begin{minipage}[t]{0.49\textwidth}
    \centering
    \includegraphics[width=\linewidth]{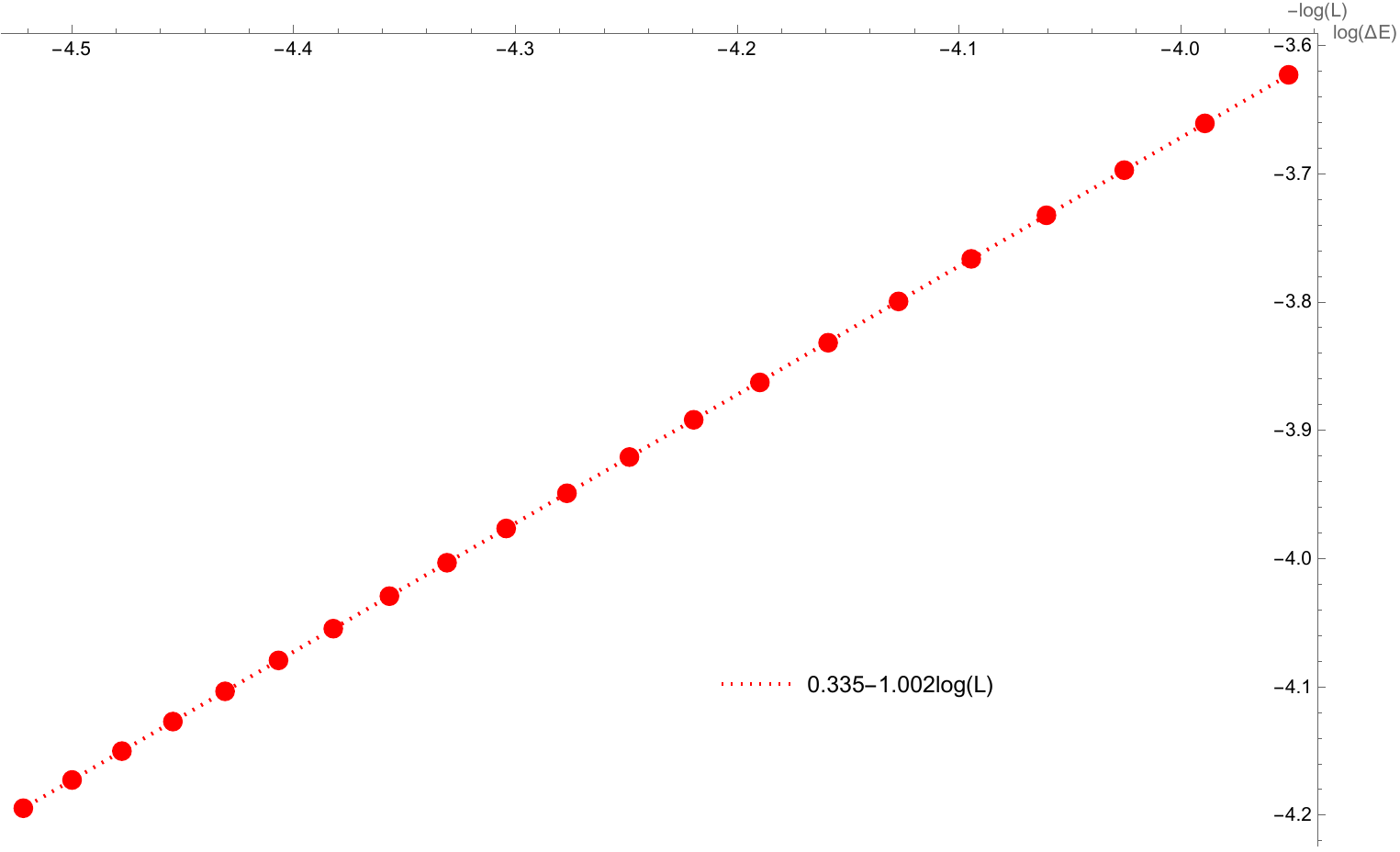}
  \end{minipage}

  \caption{Half-chain entanglement entropy and gap for the anti-ferromagnetic case $J<0$. For both plots we take even $52\leq L\leq 92$.}
  \label{fig:af-ee-gap}
\end{figure}

\begin{figure}[t]
  \centering
  \begin{minipage}[t]{0.49\textwidth}
    \centering
    \includegraphics[width=\linewidth]{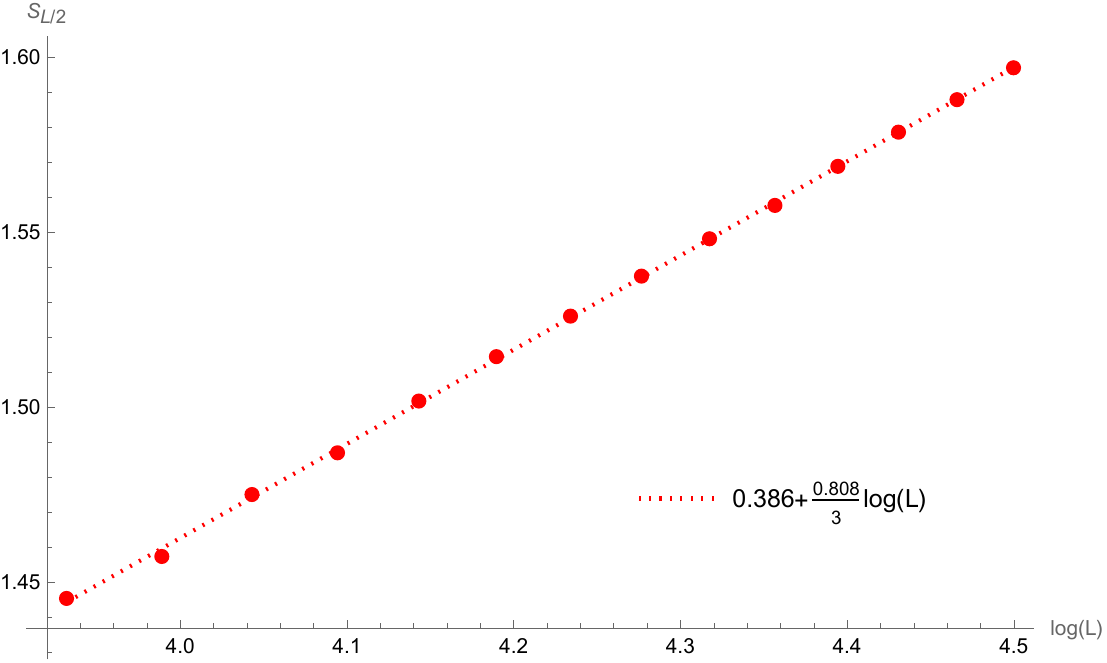}
  \end{minipage}\hfill
  \begin{minipage}[t]{0.49\textwidth}
    \centering
    \includegraphics[width=\linewidth]{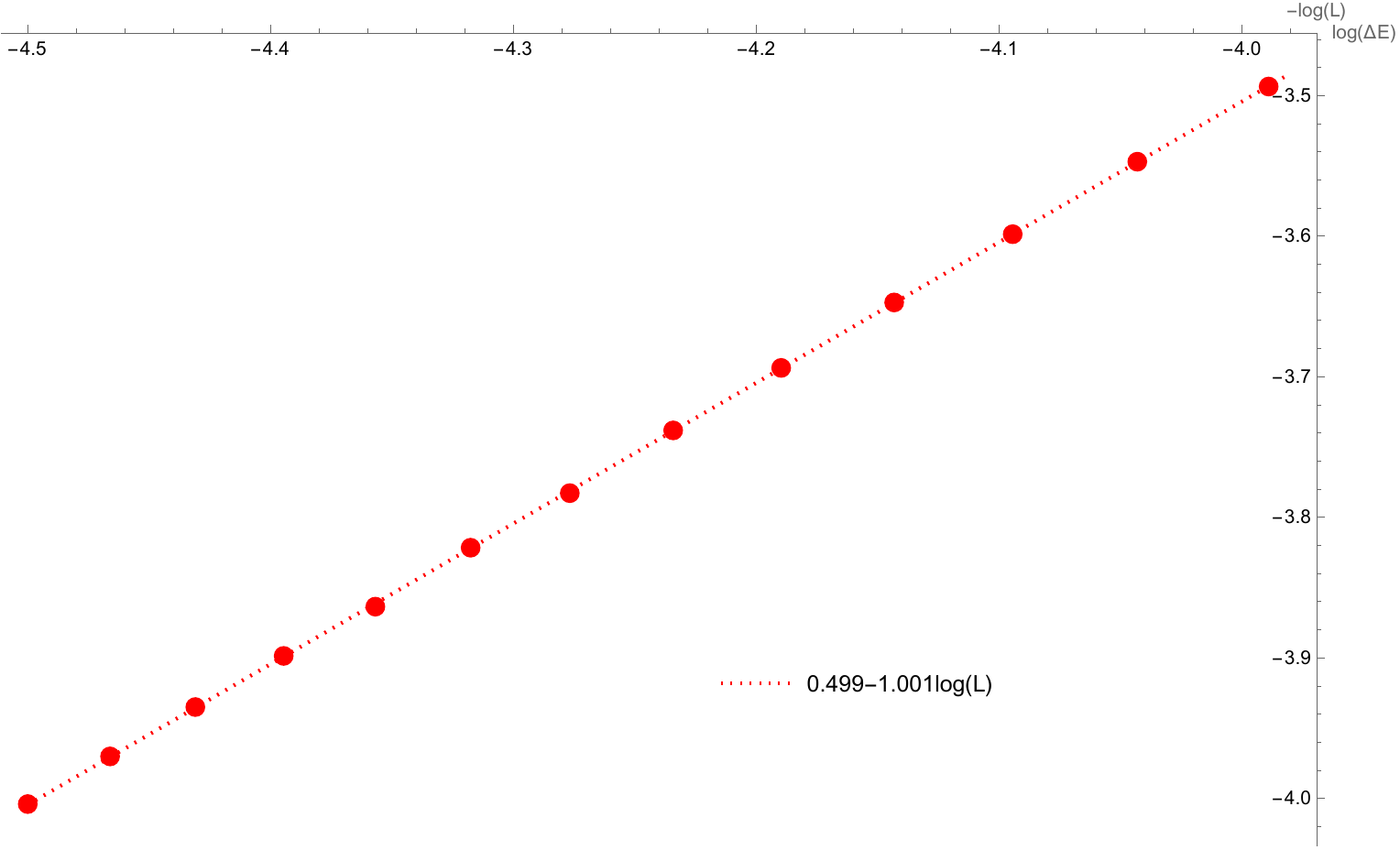}
  \end{minipage}

  \caption{Half-chain entanglement entropy and gap for the ferromagnetic case $J>0$. For both plots we take $L\equiv 0$ mod $3$, with $51\leq L \leq 90$.}
  \label{fig:f-ee-gap}
\end{figure}

We present the calculation of the quantities $\Delta E_1$ and $S_{L/2}$ for \eqref{eq:Hgolden} with $J<0$ and $J>0$ in figures \ref{fig:af-ee-gap} and \ref{fig:f-ee-gap}. The DMRG calculations were performed using ITensor \cite{Fishman_2022} on the penalized Hamiltonian \eqref{eq:Hdeformed}. We used a penalty $U=4$, maximum bond dimension $D=1000$, and $m=50$ sweeps. For $J<0$ we consider $L$ even, and see there is strong evidence for critical behavior with both \eqref{eq:loggap} and \eqref{eq:EEdef} verified, with $c\sim 7/10$. For $J>0$ it is necessary to consider $L\equiv 0$ mod 3 due to an emergent $\mathbb{Z}_3$ symmetry in the continuum limit.\footnote{For odd $L$ we compute $S_{\lceil L/2\rceil}$.} In this case we also see evidence of critical behavior with $c \sim 4/5$. We stress that these are only necessary conditions for criticality, and that for a CFT complete matching it is necessary to map the energy levels $E_i$ to scaling dimensions $\Delta_i$ via \eqref{eq:EiCFT}.

\par\medskip
\noindent\textbf{Non-Hermitian models.}\quad
In the previous sections we have focused on the unitary solution of the Fibonacci pentagon equations, which leads to the golden chain. There is another independent solution, where the non-trivial $F$-symbol is given by
\begin{equation}
F^{\tau\tau\tau}_\tau = \begin{pmatrix} -\varphi & -i\varphi^{1/2} \\ -i\varphi^{1/2} & \varphi \end{pmatrix}.
\end{equation}
The corresponding anyonic chain is non-Hermitian, and has been dubbed the Lee-Yang chain \cite{Ardonne_2011}. Several of the algebraic properties of the golden chain carry over to this model. It is still an integrable model, and satisfies the Temperley-Lieb algebra with parameter $\delta = -\varphi^{-1}$.

However, the physics of this model is very different with respect to the golden chain. For $J<0$ the model is described in the continuum limit by the non-unitary minimal model $\mathcal{M}(3,5)$, with $c=-3/5$. For $J>0$ the model is the non-unitary Lee-Yang theory $\mathcal{M}(2,5)$ with $c=-22/5$. In \cite{Ardonne_2011} this is shown by directly mapping the energy spectra to the CFT scaling dimensions using \eqref{eq:EiCFT}. We note that a direct verification of the central charge using entanglement entropy formulas such as \eqref{eq:EEdef} is subtle in the non-Hermitian case. In \cite{Bianchini:2014uta, Bianchini:2015uea} the use of an `effective' central charge $c_{\text{eff}}=c-24\Delta > 0$ in \eqref{eq:EEdef} is proposed, where $\Delta$ is the lowest holomorphic conformal dimension in the theory. Details and explicit numerics in this case can be found in \cite{PhysRevLett.119.040601}.

\section{Rydberg-blockaded chains and integrability}\label{sec:rydberg}
There are several interesting models acting on non-factorizable Hilbert spaces beyond anyonic chains. One of the most prominent examples are models on so-called \textit{Rydberg-blockaded} chains. These are models acting on a spin-$\frac{1}{2}$ Hilbert space $(\mathbb{C}^2)^{\otimes L} = \text{span}(\ket{\uparrow}, \ket{\downarrow})^{\otimes L}$. The Rydberg-blockade is the constraint that neighboring down spins on the chain are forbidden. This is precisely the constraint arising from the Fibonacci fusion category with the replacement $1\rightarrow \ket{\downarrow}$ and $\tau \rightarrow \ket{\uparrow}$. Thus the dimension of the Hilbert space scales as $\varphi^L$ as before. However, in this case the Hamiltonian need not be a fusion category projector. Therefore there are a whole host of models to study, with many interesting properties. One of these is the so-called PXP model:
\begin{equation}
H_{\text{PXP}} = -\sum_{i=1}^L P_i X_{i+1}P_{i+2},
\end{equation}
which possess several intriguing properties such as quantum many body scars \cite{Turner_2018, Turner:2018yco} and near-integrable deformations \cite{Khemani:2019huc}. Other models of note include the two-parameter hard boson model of \cite{Fendley_2004}, which contains a one-parameter integrable line. This integrable line itself includes the golden chain, discussed in the previous section, as a critical point. A closely related model was proposed by Lesanovsky  \cite{Lesanovsky_2012, PhysRevA.86.041601}, which despite being non-integrable admits several exact results for spectral data \cite{Mark_2020}. Higher-range models on a Rydberg-blockaded Hilbert space include the constrained XXZ model \cite{constrained1, constrained2, constrained3} which furnishes a range-$r$ integrable model for $r=2,3,\dots$. We note that the case $r = 2$ corresponds to the undeformed XXZ chain, $r=3$ is a model on a Rydberg-blockaded chain, and $r\geq 3$ deals with more general constraints.

It is clear that there is a vast array of interesting models on constrained Hilbert spaces, even just looking at the Rydberg-blockaded case. In particular, there are many \textit{integrable} models. While there is no rigorous definition for quantum integrability, commonly accepted definitions include Poissonian level statistics or the existence of an infinite number of independent operators commuting with the Hamiltonian. In this paper we will focus on the latter definition. In this case the formulation of integrability typically relies on the factorized nature of the Hilbert space, in order to define the usual objects of integrability such as the Lax operator, Monodromy matrix, and $R$-matrix. However, we have seen that there exist integrable models on non-factorizable Hilbert spaces. In this section we show how it is possible to adapt the usual integrability objects to the Rydberg-blockaded case \cite{Corcoran:2024ofo}, which generalizes naturally to more general constraints. We begin by viewing the models on the factorized space $(\mathbb{C}^2)^{\otimes L}$, equipped with an appropriate projection operator $\Pi$. We then show how the formalism of medium-range integrability \cite{Gombor:2021nhn} can be adapted to this case. Finally, we describe
how the boost-operator method \cite{DeLeeuw:2019gxe} for classifying integrable models can be straightforwardly adapted to the constrained Hilbert space, and enumerate the range-3 and range-4 models in the case of Rydberg atoms. We will apply this method in later sections to identify integrable models on spaces with more general fusion category constraints. We note that there have been recent studies of PXP-like models on more general constrained Hilbert spaces \cite{Mukherjee:2021exb, Mohapatra:2023ive}, not all related to fusion categories, and the methods of these sections also apply to these cases.

\subsection{Constrained Hilbert spaces}
In this section we briefly discuss the view of constrained Hilbert spaces as the projection of factorized Hilbert spaces, which is useful for the modification of the usual formulation of medium-range integrability \cite{Gombor:2021nhn}. We focus on the Rydberg blockade, although the arguments also apply to other local constraints. We discuss some subtleties of this projection, such as the definition of `range' and operatorial gauge symmetries. These are relevant when it comes to the classification of integrable models on the constrained Hilbert space.

\par\medskip
\noindent\textbf{Hilbert space and operators.}\quad
We consider a spin chain with local Hilbert space $V_j$ and length $L$. The physical space that we want to consider is a subspace of the total Hilbert space $V=\bigotimes_{j=1}^L V_j$. Specifically, we assume that there is a projection operator $\Pi:V\rightarrow V$ such that
\begin{align}
&\Pi |\mathrm{physical}\rangle =  |\mathrm{physical}\rangle ,
&& \Pi |\mathrm{unphysical}\rangle = 0,
\end{align}
so that our physical Hilbert space is given by
\begin{align}
V_\Pi  \coloneqq\Big\{ \ket{v} \in V \, \Big| \, \Pi \ket{v} =\ket{v}\Big\}.
\end{align}
In the Rydberg atom case we have $V_j = \mathbb{C}^2$ and the range-2 local constraint
\begin{equation}\label{eq:constraint}
\Pi = \prod_{i=1}^L \Pi_{i,i+1} = \prod_{i=1}^L (1-N_iN_{i+1}),
\end{equation}
where $N_i$ is the projector onto $\ket{\downarrow}_i$, so that $\text{dim} V_\Pi \sim \varphi^L$. We consider operators $\mathcal{O}:V\rightarrow V$ which satisfy
\begin{align}
 [\mathcal{O} ,\Pi] = 0.
\end{align}
This is a natural constraint since this implies that physical and unphysical states do not mix under action of such an operator and the operator can naturally be restricted to $V_\Pi$
\begin{align}
\mathcal{O}^{(\Pi)}\coloneqq \Pi\mathcal{O}\Pi = \Big( \Pi\mathcal{O} = \mathcal{O}\Pi\Big).
\end{align}

\par\medskip
\noindent\textbf{Range and gauge symmetry.}\quad
The Rydberg projector \eqref{eq:constraint} destroys the product form of the Hilbert space $V$. Because of this, it is subtle to define the notion of range of an operator. For example, on the constrained Hilbert space $V_{\Pi}$ there can be operator densities of apparently different range, which lead to the same total operator on the constrained Hilbert space $V_{\Pi}$. One example of this phenomenon is
\begin{equation}\label{eq:example}
\Pi\left[\sum_{i=1}^L P_i N_{i+1}P_{i+2}\right]\Pi = \Pi \left[\sum_{i=1}^L -P_{i}P_{i+1}+P_{i}\right]\Pi,
\end{equation}
so that this operator can apparently be equivalently be represented in terms of a range 3 or a range 2 operator density. \eqref{eq:example} is just the statement that on the Rydberg-blockaded chain one can count the number of excitations $\ket{\uparrow\downarrow\uparrow}$ directly with the operator $PNP$, or indirectly by counting the total number of up spins $\ket{\uparrow}$, and subtracting the number of adjacent pairs $\ket{\uparrow\uparrow}$. Identities such as \eqref{eq:example} lead to a large gauge freedom in defining Hamiltonian density operators on the constrained space. Such identities can be non-trivial and must be taken into account when counting the number of independent operators at each range.

For the purposes of classifying integrable models it is useful to have a consistent definition of range for operators on the constrained Hilbert space $V_\Pi$. To this end, it can be shown that any operator on $V_\Pi$ can be written as a sum of operators of the form
\begin{align}\label{eq:range}
\mathcal{O}_r^{(\Pi)}\coloneqq \Pi\Big[ \sum_{i=1}^L P_{i}  \mathcal{O}_{i+1,i+2,\dots,i+r-2}  P_{i+r-1}\Big] \Pi,
\end{align}
where $ \mathcal{O}_{i+1,i+2,\dots,i+r-2}$ is a usual operator density of range $r-2$. We will define the range of a constrained operator $\mathcal{O}^{(\Pi)}$ to be the minimal $r$ such that $\mathcal{O}^{(\Pi)}$ admits a representation as
\begin{equation}\label{eq:range2}
\mathcal{O}^{(\Pi)} = \sum_{r'=1}^r \mathcal{O}_{r'}^{(\Pi)}.
\end{equation}

\subsection{Integrability}
Due to the projection \eqref{eq:constraint}, there are no interesting nearest-neighbor models on the Rydberg-blockaded Hilbert space, and so we must consider models of range $r\geq 3$. It is therefore necessary to consider the \textit{medium-range} formulation of integrability \cite{Gombor:2021nhn}. Medium-range models are those with a finite interaction range $r\geq 3$. This is in contrast to nearest-neighbor models with $r=2$, and long-range models with $r=\infty$. In this section we give a lightning review of this formulation of integrability, and how this can be modified to the case of constrained Hilbert spaces

\par\medskip
\noindent\textbf{Medium-range integrability.}\quad
We start with a range-$r$ Hamiltonian on a Hilbert space $V = (\mathbb{C}^2)^{\otimes L}$
\begin{equation}\label{eq:Ham}
H = \sum_{i=1}^L \mathcal{H}_{i,i+1,\dots,i+1-r},
\end{equation}
where $r \geq 3$, assuming periodic boundary conditions. One definition of integrability is the existence of an infinite number of independent mutually commuting charges:
\begin{equation}\label{eq:integrability}
[Q_i, Q_j] = 0, \qquad Q_i:V\rightarrow V, \qquad i,j = 1,2,\dots,
\end{equation}
where $\text{exp}(Q_1) \sim U$ generates shifts along the chain and $Q_2 = H$. One way to ensure \eqref{eq:integrability} is to construct a Lax operator $\mathcal{L}_{Aj}(u): V_A \otimes V_j \rightarrow V_A \otimes V_j $, related to $\mathcal{H}$ in a way specified later, and an $R$-matrix $R_{AB}(u,v):V_A\otimes V_B\rightarrow V_A\otimes V_B$ which satisfy appropriate algebraic relations. Here $V_j \simeq \mathbb{C}^2$ is the \textit{physical space} and $V_A \simeq (\mathbb{C}^2)^{\otimes (r-1)}$ is the \textit{auxiliary space}. We note that in contrast to the nearest-neighbor case, the dimension of the physical and auxiliary spaces differ. If the $R\mathcal{L}\mathcal{L}$ equation
\begin{equation}\label{eq:RLL}
R_{AB}(u,v) \mathcal{L}_{Ai}(u)\mathcal{L}_{Bi}(v) = \mathcal{L}_{Bi}(v) \mathcal{L}_{Ai}(u)R_{AB}(u,v),
\end{equation}
and Yang--Baxter equation
\begin{equation}\label{eq:YBE}
R_{AB}(u,v)R_{AC}(u,w)R_{BC}(v,w)=R_{BC}(v,w)R_{AC}(u,w)R_{AB}(u,v).
\end{equation}
are satisfied by $\mathcal{L}_{Aj}(u)$ and $R_{AB}(u,v)$, the existence of the infinite set $Q_i$ satisfying \eqref{eq:integrability} follows. In particular one can define a transfer matrix
\begin{equation}
t(u) = \text{tr}_A [\mathcal{L}_{AL}(u)\mathcal{L}_{A,L-1}(u)\cdots \mathcal{L}_{A1}(u)],
\end{equation}
from which the charges can be extracted via
\begin{equation}
Q_j = \frac{d^{j-1}}{d u^{j-1}} \log t(u)\Big|_{u=0}.
\end{equation}
The commutation $[t(u),t(v)] = 0$ follows from \eqref{eq:RLL}, and \eqref{eq:YBE} is required for consistency. This then implies the integrability condition \eqref{eq:integrability}. Identifying $Q_2$ with the Hamiltonian \eqref{eq:Ham} leads to the relation
\begin{equation}
\mathcal{H}_{12\dots r} =\partial_u\check{\mathcal{L}}_{12\dots r}(u)\Big|_{u=0},
\end{equation}
where the checked Lax operator is defined
\begin{equation}
\check{\mathcal{L}}_{12\dots r}(u) \coloneqq \mathcal{P}_{r-1,r}\mathcal{P}_{r-2,r}\cdots \mathcal{P}_{1,r}\mathcal{L}_{12\dots r}(u),
\end{equation}
and $\mathcal{P}_{i,j}$ is the permutation operator on $V_i\otimes V_j$. $Q_3$ can then be expressed in terms of the Hamiltonian density as
\begin{equation}\label{eq:Q3defGeneral}
Q_3 = \sum_{i=1}^L\left( [\mathcal{H}_{i,i+1,\dots ,i+r-1},\mathcal{H}_{i+1,i+2,\dots, i+r}+\cdots +\mathcal{H}_{i+r-1,i+r,\dots,i+2r-2}]+q_{i,i+1,\dots i+r-1}\right),
\end{equation}
where $q_{i,i+1,\dots i+r-1}$ is a range-$r$ operator related to the Hamiltonian and derivatives of the Lax operator. We note from \eqref{eq:Q3defGeneral} that $Q_3$, defined in this way, is an operator of range $2r-1$.

\par\medskip
\noindent\textbf{Integrability with constraints.}\quad
For an integrable spin chain on a constrained Hilbert space $V_\Pi$

\begin{equation}
Q_2^{(\Pi)} = \Pi \left(\sum_{i=1}^{L}\mathcal{H}_{i,i+1,\dots,i+r-1}\right) \Pi,
\end{equation}
 we require the existence of an infinite number of independent charges which commute on the constrained space:

\begin{equation}\label{eq:integrability2}
[Q_i^{(\Pi)}, Q_j^{(\Pi)}] = 0, \qquad Q_i:V_\Pi\rightarrow V_\Pi, \qquad i,j = 1,2,\dots.
\end{equation}

A possible formulation of integrability in this case would be to find a Lax operator $\mathcal{L}_{Aj}$ satisfying (schematically) $\Pi\left(\sum\partial_u \check{\mathcal{L}}\big|_{u=0}\right)\Pi =Q_2^{(\Pi)}$ and the $R\mathcal{L}\mathcal{L}$ relation \eqref{eq:RLL}, and construct a transfer matrix acting on physical states 
\begin{equation}\label{eq:tproj}
t^{(\Pi)}(u) = \Pi  \text{tr}[\mathcal{L}_{AL}(u)\mathcal{L}_{A,L-1}(u)\cdots \mathcal{L}_{A1}(u)] \Pi.
\end{equation}
However, as discussed in detail in \cite{Corcoran:2024ofo}, such a transfer matrix would necessarily generate charges which also commute on the full, unconstrained Hilbert space. We know however that there exist some models which are integrable only after restriction to a certain projected subspace, and so such a formulation cannot work. One solution is to use a \textit{projected} Lax operator
\begin{equation}\label{eq:Ltilde}
\tilde{\mathcal{L}}_{Aj} = \Pi_A \mathcal{L}_{Aj},
\end{equation}
where we inserted an open projector $\Pi_A = \Pi_{a_1a_2}\cdots \Pi_{a_{r-1}a_r}$ on the auxiliary space indices. This projected Lax operator directly generates a transfer matrix acting on the constrained subspace
\begin{equation}
\tilde{t}(u) = \text{tr}_A[\tilde{\mathcal{L}}_{AL}(u)\tilde{\mathcal{L}}_{A,L-1}(u)\cdots \tilde{\mathcal{L}}_{A1}(u)] = t^{(\Pi)}(u).
\end{equation}
This transfer matrix can be shown to generate charges satisfying the integrability condition \eqref{eq:integrability2} provided certain projected versions of the $R\mathcal{L}\mathcal{L}$ and Yang--Baxter equations are satisfied, see \cite{Corcoran:2024ofo} for details. The key point is that the Hamiltonian density can be extracted from $\tilde{\mathcal{L}}_{Aj}(u)$ via\footnote{We note that there are gauge freedoms arising from adding operators to $\mathcal{H}$ which vanish on the projected subspace $V_\Pi$, and also operators which vanish when summed along a periodic chain.}
\begin{equation}
\mathcal{H}_{12\cdots r} =\mathcal{P}_{r-1,r}\mathcal{P}_{r-2,r}\cdots \mathcal{P}_{1,r} \partial_u\tilde{\mathcal{L}}_{12\dots r}(u)\Big|_{u=0},
\end{equation}
and the charge $Q_3$ can be obtained from $\mathcal{H}$ analogously to the unconstrained case:
\begin{equation}\label{eq:Q3def2}
Q_3^{(\Pi)} = \Pi\left(\sum_{i=1}^L\left( [\mathcal{H}_{i,i+1,\dots ,i+r-1},\mathcal{H}_{i+1,i+2,\dots, i+r}+\cdots +\mathcal{H}_{i+r-1,i+r,\dots,i+2r-2}]+q_{i,i+1,\dots i+r-1}\right)\right) \Pi.
\end{equation}

\subsection{Classification of integrable constrained models}\label{sec:cboost}
Although we take an infinite set of commuting charges \eqref{eq:integrability} as our definition of integrability, it is believed that the commutation of just the first two non-trivial charges
\begin{equation}\label{eq:resh}
[Q_2, Q_3] = 0,
\end{equation}
is sufficient. This is called the Reshetikhin condition \cite{KulishSklyanin1982,Grabowski:1994rb} and solving this has led to a surge of new classification results in integrable spin chains in recent years \cite{DeLeeuw:2019gxe, DeLeeuw:2019fdv, deLeeuw:2020xrw, DeLeeuw:2020ahx, deLeeuw:2020ahe, Corcoran:2023zax, deLeeuw:2024zqj}. For the unconstrained case, the idea is the following. One should first make an ansatz for a Hamiltonian density $\mathcal{H}$, which may be completely general or have certain specified symmetries. One should then form the higher charge $Q_3$ using \eqref{eq:Q3def}, and solve \eqref{eq:resh} to get a list of potentially integrable Hamiltonians. For each of these potentially integrable Hamiltonians, one should then prove integrability by constructing an $R$-matrix, for example by solving the Sutherland equations \cite{10.1063/1.1665111}. This classification procedure was adapted to the constrained case in \cite{Corcoran:2024ofo}, for which the steps are explicitly as follows:
\begin{itemize}
\item Choose a Hilbert space $\mathbb{C}^n$ and local constraint $\Pi_{i,i+1}$.\footnote{We note that higher-range local constraints are possible.}

\item Choose an ansatz for a Hamiltonian $\mathcal{H}_{i,i+1,\dots,i+r-1}$ of range $r$. For models with fusion category symmetry this would be a linear combination of fusion category projectors.

\item Form the operators $Q_{2}^{(\Pi)}$ and $Q_{3}^{(\Pi)}$ using \eqref{eq:Q3def2}. One should choose a spin chain of high enough length to ensure there are no cancellations which do not occur generically. For $r=3$ length $L=6$ is sufficient.

\item Impose the Reshetikhin condition $[Q_2^{(\Pi)},Q_3^{(\Pi)}] = 0$. This gives a list of equations for the parameters in the Hamiltonian ansatz $h_i$ and the extra parameters $q_i$ in $Q_3^{(\Pi)}$.

\item Solve the equations to get a list of potentially integrable Hamiltonians, and reduce these to an independent set.

\item Verify the integrability of each of these Hamiltonians by the construction of an appropriate Lax operator $\tilde{\mathcal{L}}_{Aj}$.
\end{itemize}

In practice the main technical hurdle in this approach is solving the large system of cubic equations in the variables $h_i$, which becomes intractable if there are too many variables under consideration. Although the equations are linear in the variables $q_i$, they still add complexity, and in many cases it is possible to consider the case $q_i=0$. In the unconstrained case it is possible to show that models derived from $R$-matrices of difference form $R(u-v)$ satisfy the Reshitkhin condition with $q_i=0$. Therefore models with $q_i\neq 0$ are referred to as being of \textit{non-difference} form.

As an example of this method, we consider a local Hilbert space $\mathbb{C}^2$ with the Rydberg blockade constraint. We then classify all space- and time-reflection invariant integrable Hamiltonians of range $3$ and $4$. In \cite{Corcoran:2024ofo} a partial classification for range 5 is performed. In later sections we apply the constrained boost operator method on Hilbert spaces of larger local dimension, with constraints arising from various fusion categories. We would like to point out that, due to the constraint, we find models that are not simply integrable models on the full space that are restricted to the constrained space.

\par\medskip
\noindent\textbf{Range 3.}\quad
We consider range-$3$ models on the Rydberg atom Hilbert space which are compatible with the Rydberg constraint and space/time-reflection invariant. This leads to a Hamiltonian density ansatz with three free parameters:
\begin{equation}\label{eq:Hansatz3}
\mathcal{H}_{123}= h_{12} P_1 X_2 P_3+ h_{11}P_1P_2P_3 +h_{22}P_1N_2P_3.
\end{equation}
It turns out that all integrable models in this case are of difference form, and so we take $q_i=0$. We compute $Q_3^{(\Pi)}$ using \eqref{eq:Q3def2} and compute the commutator $[Q_2^{(\Pi)},Q_3^{(\Pi)}]$ on a chain of length $L=6$. The only non-vanishing entries take the form
\begin{equation}
\pm h_{12} (2 h_{11}^2 + h_{12}^2 - h_{11}h_{22}).
\end{equation}
The case $h_{12}=0$ is diagonal and so trivially integrable; therefore we normalize our Hamiltonian and set $h_{12}=1, h_{11} = z$. Thus the most general integrable range-3 Hamiltonian in this case is the one-parameter family

\begin{equation}
\mathcal{H}_{123}= P_1 X_2 P_3+ zP_1P_2P_3 +\frac{1 + 2 z^2}{z}P_1N_2P_3.
\end{equation}
This model has been studied from various different perspectives \cite{Fendley_2004,Bianchini:2014bfa}, and reduces to the golden chain when $h_{11} = \varphi^{5/2}$.

\par\medskip
\noindent\textbf{Range 4.}\quad
In the range-4 case, imposing space/time-reflection invariance leads to a Hamiltonian with six free parameters:
\begin{align}\label{eq:Hansatz4}
\mathcal{H}_{1234}=& h_{11} \Piu_1 \Piu_2 \Piu_3 + h_{12}\Piu_1X_2\Piu_3 + h_{22}\Piu_1\Pid_2\Piu_3 +g_{11}\Piu_1 \Piu_2 \Piu_3\Piu_4\notag\\+& g_{23}\Piu_1\left(\frac{X_2X_3+Y_2Y_3}{2}\right) \Piu_4 + g_{12}\Piu_1(X_2\Piu_3+\Piu_2X_3)\Piu_4.
\end{align}
We note that this Hamiltonian has two interacting terms: one `hopping' term $P(XX+YY)P \sim P(\sigma^+\sigma^- + \sigma^-\sigma^+)P$ and one term which violates spin conservation $P(XP+PX)P$. We compute the commutator $[Q_2^{(\Pi)},Q_3^{(\Pi)}]$ on a chain of length $L=8$. In this case there is a single solution to the integrability condition with $q_i = 0$:
\begin{align}
\mathcal{H}^{\text{XXZ}}_{1234}=& \Piu_1\left(\frac{X_2X_3+Y_2Y_3}{2}\right) \Piu_4 +z( \Piu_1 \Piu_2 \Piu_3 + \Piu_1\Pid_2\Piu_3).
\end{align}
This model is equivalent to the constrained XXZ model \cite{constrained1, constrained2, constrained3}, and is exactly solvable by coordinate Bethe ansatz techniques.

There is one further solution to the integrability condition with $q_i \neq 0$. This was dubbed the `double golden chain' \cite{Corcoran:2024ofo}, and takes the explicit form

\begin{align}\label{eq:HDGC}
\mathcal{H}^{\text{DGC}}_{1234} =P_1\left(\frac{X_2X_3+Y_2Y_3}{2}+z(X_2P_3+P_2X_3)+\frac{2z^2}{1-z^2}P_2P_3\right)P_4\\+P_1\left(-\frac{1+z^2}{z}X_2+(z^2-1)P_2+\frac{1-7z^4+2z^6}{z^2(-1+z^2)}N_2\right)P_3 \notag.
\end{align}
The exact form of the operator coefficients $q_i$ are unimportant for the physics of the model. There are two critical points at $z = \phi^{-1/2}$ and $z = \phi^{3/2}$, although a detailed study of the nature of these points was not carried out. The ability of the boost operator procedure in the constrained case to generate all previously-known integrable models in the Rydberg-blockaded case, as well as new ones, underlines the usefulness and applicability of the method. 

\section{Beyond Fibonacci $-$ $\mathfrak{su}(2)_k$}\label{sec:su2k}
In previous sections we introduced anyonic chains and demonstrated their construction in the simplest case of the Fibonacci fusion category. We proceeded to review medium-range integrability in the context of constrained Hilbert spaces and how the boost operator method can be used to generate and classify integrable models on these spaces. For the remainder of the paper, we review the construction of anyonic chains built from fusion categories of higher rank. For each case, we use the boost operator method to classify all integrable linear combinations of fusion category projectors. We note that unless otherwise specified, we consider integrable models of difference form, i.e.\ we take $q=0$ in \eqref{eq:Q3def2}.

The first natural case to study is $\mathfrak{su}(2)_k$, which is an extension of Fibonacci and from which a whole host of anyonic chains can be defined. This is a (categorifiable) fusion algebra with $k+1$ objects $\boldsymbol{0}, \boldsymbol{\frac{1}{2}}, \boldsymbol{1},\dots, \boldsymbol{\frac{k}{2}}$, and represents a truncation of the usual angular momentum algebra:

\begin{equation}\label{eq:su2k}
\boldsymbol{j_1} \otimes \boldsymbol{j_2} = |\boldsymbol{j_1}-\boldsymbol{j_2}|\oplus |\boldsymbol{j_1}-\boldsymbol{j_2}| + 1\oplus \dots\oplus \text{min}(\boldsymbol{j_1}+\boldsymbol{j_2}, \boldsymbol{k}-(\boldsymbol{j_1}+\boldsymbol{j_2})). 
\end{equation}
For example, for $k = 3$ we have the multiplication table

\begin{table}[H]
\centering
\renewcommand{\arraystretch}{1.3}
\[
\begin{array}{c|cccc}
\otimes & \boldsymbol{0} & \boldsymbol{\frac{1}{2}} & \boldsymbol{1} & \boldsymbol{\frac{3}{2}} \\ \hline
\boldsymbol{0} & \boldsymbol{0} & \boldsymbol{\frac{1}{2}} & \boldsymbol{1} & \boldsymbol{\frac{3}{2}} \\
\boldsymbol{\frac{1}{2}} & \boldsymbol{\frac{1}{2}} & \boldsymbol{0}\oplus\boldsymbol{1} & \boldsymbol{\frac{1}{2}}\oplus\boldsymbol{\frac{3}{2}} & \boldsymbol{1} \\
\boldsymbol{1} & \boldsymbol{1} & \boldsymbol{\frac{1}{2}}\oplus\boldsymbol{\frac{3}{2}} & \boldsymbol{0}\oplus\boldsymbol{1} & \boldsymbol{\frac{1}{2}} \\
\boldsymbol{\frac{3}{2}} & \boldsymbol{\frac{3}{2}} & \boldsymbol{1} & \boldsymbol{\frac{1}{2}} & \boldsymbol{0}
\end{array}
\]
\caption{Fusion rules for $\mathfrak{su}(2)_3$.} \label{tab:su23}
\end{table}
We see that in this example, and indeed for all $k$, the integer sector $\mathfrak{psu}(2)_k$ is closed. Furthermore the $\{\boldsymbol{0},\boldsymbol{\frac{1}{2}}\} \simeq \mathbb{Z}_2$ sector is closed. In fact, for all odd $k$ the fusion algebra decomposes as
\begin{equation}\label{eq:odddecomp}
\mathfrak{su}(2)_{2k+1} = \mathfrak{psu}(2)_{2k+1} \times \mathbb{Z}_2, \qquad k=1,2,\dots,
\end{equation}
and so all non-trivial dynamical information is contained in $ \mathfrak{psu}(2)_{2k+1}$. For example, in table \ref{tab:su23} we see that the integer sector is equivalent to the fusion algebra of the Fibonacci fusion category, and so we have $\mathfrak{psu}(2)_3 \simeq \text{Fib}$. There is a subtle even/odd effect in the physical properties of anyonic chains based on $\mathfrak{su}(2)_k$, partly originating from \eqref{eq:odddecomp}. For example, while $\mathfrak{psu}(2)_4$ and $\mathfrak{psu}(2)_5$ have the same objects $\boldsymbol{0}, \boldsymbol{1},\boldsymbol{2}$, the truncation rule \eqref{eq:su2k} leads to distinct fusion rules, shown in table \ref{tab:psu245}, and quantum dimensions.
\begin{table}[H]
\centering
\renewcommand{\arraystretch}{1.3}

\begin{minipage}{0.45\textwidth}
\centering
\[
\begin{array}{c|ccc}
\otimes & \boldsymbol{0} & \boldsymbol{1} & \boldsymbol{2} \\ \hline
\boldsymbol{0} & \boldsymbol{0} & \boldsymbol{1} & \boldsymbol{2} \\
\boldsymbol{1} & \boldsymbol{1} & \boldsymbol{0}\oplus\boldsymbol{1}\oplus\boldsymbol{2} & \boldsymbol{1} \\
\boldsymbol{2} & \boldsymbol{2} & \boldsymbol{1} & \boldsymbol{0}
\end{array}
\]
\end{minipage}
\hfill
\begin{minipage}{0.45\textwidth}
\centering
\[
\begin{array}{c|ccc}
\otimes & \boldsymbol{0} & \boldsymbol{1} & \boldsymbol{2} \\ \hline
\boldsymbol{0} & \boldsymbol{0} & \boldsymbol{1} & \boldsymbol{2} \\
\boldsymbol{1} & \boldsymbol{1} & \boldsymbol{0}\oplus\boldsymbol{1}\oplus\boldsymbol{2} & \boldsymbol{1}\oplus\boldsymbol{2} \\
\boldsymbol{2} & \boldsymbol{2} & \boldsymbol{1}\oplus\boldsymbol{2} & \boldsymbol{0}\oplus\boldsymbol{1}
\end{array}
\]
\end{minipage}

\caption{Fusion rules for $\mathfrak{psu}(2)_4$ (left) and $\mathfrak{psu}(2)_5$ (right).} \label{tab:psu245}
\end{table}

The fusion rules \eqref{eq:su2k} are invariant under the automorphism 
\begin{equation}\label{eq:auto}
\boldsymbol{j}\rightarrow \boldsymbol{\frac{k}{2}}-\boldsymbol{j}. 
\end{equation}
This leads to a vast reduction in the number of independent anyonic chains to consider. However, as demonstrated in section \ref{sec:su23}, the exact mapping between chains with external object $\boldsymbol{j}$ and $\boldsymbol{\frac{k}{2}}-\boldsymbol{j}$ can be subtle. This automorphism also leads to a pyramid effect in the quantum dimensions of the objects, which start from $1$ at the object $\boldsymbol{0}$, increase until the object `$\boldsymbol{\frac{k}{4}}$', and decrease symmetrically to $1$ at the object $\boldsymbol{\frac{k}{2}}$.

The $F$-symbols of all $\mathfrak{su}(2)_k$ fusion categories are known, and can be expressed in terms of $q$-deformed factorials \cite{KirillovReshetikhin1988}. In this section for numerical calculations we use the $F$-symbols of Anyonica \cite{vercleyen_anyonica_2025}. Anyonic chains based on $\mathfrak{su}(2)_k$ up to spin-$1$ were analyzed in \cite{Gils_2013}, with several extensions/clarifications given in \cite{Vernier_2017}. In the following sections we review the properties of anyonic chains based on $\mathfrak{su}(2)_k$ for $2\leq k \leq 7$, with a focus on integrable combinations of projectors. We begin with the simplest example $k = 2$ which leads to the Ising anyonic chain, describing the Ising CFT in the continuum limit. We briefly describe $k=3$ and its precise relation to the golden chain, in order to demonstrate the how the automorphism \eqref{eq:auto} leads to equivalent models. The case $k=4$ suffices to demonstrate the general behavior of anyonic chains with external object $a = \boldsymbol{\frac{1}{2}}$. It is also possible to construct a spin-1 model in this case, but it enjoys a symmetry enhancement due to the automorphism \eqref{eq:auto}. We therefore use the case $k=5$ to review the general properties of spin-1 anyonic chains. $k=6$ is the first case where it is possible to construct a spin-$\frac32$ chain, which has previously not been considered. We construct all integrable combinations of projectors in this case. Since this case also contains enhanced symmetry due to \eqref{eq:auto}, we conclude by considering spin-$\frac32$ chains for $k=7$.

\subsection{$\mathfrak{su}(2)_2 = \text{Ising}$}\label{sec:su22}
\begin{figure}[H]
\centering

\begin{tikzpicture}[scale=1]
\draw (0,0) circle [radius=0.4] node {$1$};
\draw (1.8,0) circle [radius=0.4] node {$\sigma$};
\draw (3.6,0) circle [radius=0.4] node {$\psi$};

\draw (0.4,0) -- (1.4,0);
\draw (2.2,0) -- (3.2,0);

\end{tikzpicture}
\caption{Adjacency graph for the Ising fusion category with external object $\sigma$.}
\label{fig:ising_sigma_adjacency}
\end{figure}
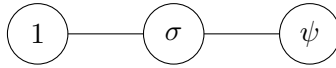

\begin{figure}[t]
  \centering
  \begin{minipage}[t]{0.49\textwidth}
    \centering
    \includegraphics[width=\linewidth]{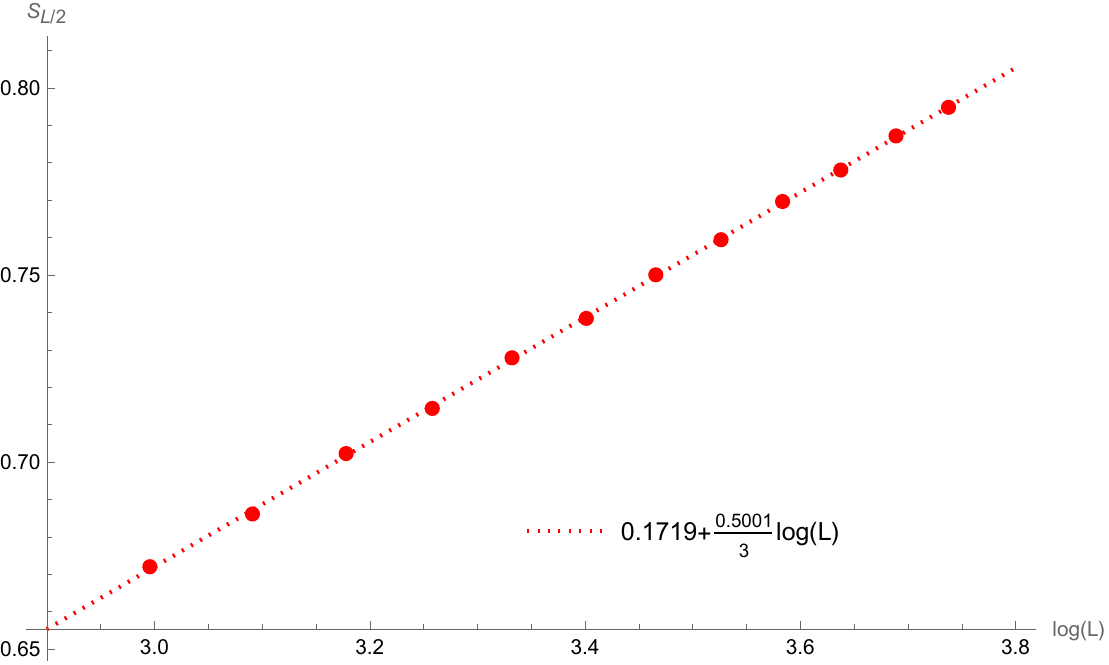}
  \end{minipage}\hfill
  \begin{minipage}[t]{0.49\textwidth}
    \centering
    \includegraphics[width=\linewidth]{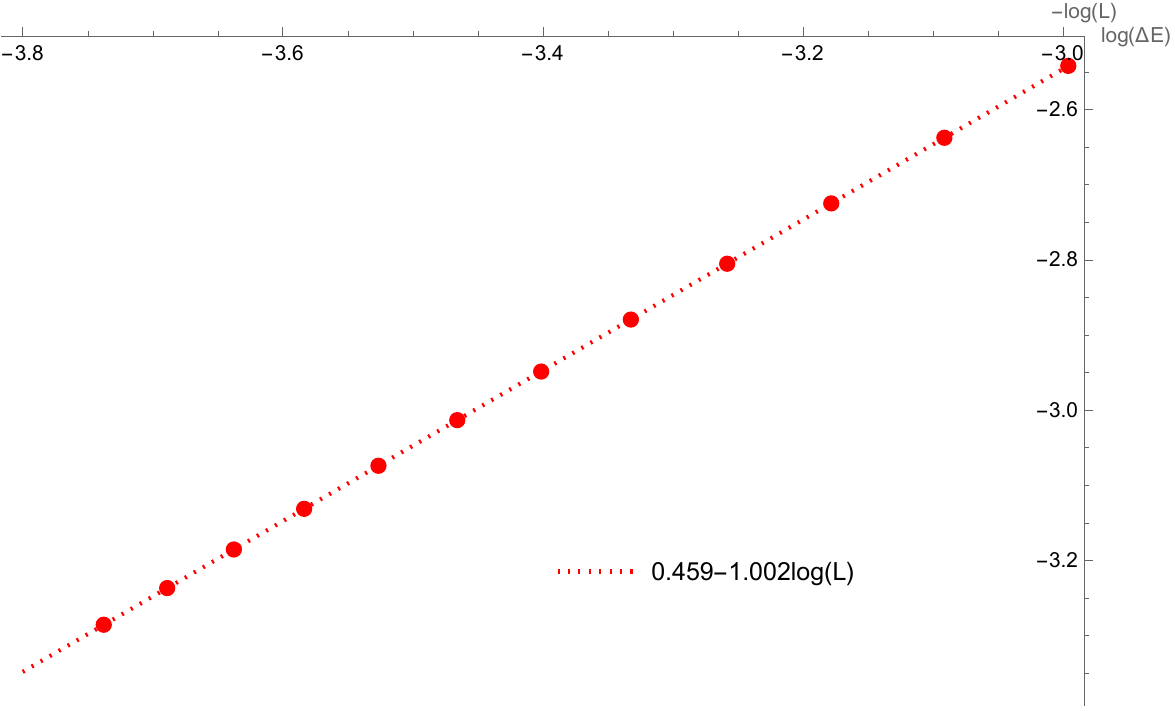}
  \end{minipage}

  \caption{Half-chain entanglement entropy and gap for Ising anyonic chain \eqref{eq:Hisingdens}. For both plots we take even $20\leq L\leq 42$.}
  \label{fig:ising-ee-gap}
\end{figure}

The simplest non-trivial case is $k=2$ where we have $\text{Ising} = \{\boldsymbol{0}, \boldsymbol{\frac{1}{2}}, \boldsymbol{1}\}$. The fusion rules are shown in table \ref{tab:ising_fusion}, in the more familiar notation where $\boldsymbol{0}\coloneqq 1, \boldsymbol{\frac{1}{2}}\coloneqq \sigma,$ and $\boldsymbol{1}\coloneqq \psi$.
\begin{table}[H]
\centering
\renewcommand{\arraystretch}{1.15}
\setlength{\tabcolsep}{6pt}
\[
\begin{array}{c|ccc}
\otimes & 1& \sigma & \psi \\ \hline
1 & 1& \sigma & \psi \\
\sigma &\sigma & 1 \oplus \psi& \sigma\\
\psi &\psi & \sigma & 1
\end{array}
\]
\caption{Fusion rules for $\mathfrak{su}(2)_2 = \mathrm{Ising}$}
\label{tab:ising_fusion}
\end{table}
The quantum dimensions of these objects are
\begin{equation}
(d_{1}, d_{\sigma}, d_{\psi}) = (1, \sqrt{2}, 1).
\end{equation}
There are two non-equivalent solutions to the pentagon equations in this case, and both give rise to the same (Hermitian) anyonic chain. We note that the only non-trivial choice of external object in this case is $a = \sigma$, which leads to adjacency rules summarized in figure \ref{fig:ising_sigma_adjacency}. Since $\sigma\otimes \sigma \rightarrow 1 \oplus \psi$, there are only two fusion channels which are described by projectors $\mathrm{P}^{\sigma}_1$ and $\mathrm{P}^{\sigma}_\psi$, of which only one is independent due to $\mathrm{P}^{\sigma}_{1,i}+\mathrm{P}^{\sigma}_{\psi,i}=1$.
The corresponding Hilbert space consists of states $\ket{x_1,x_2,\dots,x_L}$ which satisfy the constraint $x_{i+1}\in \sigma \otimes x_i$. This constraint can only be satisfied on the periodic chain for even $L$, where we have $x_i = \sigma$ on either all even or all odd sites. On the remaining sites in each case we can freely pick between $x_i = 1$ or $x_i = \psi$. This leads to a $\mathbb{Z}_2$ decomposition of the Ising Hilbert space $V^L_{\mathrm{Ising}} \simeq V^L_{\text{odd }\sigma}\oplus V^L_{\text{even }\sigma}$. Overall we have 
\begin{equation}
\dim V^L_{\mathrm{Ising}} =
\begin{cases}
0 & L \text{ odd},\\[4pt]
2 \times 2^{\frac{L}{2}} & L \text{ even},
\end{cases}
\end{equation}
which agrees with the general asymptotic formula $\text{dim } V \sim (d_\sigma)^L = 2^{L/2}$ for even $L$. Since there is a single independent projector $\mathrm{P}^{\sigma}_1$ which projects onto the identity fusion channel, due to the results described in section \ref{sec:TL} this model is Temperley-Lieb with $\delta = \sqrt{2}<2$ and therefore integrable. This operator takes the explicit form $\mathrm{P}^{\sigma}_1 = -\sum_{i=1}^L \mathcal{H}^{\text{Ising}}_{i,i+1,i+2}$, where the Hamiltonian density is
\begin{equation}\label{eq:Hisingdens}
\mathcal{H}^{\text{Ising}} = P_1P_{\sigma}P_1 + P_{\psi}P_{\sigma}P_{\psi} + \frac{1}{2}(P_{\sigma}P_1P_{\sigma}+P_{\sigma}P_{\psi}P_{\sigma}-P_{\sigma}X_{1\psi}P_{\sigma}),
\end{equation}
where we suppresed site indices for brevity. In \eqref{eq:Hisingdens} $P_x$ is the projector onto the object $x\in\text{Ising}$ and $X_{1 \psi}$ maps $\ket{1}\leftrightarrow \ket{\psi}$. There is a $\mathbb{Z}_2$ degeneracy in the spectrum of $\mathrm{P}^{\sigma}_1$, and for $L\equiv 0$ mod 4 the spectrum is symmetric about $0$ after an appropriate shift. For example, the $L=4$ (shifted) spectrum reads
\begin{equation}
\text{spec}(\mathrm{P}^{\sigma}_1)_{L=4}=(\sqrt{2},\sqrt{2},1,1,-1,-1,-\sqrt{2},-\sqrt{2}).
\end{equation}
This means that as $L\rightarrow \infty$ the low-lying spectrum of $\mathrm{P}^{\sigma}_1$ is identical in both the ferromagnetic and antiferromagnetic regimes. Indeed, in both cases the model is described by the Ising CFT, with $c=1/2$. We demonstrate this using the numerical methods described in section \ref{sec:goldennumerics} to calculate the gap and the half-chain entanglement entropy for various values of $L$, see figure \ref{fig:ising-ee-gap}. We see that even up to a relatively modest $L=42$ the formulas \eqref{eq:loggap} and \eqref{eq:EEdef} are verified, with $c\sim \frac{1}{2}$.

\subsection{$\mathfrak{su}(2)_3 = \mathbb{Z}_2 \times \text{Fib}$}\label{sec:su23}
This case contains no new physics beyond the Fibonacci fusion category discussed in section \ref{sec:fib}. Indeed, the only non-trivial choices of external object are $a = \boldsymbol{\frac12}$ and $a = \boldsymbol{1}$, see table \ref{tab:su23}. The adjacency diagrams for the corresponding constrained Hilbert spaces are shown in figures \ref{fig:su2_3_half_adjacency} and \ref{fig:su2_3_one_adjacency} respectively. Both choices satisfy the fusion rule $a \otimes a = \boldsymbol{0}\oplus \boldsymbol{1}$, and so there is only a single independent projector $\mathrm{P}^{a}_{\boldsymbol{0}}$ in each case. In fact, this operator has the same spectrum for both choices of external object due to the algebra automorphism \eqref{eq:auto}, and is essentially two independent copies of the golden chain Hamiltonian.

The choice $a = \boldsymbol{1}$ leads to a Hilbert space which decomposes into two independent Rydberg atom Hilbert spaces $V^{\boldsymbol{1}} = V_{\text{Rydberg}}^{(0,1)}\oplus V_{\text{Rydberg}}^{(1/2,3/2)} $, and so has dimension $\sim 2 \varphi^L$. On this Hilbert space the projector $\mathrm{P}^{\boldsymbol{1}}_{\boldsymbol{0}}$ acts block diagonally as two copies of the golden chain. The choice $a = \boldsymbol{\frac12}$ leads to a Hilbert space of states with strictly alternating integer and half-integer labels. Similarly to the Ising case above, this leads to two `staggering' sectors which due to the constraint can only be formed for even $L$ on a periodic chain. In each staggering sector, after applying \eqref{eq:auto} on every other site, this basis is isomorphic to a single Rydberg Hilbert space. Therefore this Hilbert space also has dimension $\sim 2 \varphi^L$ for even $L$. The projector $\mathrm{P}^{\boldsymbol{1/2}}_{\boldsymbol{0}}$ acts as the golden chain on each staggering sector.

\begin{figure}[H]
\centering

\begin{tikzpicture}[scale=1]
\draw (0,0) circle [radius=0.4] node {$\boldsymbol{0}$};
\draw (1.8,0) circle [radius=0.4] node {$\boldsymbol{\frac12}$};
\draw (3.6,0) circle [radius=0.4] node {$\boldsymbol{1}$};
\draw (5.4,0) circle [radius=0.4] node {$\boldsymbol{\frac32}$};

\draw (0.4,0) -- (1.4,0);
\draw (2.2,0) -- (3.2,0);
\draw (4.0,0) -- (5.0,0);

\end{tikzpicture}
\caption{Adjacency graph for the $\mathfrak{su}(2)_3$ fusion category with external object $\boldsymbol{\frac12}$.}
\label{fig:su2_3_half_adjacency}
\end{figure}

\begin{figure}[H]
\centering

\begin{tikzpicture}[scale=1]
\draw (0,0) circle [radius=0.4] node {$\boldsymbol{0}$};
\draw (1.8,0) circle [radius=0.4] node {$\boldsymbol{1}$};

\draw (4.8,0) circle [radius=0.4] node {$\boldsymbol{\frac12}$};
\draw (6.6,0) circle [radius=0.4] node {$\boldsymbol{\frac32}$};

\draw (0.4,0) -- (1.4,0);
\draw (5.2,0) -- (6.2,0);

\draw (2.0,0.346) to[out=30,in=-30,looseness=5] (2.0,-0.346);

\draw (4.6,0.346) to[out=150,in=210,looseness=5] (4.6,-0.346);

\end{tikzpicture}
\caption{Adjacency graph for the $\mathfrak{su}(2)_3$ fusion category with external object $\boldsymbol{1}$.}
\label{fig:su2_3_one_adjacency}
\end{figure}

\subsection{$\mathfrak{su}(2)_4$}\label{sec:su24}

\begin{table}[H]
\centering
\renewcommand{\arraystretch}{1.3}
\[
\begin{array}{c|ccccc}
\otimes & \boldsymbol{0} & \boldsymbol{\frac{1}{2}} & \boldsymbol{1} & \boldsymbol{\frac{3}{2}} & \boldsymbol{2} \\ \hline
\boldsymbol{0}
& \boldsymbol{0}
& \boldsymbol{\frac{1}{2}}
& \boldsymbol{1}
& \boldsymbol{\frac{3}{2}}
& \boldsymbol{2}
\\
\boldsymbol{\frac{1}{2}}
& \boldsymbol{\frac{1}{2}}
& \boldsymbol{0}\oplus\boldsymbol{1}
& \boldsymbol{\frac{1}{2}}\oplus\boldsymbol{\frac{3}{2}}
& \boldsymbol{1}\oplus\boldsymbol{2}
& \boldsymbol{\frac{3}{2}}
\\
\boldsymbol{1}
& \boldsymbol{1}
& \boldsymbol{\frac{1}{2}}\oplus\boldsymbol{\frac{3}{2}}
& \boldsymbol{0}\oplus\boldsymbol{1}\oplus\boldsymbol{2}
& \boldsymbol{\frac{1}{2}}\oplus\boldsymbol{\frac{3}{2}}
& \boldsymbol{1}
\\
\boldsymbol{\frac{3}{2}}
& \boldsymbol{\frac{3}{2}}
& \boldsymbol{1}\oplus\boldsymbol{2}
& \boldsymbol{\frac{1}{2}}\oplus\boldsymbol{\frac{3}{2}}
& \boldsymbol{0}\oplus\boldsymbol{1}
& \boldsymbol{\frac{1}{2}}
\\
\boldsymbol{2}
& \boldsymbol{2}
& \boldsymbol{\frac{3}{2}}
& \boldsymbol{1}
& \boldsymbol{\frac{1}{2}}
& \boldsymbol{0}
\end{array}
\]
\caption{Fusion rules for $\mathfrak{su}(2)_4$.}
\end{table}
\noindent The $k = 4$ case is the first where it is possible to choose a truly `spin-1' external object. Here there are five objects $\boldsymbol{0}, \boldsymbol{\frac{1}{2}}, \boldsymbol{1}, \boldsymbol{\frac{3}{2}}, \boldsymbol{2}$, whose quantum dimensions are
\begin{equation}
(d_{\boldsymbol{0}}, d_{\boldsymbol{1/2}}, d_{\boldsymbol{1}}, d_{\boldsymbol{3/2}}, d_{\boldsymbol{2}}) = (1, \sqrt{3}, 2, \sqrt{3}, 1).
\end{equation}
In this case there are two independent non-trivial anyonic chains to consider, one from considering a spin-1/2 external object $a=\boldsymbol{\frac{1}{2}}$, and one from considering the spin-1 external object $a=\boldsymbol{1}$. We note that the choice $a = \boldsymbol{\frac{3}{2}}$ leads to equivalent physics to $a = \boldsymbol{\frac12}$ due to \eqref{eq:auto}, similarly to the case described in section \ref{sec:su23}. 
\par\medskip
\noindent\textbf{$a = \boldsymbol{\frac12}$.}\quad
 Due to the fusion rule $ \boldsymbol{\frac12}\otimes  \boldsymbol{\frac12} =  \boldsymbol{0} \oplus \boldsymbol{1}$, this choice of external object leads to an anyonic chain with two projectors
\begin{equation}
H^{\boldsymbol{1/2}}_{\mathfrak{su}(2)_4}=\alpha \mathrm{P}^{\boldsymbol{1/2}}_{\boldsymbol{0}} + \beta \mathrm{P}^{\boldsymbol{1/2}}_{\boldsymbol{1}}.
\end{equation}
The constrained Hilbert space can be described by the adjacency diagram in figure \ref{fig:su2_4_half_adjacency}. Similarly to the cases above, the Hilbert space consists of alternating integer/half-integer labels, and so can only be formed on the periodic chain for even $L$. In this case the dimension of the Hilbert space is $2(3)^{L/2}+2$. Due to the constraint $ \mathrm{P}^{\boldsymbol{1/2}}_{\boldsymbol{0},i} +  \mathrm{P}^{\boldsymbol{1/2}}_{\boldsymbol{1},i} = 1$ there is only a single independent projector $\mathrm{P}^{\boldsymbol{1/2}}_{\boldsymbol{0}}$. Since this projects onto the identity fusion channel it is Temperley-Lieb with $\delta = \sqrt{3}<2$ and integrable. In the antiferromagnetic regime this operator corresponds to the minimal model $\mathcal{M}(5,6)$ with $c=4/5$. In the ferromagnetic regime it corresponds to the $\mathbb{Z}_4$ parafermion theory with $c = 1$.
 
 These results generalize straightforwardly to arbitrary $k$ (odd or even). For all $k$, taking the external object $a = \boldsymbol{\frac12}$ leads to a constrained Hilbert space with the $A_{k+1}$ adjacency graph and a single independent integrable projector $\mathrm{P}_{\boldsymbol{0}} $, which is Temperley-Lieb with $\delta = 2\cos \frac{\pi}{k+2}<2$. In the antiferromagnetic regime it corresponds to the minimal model $\mathcal{M}(k+1,k+2)$ with $c=1-\frac{6}{(k+1)(k+2)}$, and in the ferromagnetic regime it corresponds to $\mathbb{Z}_k$ parafermions with $c = \frac{2(k-1)}{k+2}$ \cite{Gils_2013}. We note that both of these agree with the Ising result for $k=2$, giving $c=1/2$.

\begin{figure}[H]
\centering

\begin{tikzpicture}[scale=1]
\draw (0,0) circle [radius=0.4] node {$\boldsymbol{0}$};
\draw (1.8,0) circle [radius=0.4] node {$\boldsymbol{\frac12}$};
\draw (3.6,0) circle [radius=0.4] node {$\boldsymbol{1}$};
\draw (5.4,0) circle [radius=0.4] node {$\boldsymbol{\frac32}$};
\draw (7.2,0) circle [radius=0.4] node {$\boldsymbol{2}$};

\draw (0.4,0) -- (1.4,0);
\draw (2.2,0) -- (3.2,0);
\draw (4.0,0) -- (5.0,0);
\draw (5.8,0) -- (6.8,0);

\end{tikzpicture}
\caption{Adjacency graph for the $\mathfrak{su}(2)_4$ fusion category with external object $\boldsymbol{\frac12}$.}
\label{fig:su2_4_half_adjacency}
\end{figure}

\par
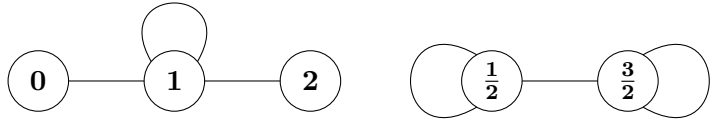
\begin{figure}[H]
\centering

\begin{tikzpicture}[scale=1]
\draw (0,0) circle [radius=0.4] node {$\boldsymbol{0}$};
\draw (1.8,0) circle [radius=0.4] node {$\boldsymbol{1}$};
\draw (3.6,0) circle [radius=0.4] node {$\boldsymbol{2}$};

\draw (6.0,0) circle [radius=0.4] node {$\boldsymbol{\frac12}$};
\draw (7.8,0) circle [radius=0.4] node {$\boldsymbol{\frac32}$};

\draw (0.4,0) -- (1.4,0);
\draw (2.2,0) -- (3.2,0);
\draw (6.4,0) -- (7.4,0);

\draw (1.494,0.257) to[out=120,in=60,looseness=5] (2.106,0.257);

\draw (5.8,0.346) to[out=150,in=210,looseness=5] (5.8,-0.346);

\draw (8.0,0.346) to[out=30,in=-30,looseness=5] (8.0,-0.346);

\end{tikzpicture}
\caption{Adjacency graph for the $\mathfrak{su}(2)_4$ fusion category with external object $1$.}
\label{fig:su2_4_one_adjacency}
\end{figure}

\medskip
\noindent\textbf{$a = \boldsymbol{1}$.}\quad
This choice of external object leads to a constrained Hilbert space described by the adjacency graph in figure \ref{fig:su2_4_one_adjacency}. We note that there are two connected components, and so the Hilbert space decomposes $V^{\boldsymbol{1}}_{\mathfrak{su}(2)_4} = V_{\text{int}}\oplus V_{\text{half-int}}$ into integer and half-integer sectors. The half-integer sector is fully connected with two nodes, and so $\text{dim }V_{\text{half-int}}=2^L$. The integer sector is closed, and corresponds to the $\mathfrak{psu}(2)_4$ adjacency graph. Here we have $\text{dim }V_{\text{int}} = 2^L + (-1)^L$. For example, for $L=3$ on a periodic chain $V_{\text{half-int}}$ consists of states $\ket{x_1x_2x_3}$ with $x_i \in \{\boldsymbol{\frac12},\boldsymbol{\frac32}\}$, and we have
\begin{equation}
V_{\text{int}} = \text{span}(\ket{\boldsymbol{0}\boldsymbol{1}\boldsymbol{1}},\ket{\boldsymbol{1}\boldsymbol{0}\boldsymbol{1}},\ket{\boldsymbol{1}\boldsymbol{1}\boldsymbol{0}},\ket{\boldsymbol{1}\boldsymbol{1}\boldsymbol{1}},\ket{\boldsymbol{1}\boldsymbol{1}\boldsymbol{2}},\ket{\boldsymbol{1}\boldsymbol{2}\boldsymbol{1}},\ket{\boldsymbol{2}\boldsymbol{1}\boldsymbol{1}}).
\end{equation}
Due to the fusion rule $\boldsymbol{1}\otimes \boldsymbol{1} = \boldsymbol{0} \oplus \boldsymbol{1} \oplus \boldsymbol{2}$ there are three projectors $\mathrm{P}^{\boldsymbol{1}}_{\boldsymbol{0}}$, $\mathrm{P}^{\boldsymbol{1}}_{\boldsymbol{1}}$, $\mathrm{P}^{\boldsymbol{1}}_{\boldsymbol{2}}$, which are dependent due to \eqref{eq:sumProjectors}. The general anyonic chain is a linear combination of the two independent projectors
\begin{equation}
H^{\boldsymbol{1}}_{\mathfrak{su}(2)_4} = \alpha  \mathrm{P}^{\boldsymbol{1}}_{\boldsymbol{1}} + \beta \mathrm{P}^{\boldsymbol{1}}_{\boldsymbol{2}}.
\end{equation}
Due to the Hilbert space decomposition, this operator can be written in block diagonal form
\begin{equation}\label{eq:Hblock}
H^{\boldsymbol{1}}_{\mathfrak{su}(2)_4}  = \begin{pmatrix}H^{\text{int}} & 0 \\ 0 & H^{\text{half-int}}  \end{pmatrix},
\end{equation}
where the operators $H^{\text{int}}$ and $H^{\text{half-int}}$ can be considered independently. We have the isomorphism $\mathfrak{psu}(2)_4 \simeq \text{Rep}(D_3)$, the algebra describing fusion of the irreps of the dihedral group $D_3$. Therefore we can identify $H_{\text{int}}$ with the $D_3$ anyonic chain, which has been discussed in detail in \cite{martina2010chainstronglycorrelatedsu24,Finch:2013nza,Braylovskaya:2016btd}, and also recently for example in \cite{Eck:2025ldx}.

This anyonic chains \eqref{eq:Hblock} have a large amount of symmetry, for one because the projectors $\mathrm{P}^{\boldsymbol{1}}_{\boldsymbol{1}} $ and  $\mathrm{P}^{\boldsymbol{1}}_{\boldsymbol{2}}$ are mapped into each other under \eqref{eq:auto}. Using the constrained boost operator formalism discussed in section \ref{sec:cboost}, we verify that $H^{\boldsymbol{1}}_{\mathfrak{su}(2)_4} $ is integrable for any choice of $\alpha, \beta$. This can also be shown via a direct mapping to the XXZ spin chain in the $D_3$ case \cite{Braylovskaya:2016btd}.

There are subtle differences between the operators $H^{\text{int}}$ and $H^{\text{half-int}}$, for example the slight mismatch in number of states between odd and even $L$. However, much of the spectrum overlaps, and there are similarities in the physical properties of both models. Parametrizing $\beta = \cos \theta$ and $\alpha = -\sin \theta$, one can study the physical properties of the model for $\theta \in [0,2\pi)$. As discussed in detail in \cite{Gils_2013}, both models possess similar extended gapless phases with central charge $c=1$. Several points can be mapped to known conformal field theories, including Ising$^2$, 4-state Potts, and parafermions. The Temperley-Lieb points have $\delta = 2$ in both cases and correspond to $\theta = -\frac{\pi}{4}$ and $\theta = \frac{3\pi}{4}$. In the integer sector the $\theta = \frac{3\pi}{4}$ point maps to the Kosterlitz-Thouless CFT, while in the half-integer sector this point maps to a superconformal minimal model. 

\subsection{$\mathfrak{su}(2)_5 = \mathbb{Z}_2 \times \mathfrak{psu}(2)_5$}\label{sec:su25}
Due to the decomposition $\mathfrak{su}(2)_5 = \mathbb{Z}_2 \times \mathfrak{psu}(2)_5$, it suffices to consider the integer sector $ \mathfrak{psu}(2)_5 = \{\boldsymbol{0},\boldsymbol{1},\boldsymbol{2}\}$, whose fusion rules are shown in table \ref{tab:psu245}. In this case there is a single unitary solution to the pentagon equations, and the quantum dimensions of the objects are given by

\begin{equation}
(d_{\boldsymbol{0}}, d_{\boldsymbol{1}}, d_{\boldsymbol{2}}) = \left(1,1+2\cos \frac{2\pi}{7},  2\cos \frac{\pi}{7}\right) \sim (1, 2.25, 1.80).
\end{equation}
Taking the external object $a = \boldsymbol{2}$ is equivalent to taking $a = \boldsymbol{\frac 12}$ in $\mathfrak{su}(2)_5$ due to the automorphism \eqref{eq:auto}. The general properties for this case were explained in section \ref{sec:su24}, and so we omit them here.

There is one choice of external object, $a = \boldsymbol{1}$, which leads to a spin-1 anyonic chain with new properties with respect to earlier cases. The adjacency graph describing the constrained Hilbert space is shown in figure \ref{fig:psu2_5_one_adjacency}. The dimension of this Hilbert space grows as $\text{dim }V^{\boldsymbol{1}}_{\mathfrak{psu}(2)_5} \sim d_{\boldsymbol{1}}^L \sim 2.25^L$. Due to the fusion rule $\boldsymbol{1}\otimes \boldsymbol{1} = \boldsymbol{0}\oplus \boldsymbol{1}\oplus \boldsymbol{2}$, the general anyonic chain in this case is a linear combination of two independent projectors

\begin{equation}\label{eq:Hpsu25}
H^{\boldsymbol{1}}_{\mathfrak{psu}(2)_5} = -\sin \theta \mathrm{P}^{\boldsymbol{1}}_{\boldsymbol{1}} + \cos \theta \mathrm{P}^{\boldsymbol{1}}_{\boldsymbol{2}},
\end{equation}
where we picked a conventional normalization and parametrization. In order to identify the integrable points of the model, one can apply the constrained boost operator formalism described in section \ref{sec:cboost}. We find three integrable points in the range $\theta \in (-\pi, 0]$, in agreement with \cite{Vernier_2017}:
\begin{equation}\label{eq:intpoints}
\theta_{\text{TL}} = -\frac{\pi}{4}, \qquad \theta_{\text{FZ}} = -\arctan \frac{2\cos ^2 \frac{\pi}{7}}{1+2\cos \frac{2\pi}{7}}, \qquad  \theta_{\text{IK}} =  -\arctan \frac{1}{\cos \frac{2\pi}{7}-\cos \frac{4\pi}{7}}.
\end{equation}
The `ferromagnetic' integrable points can be found by sending $\theta\rightarrow \theta+\pi$, which corresponds to changing the sign of the Hamiltonian. The Temperley-Lieb points correspond to the coefficients of the projectors in \eqref{eq:Hpsu25} being equal, and so they correspond to $\pm \mathrm{P}^{\boldsymbol{1}}_{\boldsymbol{0}}$. Since these operators are Temperley-Lieb with $\delta = d_{\boldsymbol{1}}\sim 2.25 > 2$, these points are not critical, as mentioned in section \ref{sec:TL}. The points $\theta_{\text{TL}}$ and $\theta_{\text{TL}}+\pi$ both correspond to a first-order phase transitions, with a finite correlation length.

The two further integrable points are critical in both ferromagnetic and antiferromagnetic regimes. The Fateev-Zamolodchikov (FZ) and Izergin-Korepin (IK) points correspond to different representations of the Birman-Murakami-Wenzl (BMW) algebra, which is a spin-1 generalization of Temperley-Lieb. The point $\theta_{\text{FZ}}$ is described by the coset CFT $\frac{\mathfrak{su}(2)_3\times \mathfrak{su}(2)_2}{\mathfrak{su}(2)_5}$, with $c = 81/70$. The model $\theta_{\text{FZ}} + \pi$ corresponds to a $\mathbb{Z}_5$ parafermionic theory with $c = 8/7$. The point $\theta_{\text{IK}}$ is also described by the $\mathbb{Z}_5$ theory with $c=8/7$. Finally, the point $\theta_{\text{IK}}+\pi$ is critical and is described by the coset CFT $\frac{\mathfrak{su}(2)_1\times \mathfrak{su}(2)_4}{\mathfrak{su}(2)_5}$ with $c = 6/7$.

The full phase diagram of \eqref{eq:Hpsu25} is elucidated in \cite{Vernier_2017}. The point $\theta_{\text{TL}}+ \pi$ sits at the phase transition between a $\frac{\mathfrak{su}(2)_1\times \mathfrak{su}(2)_4}{\mathfrak{su}(2)_5}$ phase (which contains the point $\theta_{\text{IK}}+\pi$) and a $\mathbb{Z}_5$ parafermion phase (which contains $\pi + \theta_{\text{FZ}}$). Meanwhile, $\theta_{\text{TL}}$ sits at the boundary between unknown phases. $\theta_{\text{IK}}$ lies at the endpoint of a $\frac{\mathfrak{su}(2)_4\times \mathfrak{su}(2)_1}{\mathfrak{su}(2)_5}$ phase, and the corresponding $\mathbb{Z}_5$ CFT to which it corresponds is different in nature to $\theta_{\text{FZ}} + \pi$, which lies in an extended phase. $\theta_{\text{FZ}}$ lies at the end of a massive Haldane phase.

The phase diagram of the $\mathfrak{su}(2)_k$ anyonic chain with external object $a=\boldsymbol{1}$ is similar for higher $k>4$. There remain three integrable points in the region $\theta \in (-\pi,0]$, and the expressions \eqref{eq:intpoints} are upgraded to functions of the level $k$. The parafermion CFTs are upgraded to $\mathbb{Z}_k$, and the quotients are modified
$\frac{\mathfrak{su}(2)_3\times \mathfrak{su}(2)_2}{\mathfrak{su}(2)_5} \rightarrow \frac{\mathfrak{su}(2)_{k-2}\times \mathfrak{su}(2)_{2}}{\mathfrak{su}(2)_k}, \frac{\mathfrak{su}(2)_1\times \mathfrak{su}(2)_4}{\mathfrak{su}(2)_5}\rightarrow \frac{\mathfrak{su}(2)_{k-4}\times \mathfrak{su}(2)_{4}}{\mathfrak{su}(2)_k}, \frac{\mathfrak{su}(2)_4\times \mathfrak{su}(2)_1}{\mathfrak{su}(2)_5}\rightarrow \frac{\mathfrak{su}(2)_{k-1}\times \mathfrak{su}(2)_1}{\mathfrak{su}(2)_k}$.
The distinction in the physics between odd and even $k$ is milder than previously thought in \cite{Gils_2013}. The main odd/even distinction concerns the region near $\theta_{\text{IK}}$: for odd $k$ this region marks the termination of the $ \frac{\mathfrak{su}(2)_{k-1}\times \mathfrak{su}(2)_1}{\mathfrak{su}(2)_k}$ phase, while for even $k>4$ the corresponding region is only partially characterized, with a massive dimerized phase identified but the full phase structure still unresolved. We note that for odd $k$ due to the decomposition \eqref{eq:odddecomp} the spin-1 $\mathfrak{su}(2)_k$ anyonic chain is literally two copies of the spin-1 $\mathfrak{psu}(2)_k$ anyonic chain, as explicitly discussed in section \ref{sec:su23} for $k=3$. For even $k$ there is a slight deviation due to the decomposition of the Hamiltonian into integer/half-integer sector with different numbers of states. However, this only introduces a mild discrepancy in the spectrum and the large-$L$ behavior between both sectors is equivalent.

\begin{figure}[H]
\centering

\begin{tikzpicture}[scale=1]
\draw (0,0) circle [radius=0.4] node {$\boldsymbol{0}$};
\draw (1.8,0) circle [radius=0.4] node {$\boldsymbol{1}$};
\draw (3.6,0) circle [radius=0.4] node {$\boldsymbol{2}$};

\draw (0.4,0) -- (1.4,0);
\draw (2.2,0) -- (3.2,0);

\draw (1.5,0.25) to[out=120,in=60,looseness=5] (2.1,0.25);

\draw (3.8,0.346) to[out=30,in=-30,looseness=5] (3.8,-0.346);

\end{tikzpicture}
\caption{Adjacency graph for the $\mathfrak{psu}(2)_5$ fusion category with external object $\boldsymbol{1}$.}
\label{fig:psu2_5_one_adjacency}
\end{figure}

\subsection{$\mathfrak{su}(2)_6$}\label{sec:su26}
The case $k = 6$ contains seven objects $\boldsymbol{0}, \boldsymbol{\frac12},\boldsymbol{1},\boldsymbol{\frac32},\boldsymbol{2},\boldsymbol{\frac52},\boldsymbol{6}$, whose quantum dimensions are given by
\begin{equation}
(d_{\boldsymbol{0}}, d_{\boldsymbol{1/2}},d_{\boldsymbol{1}},d_{\boldsymbol{3/2}},d_{\boldsymbol{2}},d_{\boldsymbol{5/2}},d_{\boldsymbol{6}}) = \left(1, 1+\sqrt{2}, \sqrt{2+\sqrt{2}}, \sqrt{4+2\sqrt{2}},  \sqrt{2+\sqrt{2}},  1+\sqrt{2}, 1\right).
\end{equation}
There are three inequivalent non-trivial choices of external object to define an anyonic chain in this case. The choice $a = \boldsymbol{\frac12}$ (equivalently $a = \boldsymbol{\frac52}$) leads to a spin-$\frac12$ chain whose properties are described in section \ref{sec:su24}. The choice $a = \boldsymbol{1}$ (equivalently $a = \boldsymbol{2}$) leads to a spin-1 chain whose properties are described in section \ref{sec:su25}. We note that this case is special in the spin-1 case since the objects in the projection channel $\boldsymbol{1}$ and $\boldsymbol{2}$ are mapped into each other under \eqref{eq:auto}. This leads, for example, to the spectrum of the operator corresponding to \eqref{eq:Hpsu25} being invariant under $\theta \rightarrow \frac{3\pi}{2} - \theta$. Furthermore, the integrable points $\theta_{\text{IK}} = -\arctan \sqrt{2} $ and $\theta_{\text{FZ}} = -\arctan \frac{1}{\sqrt{2}}$ are symmetrically distributed around $\theta_{\text{TL}}$. We note that the physics of this model can be extracted by studying the integer sector $\mathfrak{psu}(2)_6$, which is isomorphic to the Haagerup--Izumi ring HI$(\mathbb{Z}_2)$.

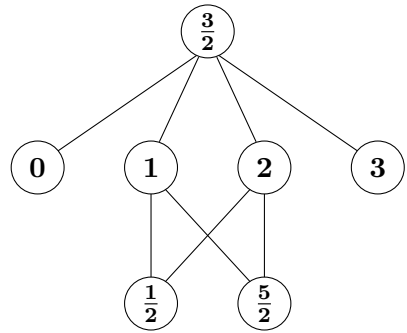
\begin{figure}[H]
\centering

\begin{tikzpicture}[scale=1]

\draw (2.25,1.8) circle [radius=0.35] node {$\boldsymbol{\frac{3}{2}}$};

\draw (0,0) circle [radius=0.35] node {$\boldsymbol{0}$};
\draw (1.5,0) circle [radius=0.35] node {$\boldsymbol{1}$};
\draw (3.0,0) circle [radius=0.35] node {$\boldsymbol{2}$};
\draw (4.5,0) circle [radius=0.35] node {$\boldsymbol{3}$};

\draw (1.5,-1.8) circle [radius=0.35] node {$\boldsymbol{\frac{1}{2}}$};
\draw (3.0,-1.8) circle [radius=0.35] node {$\boldsymbol{\frac{5}{2}}$};

\draw (2.10,1.48) -- (0.27,0.22);
\draw (2.13,1.46) -- (1.61,0.34);
\draw (2.37,1.46) -- (2.89,0.34);
\draw (2.40,1.48) -- (4.23,0.22);

\draw (1.5,-1.45) -- (1.5,-0.35);
\draw (1.69,-1.51) -- (2.81,-0.29);

\draw (2.81,-1.51) -- (1.69,-0.29);
\draw (3.0,-1.45) -- (3.0,-0.35);

\end{tikzpicture}
\caption{Adjacency graph for the $\mathfrak{su}(2)_6$ fusion category with external object $\boldsymbol{\frac{3}{2}}$.}
\label{fig:su2_6_threehalf_adjacency}
\end{figure}

The choice of external object which leads to a previously unstudied model is $a = \boldsymbol{\frac32}$. This model is particularly interesting in this case since it is a fixed point under the automorphism \eqref{eq:auto}. Therefore, similarly to the case $a = \boldsymbol{1}$ for $\mathfrak{su}(2)_4,$ we expect an enhancement of symmetry. The adjacency graph for this choice of external object is shown in figure \ref{fig:su2_6_threehalf_adjacency}. We note that the states $\ket{x_1,x_2,\dots, x_L}$ consist of alternating integer/half-integer spin objects, and so can only be formed on the periodic chain for even $L$. In this case the dimension scales $\text{dim }V^{\boldsymbol{3/2}}_{\mathfrak{su}(2)_6} \sim 2 d_{\boldsymbol{3/2}}^L$. Due to the fusion rule $\boldsymbol{\frac{3}{2}}\otimes \boldsymbol{\frac{3}{2}} = \boldsymbol{0} \oplus \boldsymbol{1}\oplus \boldsymbol{2}\oplus \boldsymbol{3}$, the general anyonic chain is a linear combination of three independent projectors

\begin{equation}\label{eq:Hsu26}
H^{\boldsymbol{3/2}}_{\mathfrak{su}(2)_6} = \alpha \mathrm{P}^{\boldsymbol{3/2}}_{\boldsymbol{1}} + \beta  \mathrm{P}^{\boldsymbol{3/2}}_{\boldsymbol{2}}  + \gamma  \mathrm{P}^{\boldsymbol{3/2}}_{\boldsymbol{3}}.
\end{equation}
Applying the constrained boost operator formalism to this Hamiltonian, we find a host of integrable models. The first solution corresponds to all parameters being equal
\begin{equation}
\alpha = \beta = \gamma, 
\end{equation}
which is the Temperley-Lieb Hamiltonian since $\mathrm{P}^{\boldsymbol{3/2}}_{\boldsymbol{1}} + \mathrm{P}^{\boldsymbol{3/2}}_{\boldsymbol{2}}  + \mathrm{P}^{\boldsymbol{3/2}}_{\boldsymbol{3}} = - \mathrm{P}^{\boldsymbol{3/2}}_{\boldsymbol{0}}$. Since the TL parameter is $\delta = d_{\boldsymbol{3/2}} > 2$ this model is not critical. Furthermore, there are two isolated integrable points which are reminiscent of the spin-1 case:
\begin{equation}
(\alpha, \beta) = (4-\sqrt{2}, 5-3\sqrt{2})\frac{\gamma}{7}, \qquad  (\alpha, \beta) =( 2+3\sqrt{2}, 3+\sqrt{2}) \frac{\gamma}{7},
\end{equation}
which are single points after scaling $\gamma \rightarrow 1$. Another solution is simply
\begin{equation}
\alpha = \beta = 0,
\end{equation}
which is expected because the $\mathrm{P}^{\boldsymbol{3/2}}_{\boldsymbol{3}} $ is equivalent to the TL projector $\mathrm{P}^{\boldsymbol{3/2}}_{\boldsymbol{0}}$ under \eqref{eq:auto}. Finally, there is a one-parameter family of integrable Hamiltonians
\begin{equation}
\alpha + \beta = \gamma.
\end{equation}
 These results for integrable spin-$\frac32$ anyonic chains are new. It would be interesting to study these models numerically, especially in the context of the full phase diagram of the model \eqref{eq:Hsu26}. This is much more challenging compared to the spin-1 case, since after normalization there are two angular parameters in the Hamiltonian. There have been studies of this phase diagram in the undeformed $\mathfrak{su}(2)$ case ($k=\infty$), where unconventionally ordered spin nematic and antinematic states were found \cite{PhysRevLett.106.097202}.

\subsection{$\mathfrak{su}(2)_7 = \mathbb{Z}_2\times \mathfrak{psu}(2)_7$}
In this case we can consider $\mathfrak{psu}(2)_7 = \{\bs{0}, \bs{1}, \bs{2},\bs{3}\}$, where the quantum dimensions are
\begin{equation}
(d_{\bs{0}},d_{\bs{1}},d_{\bs{2}},d_{\bs{3}}) = \left(1, 1+2\cos \frac{2\pi}{9}, 1+2\cos \frac \pi 9 ,2 \cos \frac \pi 9\right) \sim (1, 1.88, 2.88, 2.53).
\end{equation}
Due to \eqref{eq:auto} the choice $a = \bs{2}$ leads to a spin-$\frac32$ anyonic chain. The corresponding constrained Hilbert space of dimension $\sim d_{\bs{2}}^L$ is described by the adjacency diagram in figure \ref{fig:psu2_7_two_adjacency}. The general anyonic chain is a linear combination of three independent projectors
\begin{equation}
H^{\bs{2}}_{\mathfrak{psu}(2)_7}=\alpha \mathrm{P}^{\bs{2}}_{\bs{1}} + \beta  \mathrm{P}^{\bs{2}}_{\bs{2}} + \gamma  \mathrm{P}^{\bs{2}}_{\bs{3}}.
\end{equation}
Applying the constrained boost operator formalism to this operator leads to three integrable Hamiltonians. There is the Temperley-Lieb solution $\alpha = \beta = \gamma$, as well as two further integrable points
\begin{align}
(\alpha, \beta, \gamma) &= \left(3 + 3\cos \frac \pi 9+2\cos \frac{2\pi}{9}, 3 + 2\cos \frac \pi 9+2\cos \frac{2\pi}{9}, 2 + \cos \frac \pi 9+2\cos \frac{2\pi}{9}\right),  \\
(\alpha, \beta, \gamma) &=  \left(1, 5 + 4\cos \frac \pi 9, 4(1 + \cos \frac \pi 9+\cos \frac{2\pi}{9})\right). 
\end{align}
We expect three isolated integrable points to be the general situation for spin-$\frac32$ chains of arbitrary $k$, since the $\mathfrak{su}(2)_6$ case had extra symmetry. \footnote{We find three isolated points numerically in the $\mathfrak{psu}(2)_8$ and $\mathfrak{psu}(2)_9$ spin-$\frac32$ case.}

\begin{figure}[H]
\centering

\begin{tikzpicture}[scale=1]

\draw (0,1.5) circle [radius=0.4] node {$\boldsymbol{0}$};
\draw (2,1.5) circle [radius=0.4] node {$\boldsymbol{1}$};
\draw (0,0) circle [radius=0.4] node {$\boldsymbol{2}$};
\draw (2,0) circle [radius=0.4] node {$\boldsymbol{3}$};

\draw (0,1.1) -- (0,0.4);
\draw (1.65,1.25) -- (0.35,0.25);
\draw (2,1.1) -- (2,0.4);
\draw (0.4,0) -- (1.6,0);

\draw (2.2,1.846) to[out=30,in=-30,looseness=5] (2.2,1.154);

\draw (-0.2,0.346) to[out=150,in=210,looseness=5] (-0.2,-0.346);

\end{tikzpicture}
\caption{Adjacency graph for the $\mathfrak{psu}(2)_7$ fusion category with external object $\boldsymbol{2}$.}
\label{fig:psu2_7_two_adjacency}
\end{figure}

\section{Other fusion categories}\label{sec:other}
In this section we continue our review of anyonic chains beyond the $\mathfrak{su}(2)_k$ case. 

\subsection{$\mathfrak{so}(5)_2$}
One interesting fusion category, with applications to the topological Kondo effect, is $\mathfrak{so}(5)_2$ \cite{Finch:2014ina,Finch:2014nxa, Finch:2017puy}. There are 6 objects $\psi_i$ for $i=1,\dots,6$, with $\psi_1$ corresponding to the identity object. The fusion rules are given in table \ref{tab:so52_fusion}. There are two subalgebras, $\mathbb{Z}_2 = \{\psi_1,\psi_2\}$ and $\text{Rep}(D_5)=\{\psi_1,\psi_2,\psi_3,\psi_4\}$. There are four unitary solutions to the pentagon equations, and for concreteness we take those given in FR$^{6,0}_{9,3,1,1}$ of Anyonwiki \cite{gert_vercleyen_anyonwiki}, which is equivalent to the choice in \cite{Finch:2014ina}. The corresponding quantum dimensions of the objects are

\begin{equation}
(d_{\psi_1}, d_{\psi_2}, d_{\psi_3}, d_{\psi_4}, d_{\psi_5}, d_{\psi_6}) = (1,1,2,2,\sqrt{5},\sqrt{5}).
\end{equation}

\begin{table}[H]
\centering
\renewcommand{\arraystretch}{1.15}
\setlength{\tabcolsep}{4pt}
\[
\begin{array}{c|cccccc}
\otimes & \psi_1 & \psi_2 & \psi_3 & \psi_4 & \psi_5 & \psi_6 \\ \hline
\psi_1 & \psi_1 & \psi_2 & \psi_3 & \psi_4 & \psi_5 & \psi_6 \\
\psi_2 & \psi_2 & \psi_1 & \psi_3 & \psi_4 & \psi_6 & \psi_5 \\
\psi_3 & \psi_3 & \psi_3 & \psi_1 \oplus \psi_2 \oplus \psi_4 & \psi_3 \oplus \psi_4 & \psi_5 \oplus \psi_6 & \psi_5 \oplus \psi_6 \\
\psi_4 & \psi_4 & \psi_4 & \psi_3 \oplus \psi_4 & \psi_1 \oplus \psi_2 \oplus \psi_3 & \psi_5 \oplus \psi_6 & \psi_5 \oplus \psi_6 \\
\psi_5 & \psi_5 & \psi_6 & \psi_5 \oplus \psi_6 & \psi_5 \oplus \psi_6 & \psi_1 \oplus \psi_3 \oplus \psi_4 & \psi_2 \oplus \psi_3 \oplus \psi_4 \\
\psi_6 & \psi_6 & \psi_5 & \psi_5 \oplus \psi_6 & \psi_5 \oplus \psi_6 & \psi_2 \oplus \psi_3 \oplus \psi_4 & \psi_1 \oplus \psi_3 \oplus \psi_4
\end{array}
\]
\caption{Fusion rules for $\mathfrak{so}(5)_2$.}
\label{tab:so52_fusion}
\end{table}

\par\medskip
\noindent\textbf{Anyonic chains and integrability.}\quad
There are two non-equivalent choices of external object $a$, leading to distinct anyonic chains. Taking $a=\psi_3$ leads to a Rep$(D_5)$ chain, which is a combination of two independent projectors
\begin{equation}
 H^{\psi_3}_{\mathfrak{so}(5)_2} = \alpha_1 \mathrm{P}^{\psi_3}_{\psi_2} +\alpha_2  \mathrm{P}^{\psi_3}_{\psi_4},
\end{equation}
and acts on a constrained Hilbert space of dimension $\sim 2\times 2^L$. Similarly to the $\mathfrak{psu}(2)_4\simeq $ Rep$(D_3)$ case, this operator is integrable for all choices of $\alpha_1,\alpha_2$.

\begin{figure}[H]
\centering

\begin{tikzpicture}[scale=1]

\draw (0,0) circle [radius=0.4] node {$\psi_1$};
\draw (1.8,0) circle [radius=0.4] node {$\psi_3$};
\draw (3.6,0) circle [radius=0.4] node {$\psi_4$};
\draw (5.4,0) circle [radius=0.4] node {$\psi_2$};

\draw (1.8,-1.6) circle [radius=0.4] node {$\psi_5$};
\draw (3.6,-1.6) circle [radius=0.4] node {$\psi_6$};

\draw (0.35,-0.18) -- (1.45,-1.42); 

\draw (1.8,-0.4) -- (1.8,-1.2);     
\draw (2.15,-0.2) -- (3.25,-1.4);    

\draw (3.25,-0.2) -- (2.15,-1.4);     
\draw (3.6,-0.4) -- (3.6,-1.2);  

\draw (5.05,-0.18) -- (3.95,-1.42);   

\end{tikzpicture}
\caption{Adjacency graph for the $\mathfrak{so}(5)_2$ fusion category with external object $\psi_5$.}
\label{fig:so5_2_psi5_adjacency}
\end{figure}
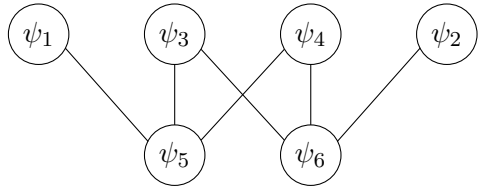

Taking the external object $a = \psi_5$ leads to a model acting on a constrained Hilbert space of dimension $\sim 2\times 5^{L/2}$, with an adjacency diagram shown in figure \ref{fig:so5_2_psi5_adjacency}. This model, studied in \cite{Finch:2014ina}, is also a linear combination of two independent projectors\footnote{We note that the choices $a=\psi_4,\psi_6$ are equivalent to $a=\psi_3,\psi_5$ under the automorphism $\psi_3\leftrightarrow \psi_4, \psi_5\leftrightarrow \psi_6$.}:
\begin{equation}
 H^{\psi_5}_{\mathfrak{so}(5)_2} = \beta_1 \mathrm{P}^{\psi_5}_{\psi_3} +\beta_2  \mathrm{P}^{\psi_5}_{\psi_4}.
\end{equation}
In this case there are three integrable points. The first is the standard TL integrable point which corresponds to $\beta_1 = \beta_2$. The other integrable points are
\begin{equation}\label{eq:so52sol}
\beta_2 = \frac{\beta_1}{-1+\sqrt{5}}, \qquad \frac{4\beta_1}{1+\sqrt{5}},
\end{equation}
which give different representations of the BMW algebra. The TL point in the antiferromagnetic regime corresponds to an XXZ spin chain with a very small gap (since $\delta = \sqrt{5} > 2$), and in the ferromagnetic regime it corresponds to a massive ferromagnetic XXZ chain with $c = 1$. The integrable points both correspond to critical models with $c=8/7$. The full phase diagram can be found in \cite{Finch:2014ina}.

\subsection{HI$(\mathbb{Z}_n)$}
A large class of fusion rings labeled by a finite group $G$ was introduced by Izumi \cite{Izumi:2001mi, grossman2019drinfeldcentersfusioncategories, Grossman_2020}, dubbed \textit{Haagerup-Izumi} fusion rings $\text{HI}(G)$. In recent years there has been a surge of interest in models defined from the Haagerup fusion ring, which is the case $G = \mathbb{Z}_3$ and the simplest example of a fusion category not related to a deformation of a Lie algebra \cite{haagerup, Grossman_2012}.\footnote{The case $G = \mathbb{Z}_2$ is isomorphic to $\mathfrak{psu}(2)_6$.} Although evidence for a $c=8$ CFT with Haagerup symmetry was first proposed in \cite{Evans:2010yr}, numerical calculations were unfeasible due to the lack of an explicit form of the $F$-symbols of this fusion category. The calculation of these $F$-symbols was completed in \cite{osborne2019fsymbols, Huang:2020lox}, which allowed for numerical searches for signs of conformal symmetry \cite{Wolf:2020qdo}. This led to the discovery of a $c=2$ CFT at the level of the anyonic chain \cite{Huang:2021nvb} and in \cite{Vanhove:2021zop} within the formalism of strange correlators. Several further models with Haagerup symmetry have been constructed recently \cite{Bottini:2025hri,Albert:2025umy,Jia:2024wnu}.

\par\medskip
\noindent\textbf{Fusion rules and anyonic chain.}\quad
The Haagerup fusion ring HI$(\mathbb{Z}_3)$ consists of six objects $\{1,a,a^2,\rho,a\rho,a^2\rho\}$. The objects $1,a, a^2$ are invertible, whereas $\rho,a\rho,a^2\rho$ are non-invertible. The non-trivial fusion rules are
\begin{equation}\label{eq:haagerupfusion}
\rho\otimes \rho = 1\oplus \rho \oplus a\rho \oplus a^2\rho, \qquad \rho\otimes a = a^2\rho,
\end{equation}
and $a^3 = 1$. The full fusion table can be populated using \eqref{eq:haagerupfusion}, for example $\rho \otimes a \rho = a^2 \rho\otimes \rho = a^2\oplus \rho \oplus  a\rho \oplus a^2\rho.$ There are two unitary solutions to the pentagon equations in this case (given in \cite{Huang:2020lox}), which correspond to the Haagerup $\mathcal{H}_2$ and $\mathcal{H}_3$ fusion categories. The corresponding anyonic chains are equivalent, so we focus on the $\mathcal{H}_3$ case. The quantum dimensions are
\begin{equation}
(d_1, d_a, d_{a^2}, d_{\rho}, d_{a \rho}, d_{a^2\rho}) = \left(1,1,1,\frac{3+\sqrt{13}}{2},\frac{3+\sqrt{13}}{2},\frac{3+\sqrt{13}}{2}\right).
\end{equation}
Taking the external object $\rho$ leads to an anyonic chain with four non-trivial projectors
\begin{equation}\label{eq:HamiltonianHag}
H^\rho_{\mathcal{H}_3} = \alpha_1 \mathrm{P}^{\rho}_{1}+\alpha_2 \mathrm{P}^{\rho}_{\rho}+\alpha_3 \mathrm{P}^{\rho}_{a\rho}+\alpha_4 \mathrm{P}^{\rho}_{a^2\rho},
\end{equation}
of which only three are independent due to the relation $\mathrm{P}^{\rho}_{1,i}+ \mathrm{P}^{\rho}_{\rho,i}+ \mathrm{P}^{\rho}_{a\rho,i}+ \mathrm{P}^{\rho}_{a^2\rho,i} = 1$. The Hamiltonian \eqref{eq:HamiltonianHag} was first studied in \cite{Wolf:2020qdo}, and acts on a Hilbert space of dimension $\sim (\frac{3+\sqrt{13}}{2})^L$. It was argued in \cite{Huang:2021nvb} that the projector $\mathrm{P}^{\rho}_{\rho}$, i.e.\ the case $\alpha_1=\alpha_3=\alpha_4=0$, corresponds to a Haagerup CFT in the long chain limit with $c=2$, although the precise CFT to which this corresponds remains unknown.

\par\medskip
\noindent\textbf{Integrable $\mathcal{H}_3$ chains.}\quad
Applying the constrained boost operator formalism, discussed in section \ref{sec:cboost}, to the Hamiltonian \eqref{eq:HamiltonianHag} leads to the result that the only integrable anyonic chain compatible with Haagerup symmetry is the TL projector $\mathrm{P}^{\rho}_1$ \cite{Corcoran:2024eeh}. Therefore, in order to find more integrable Hamiltonians on the Haagerup Hilbert space, it is necessary to break this symmetry and consider Hamiltonians beyond the projector ansatz \eqref{eq:HamiltonianHag}. This is analogous to constructing integrable models on Rydberg-blockaded chains which don't necessarily commute with the Fibonacci fusion category symmetry.

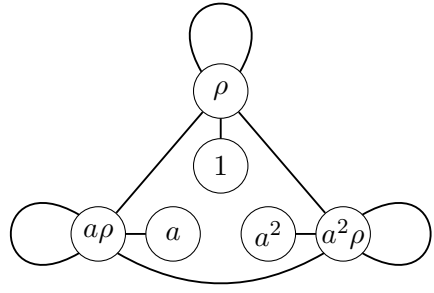
\begin{figure}
\centering
\begin{tikzpicture}[scale=0.9]

\draw (-1.8,0) circle [radius=0.4] node {$a \rho$} ;
\draw (-0.7,0) circle [radius=0.4] node {$a $};
\draw (0.7,0) circle [radius=0.4] node {$a^2 $};
\draw (1.8,0) circle [radius=0.4] node {$a^2\rho$};
\draw (0,1) circle [radius=0.4] node {$1$};
\draw (0,2.1) circle [radius=0.4] node {$\rho$};

\draw [line,] (-1.1,0) -- (-1.4,0);
\draw [line,] (1.1,0) -- (1.4,0);
\draw [line,] (0,1.4) -- (0,1.7);

\draw [line] (-1.535, 0.309167) -- (-0.265, 1.79083);
\draw [line] (1.535, 0.309167) -- (0.265, 1.79083);

\draw[line]    (-1.5172,-0.2828) to[out=-30,in=-150] (1.5172,-0.2828);

\draw[line]    (-1.8-0.2828,0.2828) to[out=-210,in=-150,looseness=7] (-1.8-0.2828,-0.2828);

\draw[line]    (1.8+0.2828,0.2828) to[out=30,in=-30,looseness=7] (1.8+0.2828,-0.2828);

\draw[line]    (0.2828,2.1+0.2828) to[out=60,in=120,looseness=7] (-0.2828,2.1+0.2828);

\end{tikzpicture}
\caption{Adjacency rules for the Haagerup Hilbert space.}
\label{fig:allowedstate}
\end{figure}

The Haagerup Hilbert space consists of states $\ket{x_1x_2\cdots x_L}$, where each $x_i\in \{1,a,a^2,\rho,a\rho,a^2\rho\}$ and the constraint $x_{i+1}\in \rho\otimes x_i$ leads to the adjacency rule summarized in figure \ref{fig:allowedstate}.
The topological symmetry $Y_a$ is the operator which maps every $x_i$ to $a\otimes x_{i}$. Imposing that a Hamiltonian $H$ commutes with $Y_a$ leads to the decomposition
\begin{equation}
H = \tilde{H} + Y_a \tilde{H}Y_{a}^{-1} + Y_a^{2} \tilde{H}Y_a^{-2}.
\end{equation}
Further imposing that $H=\sum_i\mathcal{H}_{i,i+1,i+2}$ is range 3, symmetric, and has a generalized `PXP' form $\mathcal{H}_{i,i+1,i+2} = D_i \mathcal{O}_{i+1} D_{i+2}$, there are 18 independent operators contributing to the reduced Hamiltonian $\tilde{\mathcal{H}}_{i,i+1,i+2}$. These can be expressed in terms of the local projectors
\begin{equation}
P_x \ket{y} = \delta_{xy}\ket{y}, \qquad x,y \in \{1,a,a^2,\rho,a\rho,a^2\rho\},
\end{equation}
as well as the `flips' $X:\ket{1}\leftrightarrow \ket{\rho}$, $Y:\ket{1}\leftrightarrow \ket{a\rho}$, $Z:\ket{1}\leftrightarrow \ket{a^2\rho}$, and the transposition $T:\ket{\rho}\leftrightarrow \ket{a\rho}$.
In terms of these operators, two integrable reduced Hamiltonians of note, found using the boost operator formalism, are \cite{Corcoran:2024eeh}
\begin{align}\label{eq:Hsol}
\tilde{\mathcal{H}}_{i,i+1,i+2}= &d_\rho^{-1/2} P_{\rho}(X+Y+Z)P_\rho + P_{\rho}TP_{\rho} +P_{a\rho}TP_{a\rho}+ P_{a^2\rho}TP_{a^2\rho} \notag\\ &+ d_\rho  P_1P_{\rho}P_1 + d_\rho^{-1}P_\rho P_1 P_\rho + P_\rho P_\rho P_\rho+ P_{\rho}P_{a\rho}P_{\rho}+P_{a\rho}P_{\rho}P_{a\rho} \\
\tilde{\mathcal{H}}_{i,i+1,i+2}&= \gamma^{-1/2} P_{\rho}(X+Y+Z)P_\rho + P_{\rho}TP_{\rho} +P_{a\rho}TP_{a\rho}+ P_{a^2\rho}TP_{a^2\rho}\notag\\
 +&(1+2\gamma) P_1P_{\rho}P_1 + (1+2\gamma^{-1})P_\rho P_1 P_\rho +2\gamma( P_\rho P_\rho P_\rho+ P_{\rho}P_{a\rho}P_{\rho}+P_{a\rho}P_{\rho}P_{a\rho}),
\end{align}
where $\gamma\coloneqq \frac{1+\sqrt{3}}{2}$. The Hamiltonian \eqref{eq:Hsol} is an explicit representation of $\mathrm{P}^{\rho}_1$, and is not critical since it is a Temperley-Lieb integrable model with $\delta = \frac{3+\sqrt{13}}{2}>2$ \cite{Blakeney:2025ext}. The integrable Hamiltonian \eqref{eq:Hsol} does not commute with the full Haagerup symmetry, although it has been argued to critical in the continuum limit with $c=3/2$. A one-parameter integrable family to which both of these integrable points belong, as exists for the golden chain, is yet to be found.

\par
\begin{figure}[tbp]
\centering
\begin{tikzpicture}[scale=0.95,
    every node/.style={circle, draw, inner sep=2pt, minimum size=0.8cm},
    every loop/.style={looseness=8, min distance=8mm}
]

\node (g0) at (90:1.35) {$1$};
\node (g1) at (18:1.35) {$a$};
\node (g2) at (-54:1.35) {$a^2$};
\node (g3) at (-126:1.35) {$a^3$};
\node (g4) at (162:1.35) {$a^4$};

\node[minimum size=0.9cm] (r0) at (90:3.0) {$\rho$};
\node[minimum size=0.9cm] (r1) at (18:3.0) {$a\rho$};
\node[minimum size=0.9cm] (r2) at (-54:3.0) {$a^2\rho$};
\node[minimum size=0.9cm] (r3) at (-126:3.0) {$a^3\rho$};
\node[minimum size=0.9cm] (r4) at (162:3.0) {$a^4\rho$};

\draw (g0) -- (r0);
\draw (g1) -- (r1);
\draw (g2) -- (r2);
\draw (g3) -- (r3);
\draw (g4) -- (r4);

\draw (r0) -- (r1);
\draw (r1) -- (r2);
\draw (r2) -- (r3);
\draw (r3) -- (r4);
\draw (r4) -- (r0);

\draw (r0) -- (r2);
\draw (r0) --  (r3);
\draw (r1) -- (r3);
\draw (r1) -- (r4);
\draw (r2) -- (r4);

\draw (r0) edge[loop, out=120, in=60, looseness=6] (r0);
\draw (r1) edge[loop, out=30, in=-30, looseness=6] (r1);
\draw (r2) edge[loop, out=-30, in=-110, looseness=5] (r2);
\draw (r3) edge[loop, out=-70, in=-150, looseness=5] (r3);
\draw (r4) edge[loop, out=210, in=150, looseness=6] (r4);

\end{tikzpicture}
\caption{Adjacency rules for the $\text{HI}(\mathbb{Z}_5)$ Hilbert space with external object $\rho$.}
\label{fig:HI_Z5_rho_adjacency}
\end{figure}
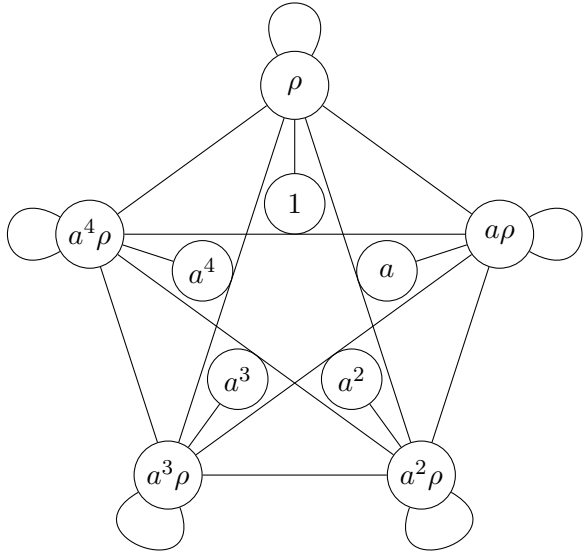

\begin{figure}[t]
  \centering
  \begin{minipage}[t]{0.49\textwidth}
    \centering
    \includegraphics[width=\linewidth]{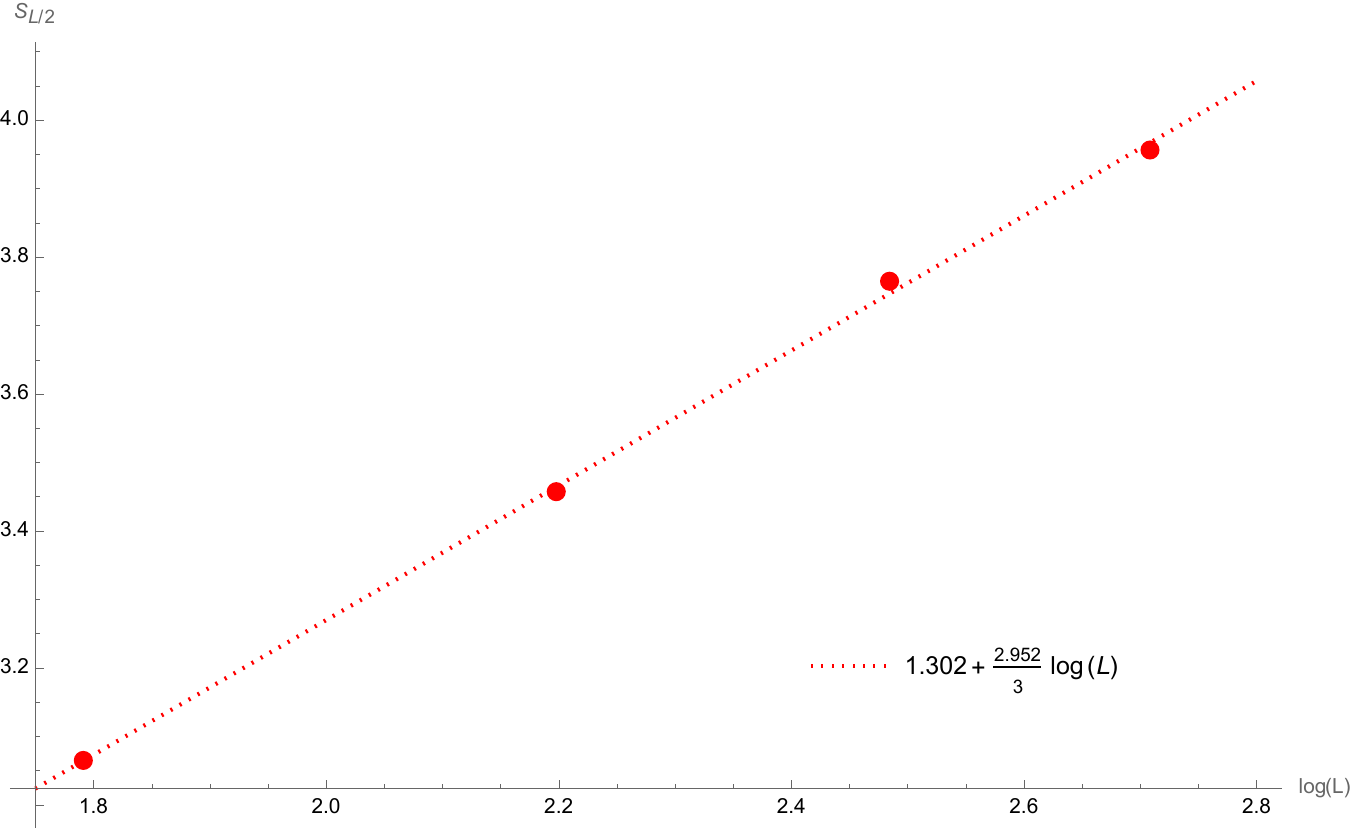}
  \end{minipage}\hfill
  \begin{minipage}[t]{0.49\textwidth}
    \centering
    \includegraphics[width=\linewidth]{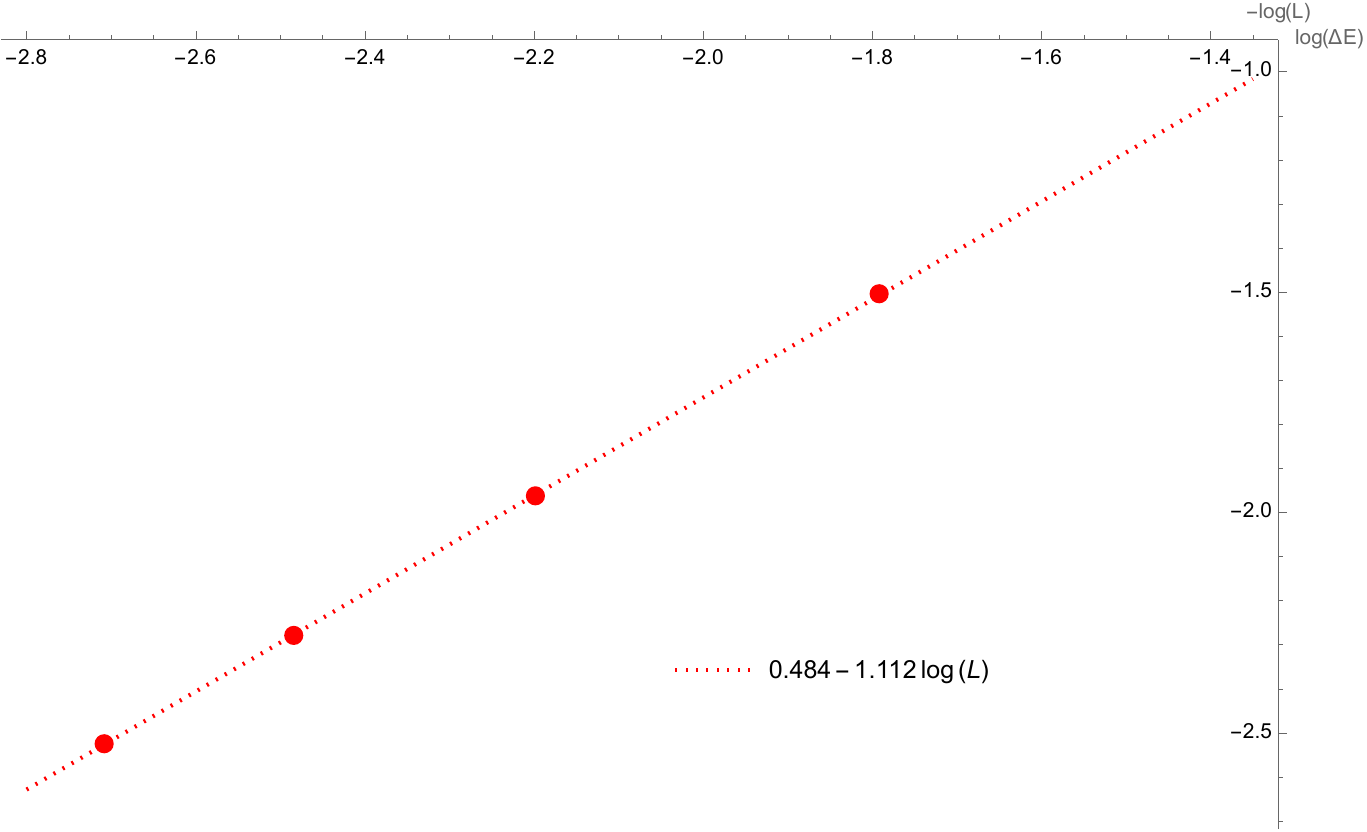}
  \end{minipage}

  \caption{Half-chain entanglement entropy and gap for the $\text{HI}(\mathbb{Z}_5)$ anyonic chain $\mathrm{P}^{\rho}_\rho$. For both plots we take $L=6,9,12,15$.}
  \label{fig:H5-ee-gap}
\end{figure}

\medskip
\noindent\textbf{Beyond $\mathcal{H}_3$.}\quad
In \cite{Huang:2020lox} the fusion rings HI($\mathbb{Z}_n$) are considered, and unitary $F$-symbols are calculated for odd $n$ up to $n=15$. This provides a wealth of new anyonic chain models to study. However, most of these are beyond the reach of current numerical methods due to the large Hilbert spaces involved. For example, for $n=5$ there are ten objects $\{1,a,a^2,a^3,a^4,\rho,a\rho,a^2\rho,a^3\rho,a^4\rho\}$. Taking the external object $\rho$ leads to a constrained Hilbert space of dimension $\sim \left(\frac{5+\sqrt{29}}{2}\right)^L$, although from the point of view of DMRG the Hilbert space is of dimension $10^L$. The adjacency diagram in this case is shown in figure \ref{fig:HI_Z5_rho_adjacency}. Although there are five independent projectors, given the results in the $n=3$ case it is most interesting to test the criticality of $\mathrm{P}^{\rho}_\rho$. Preliminary numerical evidence (up to $L=15$) from the authors suggests that, similarly to the $n=3$ case, the operator $\mathrm{P}^{\rho}_\rho$ is potentially critical with $c\sim 3$. We display the numerical results in figure \ref{fig:H5-ee-gap}. Although the finite-size effects should be large in this case, we note that the analogous numerics for $\text{HI}(\mathbb{Z}_3)$ already give a ballpark central charge estimate $c\sim 2.1$, which is not too far from the result $c=2$ suggested in \cite{Huang:2021nvb}. However, without more specialized numerics tailored to the constrained Hilbert space, it is difficult to make precise conjectures.

For even $n$, the $F$-symbols are much harder to calculate because the transparent ansatz \cite{Huang:2020lox} which reduces the number of non-trivial pentagon equations from $\mathcal{O}(n^6)$ to $\mathcal{O}(n^2)$ is no longer valid. The case $n=2$ is known due to the equivalence $\text{HI}(\mathbb{Z}_2)\simeq \mathfrak{psu}(2)_6$. However, already for $n=4$ to pentagon identities resist modern methods of solution. Solving the pentagon equations for this case, and for other groups $G$ of small order, could provide several new interesting anyonic chains.

\subsection{TY$(\mathbb{Z}_n)$}
The Tambara-Yamagami fusion category over $\mathbb{Z}_n$ consists of a non-invertible object $\rho$, together with $n$ invertible objects $\alpha_0,\dots \alpha_{n-1}$ \cite{TAMBARA1998692}. The objects $\alpha_i$ form a $\mathbb{Z}_n$ subalgebra:
\begin{equation}
\alpha_i \otimes \alpha_j = \alpha_{i+j \text{ mod } n}
\end{equation}
The object $\rho$ is stable under fusion by invertible objects $\alpha_i\otimes \rho = \rho\otimes \alpha_i = \rho$, and further satisfies
\begin{equation}
\rho\otimes\rho = \bigoplus_{i=0}^{n-1} \alpha_i.
\end{equation}
The quantum dimension of the non-invertible object $\rho$ is $\sqrt{n}$. For $n = 2$ this category is equivalent to the Ising fusion category, discussed in section \ref{sec:su22}, so we have TY$(\mathbb{Z}_2) \simeq \text{Ising}$. For $n = 3$ we have the Potts CFT, and the corresponding anyonic chain is equivalent to the $\mathbb{Z}_3$ chiral clock model \cite{Eck2025QuantumLatticeModels,Millar2019InteractingAnyons}. For higher $n$ this fusion category has been studied less, and we investigate from an integrability point of view.

\par
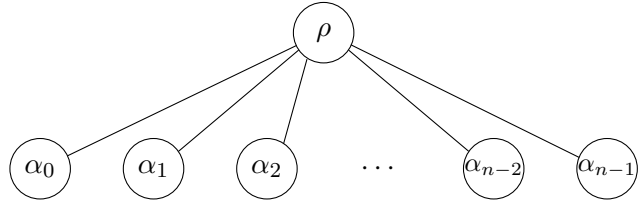
\begin{figure}[tbp]
\centering
\begin{tikzpicture}[scale=1]

\draw (3.75,1.8) circle [radius=0.4] node {$\rho$};

\draw (0,0) circle [radius=0.4] node {$\alpha_0$};
\draw (1.5,0) circle [radius=0.4] node {$\alpha_1$};
\draw (3.0,0) circle [radius=0.4] node {$\alpha_2$};
\draw (4.5,0) node {$\cdots$};
\draw (6.0,0) circle [radius=0.4] node {\small $\alpha_{n-2}$};
\draw (7.5,0) circle [radius=0.4] node {\small $\alpha_{n-1}$};

\draw (0.36,0.17) -- (3.39,1.63);
\draw (1.83,0.23) -- (3.42,1.57);
\draw (3.22,0.33) -- (3.53,1.45);
\draw (5.67,0.23) -- (4.08,1.57);
\draw (7.13,0.15) -- (4.11,1.64);

\end{tikzpicture}
\caption{Adjacency graph for TY$(\mathbb{Z}_n)$ with external object $\rho$.}
\label{fig:TY_Zn_rho_adjacency}
\end{figure}

\medskip
\noindent\textbf{Integrable anyonic chains.}\quad
The TY$(\mathbb{Z}_n)$ anyonic chain describes the fusion of $L$ copies of the object $\rho$, and is a linear combination of $n-1$ independent projectors
\begin{equation}\label{eq:HTY}
H^{\rho}_{\text{TY}(\mathbb{Z}_n)} = \sum_{i=1}^{n-1}\beta_i \mathrm{P}^{\rho}_{\alpha_i},
\end{equation}
where $\beta_i$ are coefficients and $\mathrm{P}^{\rho}_{\alpha_i}$ is the projector $\rho\otimes \rho\rightarrow \alpha_i$. The operator \eqref{eq:HTY} acts on the constrained Hilbert space described by the adjacency diagram in figure \ref{fig:TY_Zn_rho_adjacency}. Since the labels $x_i$ in a state $\ket{x_1x_2\cdots x_L}$ alternate between $\rho$ and $\alpha_i$, the Hilbert space is non-zero only for even $L$, and in this case decomposes into staggering sectors where the $\rho$ object is at either odd or even sites. In this case we have $\text{dim }V^{\rho}_{\text{TY}(\mathbb{Z}_n)} = 2\times n^{L/2}$. 

For $3\leq n \leq 5$ we apply the constrained boost operator formalism to classify integrable combinations of projectors. For each $n$ the case $\beta_1=\beta_2=\cdots=\beta_{n-1}\equiv \beta$ corresponds to the integrable TL projector $-\beta \mathrm{P}^{\rho}_{\alpha_0}$, with $\delta = \sqrt{n}$. We find that for all $n$ the individual projectors $\mathrm{P}^{\rho}_{\alpha_i}$ are also integrable. This is not surprising, as any projector $\mathrm{P}^{\rho}_{\alpha_i}$ can be mapped into the TL projector $\mathrm{P}^{\rho}_{\alpha_0}$ under the $\mathbb{Z}_n$ automorphism of the fusion rules. For $n = 3$ there are no further integrable combinations.

For $n=4$ there are two further families of integrable Hamiltonians
\begin{align}
H^{\beta,\Delta}_{n=4} = &\beta (\mathrm{P}^{\rho}_{\alpha_1} + \mathrm{P}^{\rho}_{\alpha_3}) + \Delta \mathrm{P}^{\rho}_{\alpha_2}, \\
H^{\beta_1,\beta_3}_{n=4} = & \beta_1 \mathrm{P}^{\rho}_{\alpha_1} +\beta_3 \mathrm{P}^{\rho}_{\alpha_3}.
\end{align}
The first of these models is reminiscent to a one-parameter family of integrable models originating from the Rep$(D_4)$ fusion rules, which differ only slightly from TY$(\mathbb{Z}_4)$. This model reduces to the TL projector for $\Delta = \beta$.

\begin{figure}[t]
  \centering
  \begin{minipage}[t]{0.49\textwidth}
    \centering
    \includegraphics[width=\linewidth]{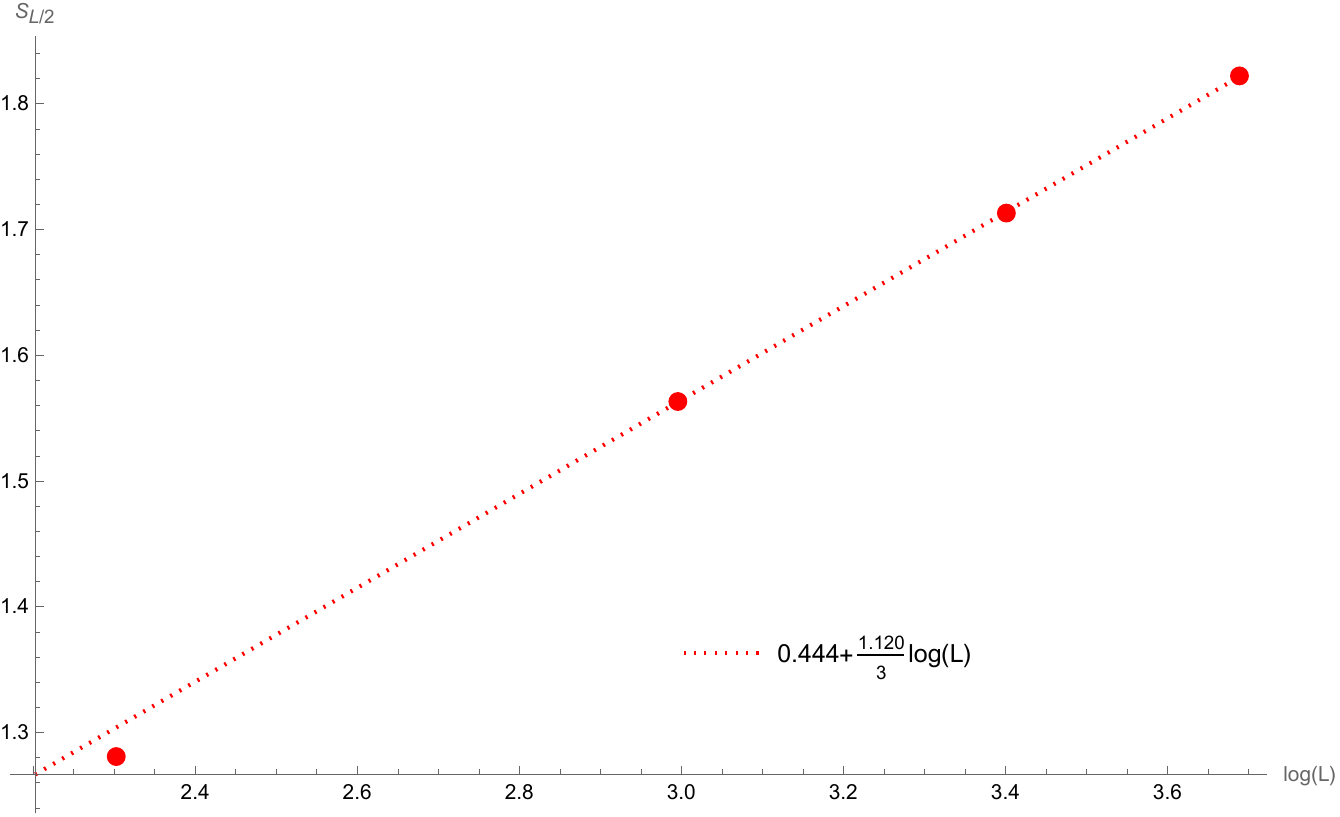}
  \end{minipage}\hfill
  \begin{minipage}[t]{0.49\textwidth}
    \centering
    \includegraphics[width=\linewidth]{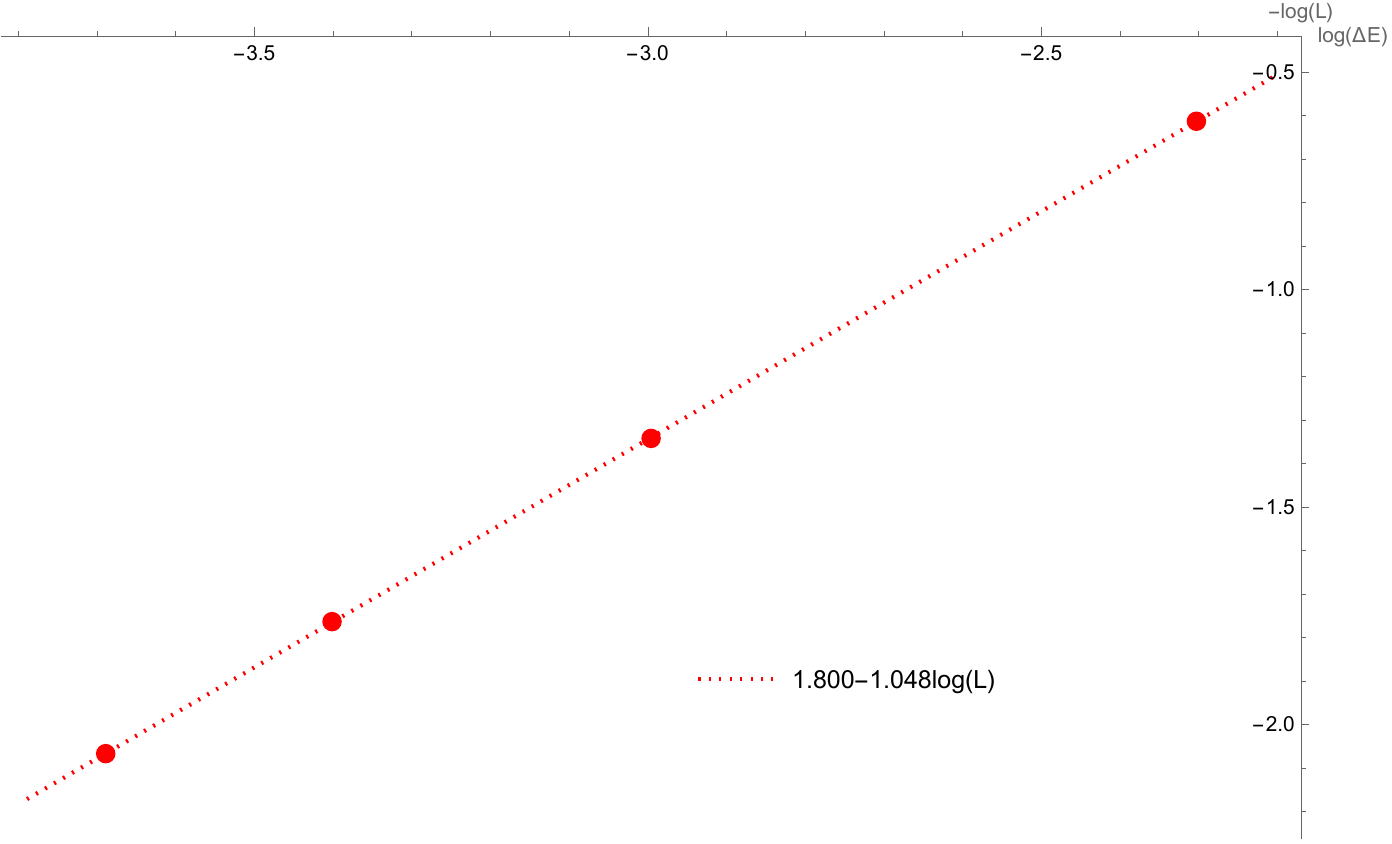}
  \end{minipage}

  \caption{Half-chain entanglement entropy and gap for the TY$(\mathbb{Z}_5)$ chain \eqref{eq:TYintegrable}, $\gamma = 3+\sqrt{5}$. We take $L = 10, 20, 30, 40$. For the entanglement entropy fit we use the last three points.}
  \label{fig:TYZ5-ee-gap}
\end{figure}

For $n=5$ the equations arising from the boost operator formalism are more involved, and we didn't attempt a full classification of integrable chains. We point out a couple solutions, however. The chain
\begin{equation}\label{eq:HTYZ5}
H^{\beta_{1,3},\beta_{2,4}}_{n=5} = \beta_{1,3}(\mathrm{P}^{\rho}_{\alpha_1}+\mathrm{P}^{\rho}_{\alpha_3})+\beta_{2,4}(\mathrm{P}^{\rho}_{\alpha_3}+\mathrm{P}^{\rho}_{\alpha_4})
\end{equation}
is integrable for the choices
\begin{equation}
\beta_{1,3} = \frac{\beta_{2,4}}{-1+\sqrt{5}},\qquad  \frac{4\beta_{2,4}}{1+\sqrt{5}}.
\end{equation}
These choices are reminiscent of the BMW integrable models in the $\mathfrak{so}(5)_2$ chain \eqref{eq:so52sol}. There are also new non-trivial integrable models given by
\begin{equation}\label{eq:TYintegrable}
 H^{\gamma}_{n=5} = \mathrm{P}^{\rho}_{\alpha_3}+\mathrm{P}^{\rho}_{\alpha_4} + \gamma \mathrm{P}^{\rho}_{\alpha_2},
\end{equation}
where $\gamma = 3+\sqrt{5}, -2-\sqrt{5}$. We did not find any integrable models in this case with all $\beta_i$ non-zero beyond \eqref{eq:HTYZ5}. We present numerics for the model \eqref{eq:TYintegrable} with $\gamma = 3+\sqrt{5}$ in figure \ref{fig:TYZ5-ee-gap}. We notice a $\mathbb{Z}_5$ behavior in the ground state of this model, and since the Hilbert space is non-zero only for even $L$ we take $L\equiv 0$ mod 10. Although further numerics are required for a precise statement, we expect this model to lie in the universality class of the $\mathbb{Z}_5$ parafermion CFT with $c = 8/7$.

\subsection{Fib $\times$ Fib}
Recently there has been interest in models constructed by considering products of fusion categories. One of the simplest examples is the choice $\text{Fib}\times \text{Fib}$ \cite{Antunes:2025huk, Blakeney:2025ext, Ferragatta:2026ejs}. In this fusion category the objects are constructed as products of Fib objects $\{1\times 1, 1\times \tau, \tau \times 1, \tau\times\tau\} \coloneqq \{\boldsymbol{1},\boldsymbol{2}, \boldsymbol{3},\boldsymbol{4}\}$. The fusion rules can be derived from the Fib rules, and are shown in table \ref{tab:fibxfib_fusion}. The $F$-symbols can be obtained as products of Fib $F$-symbols. In \cite{Antunes:2025huk} several models based on coupled golden chains, which can also be formulated in terms of $\text{Fib}^{\otimes N}$ anyonic chains, are studied. They find a rich phase diagram with several critical models, including potentially irrational CFTs. We note that models based on coupled TL models, related to those studied here, were previously studied \cite{Fendley_2008, Vernier_2014} , and several integrable/critical points were found.
\begin{table}[H]
\centering
\renewcommand{\arraystretch}{1.15}
\setlength{\tabcolsep}{4pt}
\[
\begin{array}{c|cccc}
\otimes & \boldsymbol{1} & \boldsymbol{2} & \boldsymbol{3} & \boldsymbol{4} \\ \hline
\boldsymbol{1} & \boldsymbol{1} & \boldsymbol{2} & \boldsymbol{3} & \boldsymbol{4} \\
\boldsymbol{2} & \boldsymbol{2} & \boldsymbol{1} \oplus \boldsymbol{2} & \boldsymbol{4} & \boldsymbol{3} \oplus \boldsymbol{4} \\
\boldsymbol{3} & \boldsymbol{3} & \boldsymbol{4} & \boldsymbol{1} \oplus \boldsymbol{3} & \boldsymbol{2} \oplus \boldsymbol{4} \\
\boldsymbol{4} & \boldsymbol{4} & \boldsymbol{3} \oplus \boldsymbol{4} & \boldsymbol{2} \oplus \boldsymbol{4} & \boldsymbol{1} \oplus \boldsymbol{2} \oplus \boldsymbol{3} \oplus \boldsymbol{4}
\end{array}
\]
\caption{Fusion rules for $\mathrm{Fib}\times \mathrm{Fib}$.}
\label{tab:fibxfib_fusion}
\end{table}

\par
\begin{figure}[h!]
\centering

\begin{tikzpicture}[scale=1]

\draw (0,1.5) circle [radius=0.4] node {$\boldsymbol{1}$};
\draw (2,1.5) circle [radius=0.4] node {$\boldsymbol{2}$};
\draw (0,0) circle [radius=0.4] node {$\boldsymbol{3}$};
\draw (2,0) circle [radius=0.4] node {$\boldsymbol{4}$};

\draw (0.33,1.27) -- (1.67,0.23); 
\draw (1.67,1.27) -- (0.33,0.23);  
\draw (2,1.1) -- (2,0.4);         
\draw (0.4,0) -- (1.6,0);   

\draw (2.2,0.346) to[out=30,in=-30,looseness=5] (2.2,-0.346);

\end{tikzpicture}
\caption{Adjacency graph for $\mathrm{Fib}\times\mathrm{Fib}$ with external object $\boldsymbol{4}$.}
\label{fig:fibxfib_four_adjacency}
\end{figure}
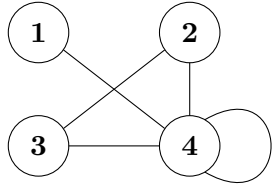

\medskip
\noindent\textbf{Integrable anyonic chains.}\quad
The quantum dimensions of the Fib$\times$Fib objects are
\begin{equation}
(d_{\boldsymbol{1}}, d_{\boldsymbol{2}}, d_{\boldsymbol{3}}, d_{\boldsymbol{4}})  = (1, \varphi, \varphi, \varphi+1),
\end{equation}
where $\varphi=\frac{1+\sqrt{5}}{2}$. The choice of external object $a = \boldsymbol{2}$ or $\boldsymbol{3}$ leads directly to the golden chain. Therefore the only new choice is $a=\boldsymbol{4}$. In this case the anyonic chain is a combination of three independent projectors

\begin{equation}\label{eq:HFibFib}
H^{\text{Fib}\times\text{Fib}} = \alpha_1 \mathrm{P}^{\boldsymbol{4}}_{\boldsymbol{4}} + \alpha_2 \mathrm{P}^{\boldsymbol{4}}_{\boldsymbol{3}} + \alpha_3\mathrm{P}^{\boldsymbol{4}}_{\boldsymbol{2}}.
\end{equation}
This Hamiltonian acts on a Hilbert space of dimension $\sim(\varphi+1)^L$, with adjacency diagram shown in figure \ref{fig:fibxfib_four_adjacency}. Applying the constrained boost operator formalism to \eqref{eq:HFibFib} leads to a host of integrable models. The first solution is 
\begin{equation}
\alpha_1=\alpha_2=\alpha_3\equiv \alpha,\qquad H^{\text{Fib}\times\text{Fib}}  = \alpha( \mathrm{P}^{\boldsymbol{4}}_{\boldsymbol{4}} +  \mathrm{P}^{\boldsymbol{4}}_{\boldsymbol{3}} +\mathrm{P}^{\boldsymbol{4}}_{\boldsymbol{2}}) = -\alpha \mathrm{P}^{\boldsymbol{4}}_{\boldsymbol{1}}
\end{equation} 
which corresponds to the TL integrable projector, studied in \cite{Antunes:2025huk, Blakeney:2025ext}. This model is gapped for all $\alpha$, however for $\alpha > 0$ the large correlation length can obscure some of the numerics.

There is also a family of integrable models
\begin{equation}
\alpha_1 = \alpha_2 + \alpha_3,
\end{equation}
which is effectively one-parameter after rescaling $\alpha_1 =\pm1$. Writing this model as
\begin{equation}\label{eq:FibFibz}
H^{\text{Fib}\times \text{Fib}}_z = - \mathrm{P}^{\boldsymbol{4}}_{\boldsymbol{4}} - z \mathrm{P}^{\boldsymbol{4}}_{\boldsymbol{2}}  - (1-z)\mathrm{P}^{\boldsymbol{4}}_{\boldsymbol{3}}, 
\end{equation}
one can study it numerically as a function of $z$. Due to the $\boldsymbol{2} \leftrightarrow \bs{3}$ automorphism of the fusion rules, it suffices to consider $z \geq \frac12$. There are two distinct regimes, one corresponding $z \in [\frac12, 1)$, and one corresponding to $z\in (1,\infty)$. The case $z = 1$ has a highly degenerate spectrum, and serves as the transition between these two phases. This behavior is explained by considering the projectors $\mathrm{P}^{\bs{4}}_{\bs{i}}$ in terms of golden chain projectors on the individual copies in $\text{Fib}\times \text{Fib}$. Indeed, we have
\begin{equation}
\mathrm{P}^{\bs{4}}_{\bs{2}} = p^L(1-p^R), \qquad \mathrm{P}^{\bs{4}}_{\bs{3}}  = (1-p^L)p^R, \qquad \mathrm{P}^{\bs{4}}_{\bs{4}}  = (1-p^L)(1-p^R),
\end{equation}
where $p^{L/R}$ is the golden chain $\mathrm{P}^{\tau}_1$ acting on the left/right copy of $\text{Fib}\times \text{Fib}$. Then restoring an overall scale $J$, up to an additive constant the cross term $p^Lp^R$ cancels and \eqref{eq:FibFibz} becomes
\begin{equation}\label{eq:FibFibz2}
H^{\text{Fib}\times \text{Fib}}_z = J(p^L(1-z) + z p^R).
\end{equation}
We see that this model is simply a sum of two independent golden chains, which depending on $J$ and $z$ are either both ferromagnetic, anti-ferromagnetic, or a mixture. For example, if $J<0$ then $z \in [\frac12, 1)$ corresponds to two golden chains with the same sign, in this case corresponding to the CFT $\text{Potts} \times \text{Potts}$ with $c=2\times 0.8 =1.6$. If $z\in (1,\infty)$ then \eqref{eq:FibFibz2} corresponds to golden chains with a different sign and this corresponds to $\text{TCI}\times \text{Potts}$ with $c = 0.7 + 0.8 = 1.5$. If $z=1$ then the model acts as the identity on the left copy of Fib, explaining the large degeneracy in the spectrum.

\begin{figure}[t]
  \centering
  \begin{minipage}[t]{0.49\textwidth}
    \centering
    \includegraphics[width=\linewidth]{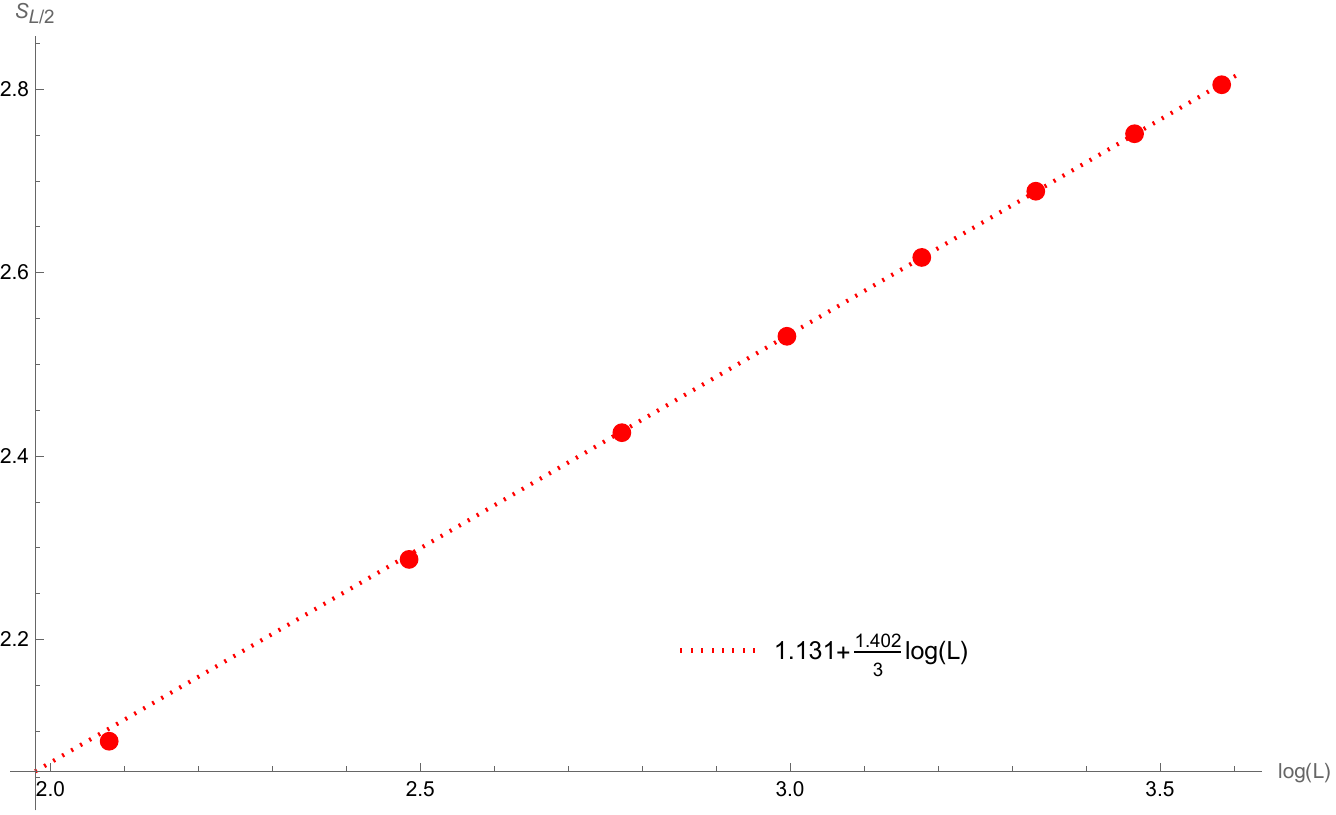}
  \end{minipage}\hfill
  \begin{minipage}[t]{0.49\textwidth}
    \centering
    \includegraphics[width=\linewidth]{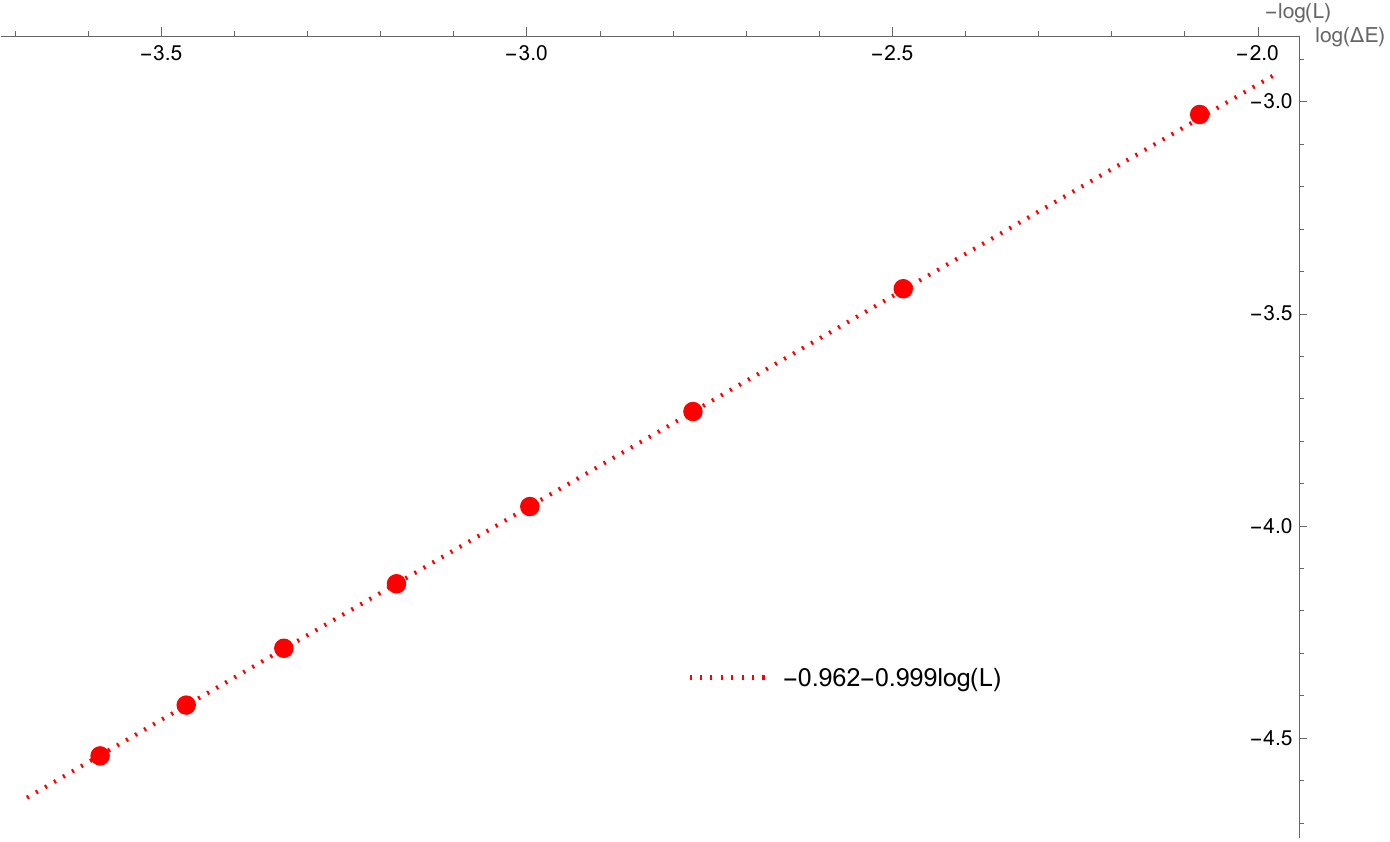}
  \end{minipage}

  \caption{Half-chain entanglement entropy and gap for the $\text{Fib}\times \text{Fib}$ chain \eqref{eq:fibfibint} with $J < 0$. We take $8 \leq L \leq 36$ with $L \equiv 0$ mod 4. For the fits we used the last five points.}
  \label{fig:fibfib-ee-gap}
\end{figure}

\begin{figure}[t]
  \centering
  \begin{minipage}[t]{0.49\textwidth}
    \centering
    \includegraphics[width=\linewidth]{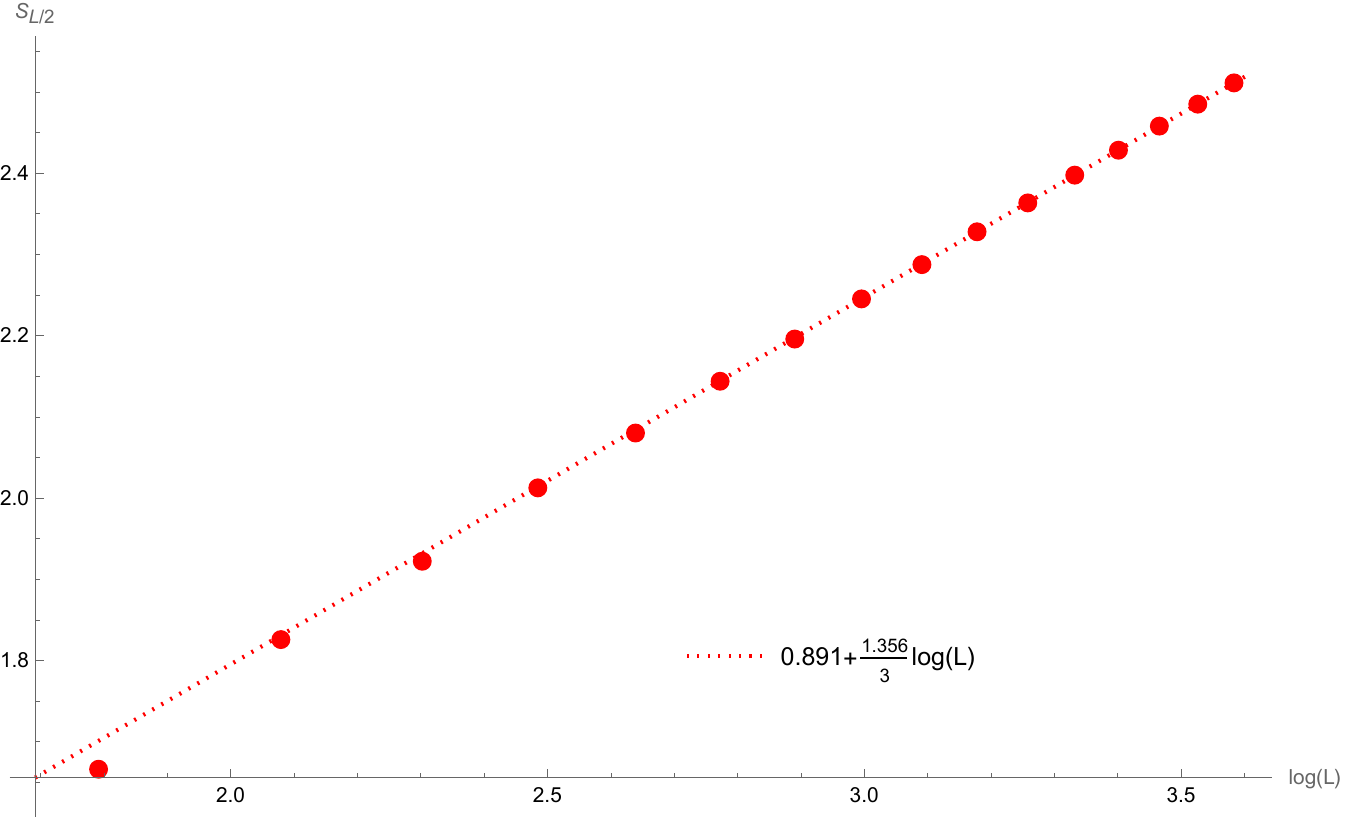}
  \end{minipage}\hfill
  \begin{minipage}[t]{0.49\textwidth}
    \centering
    \includegraphics[width=\linewidth]{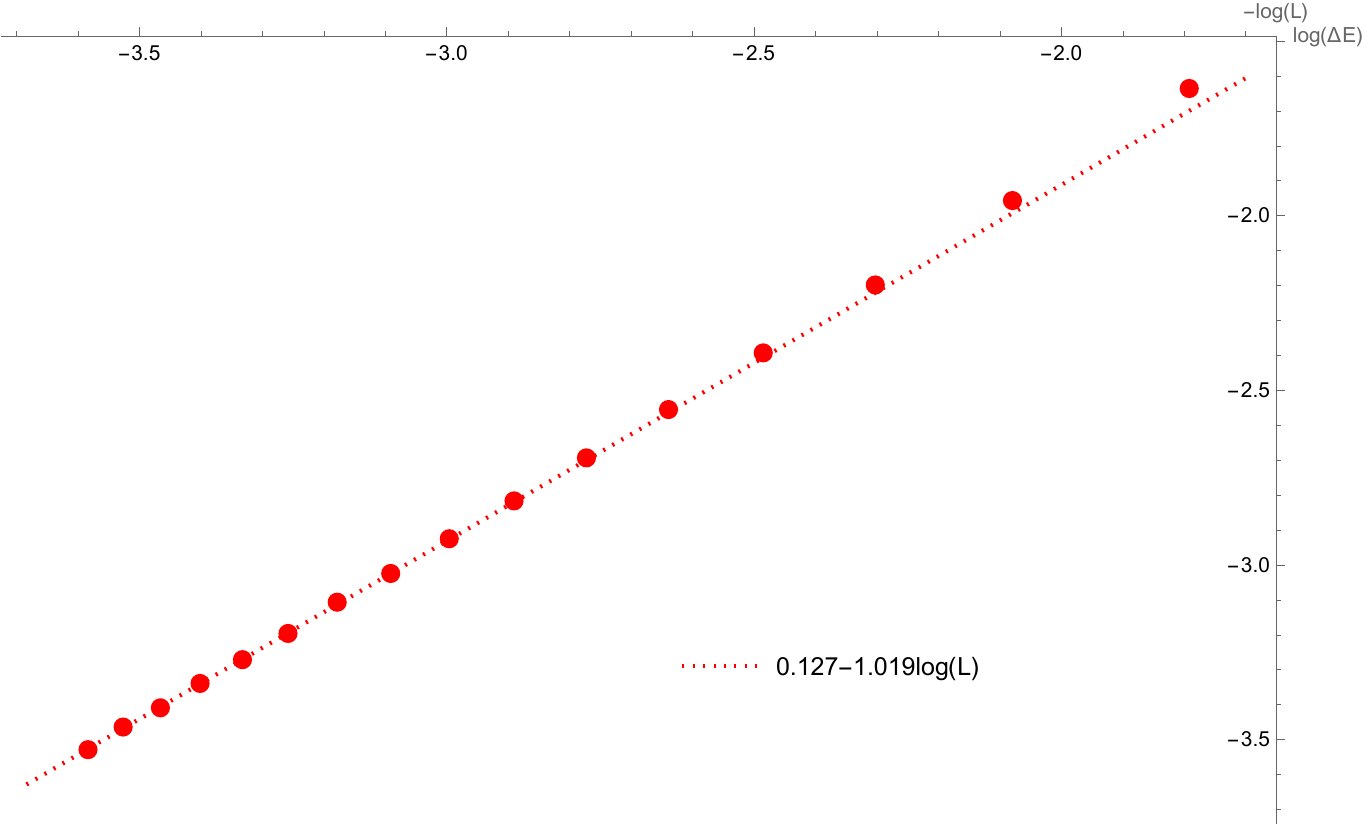}
  \end{minipage}

  \caption{Half-chain entanglement entropy and gap for the $\text{Fib}\times \text{Fib}$ chain \eqref{eq:fibfibint} with $J > 0$. We take even $6 \leq L \leq 36$. For the fits we used the last six points.}
  \label{fig:afibfib-ee-gap}
\end{figure}

Finally, there is an isolated integrable model found by choosing the parameters
\begin{equation}
\alpha_2 = \alpha_3 = \left(1+\frac{1}{\sqrt{5}}\right) \alpha_1.
\end{equation}
In terms of golden chain projectors, this model (after normalization and shifting) can be written
\begin{equation}\label{eq:fibfibint}
H^{\text{Fib}\times \text{Fib}}_{\text{int}} = J(p^Lp^R - \varphi^{-3} (p^L + p^R)),
\end{equation}
and so contains a non-trivial coupling term $p^Lp^R$. Interestingly, this Hamiltonian matches closely with the critical point found in \cite{Antunes:2025huk}, with $J/K \sim 0.236 \sim \varphi^{-3}$. \footnote{The relative sign difference originates from a sign difference in the golden chain definitions $p^{L/R}\rightarrow -p^{L/R}$.} We present some numerics for this model in figures \ref{fig:fibfib-ee-gap} and \ref{fig:afibfib-ee-gap}, to verify \eqref{eq:loggap} and \eqref{eq:EEdef}. For $J < 0$ we find evidence that this model is critical with a central charge of $c \sim 1.4$, leading to a likely CFT candidate of $\text{TCI} \times \text{TCI}$. For $J>0$ we find evidence that the model is critical with $c\sim 1.35$, confirming the results of \cite{Antunes:2025huk}. A natural candidate for this CFT is the coset
\begin{equation}
\frac{\mathfrak{su}(2)_3\times \mathfrak{su}(2)_3}{\mathfrak{su}(2)_6}.
\end{equation}

\subsection{$\text{Fib}\times \text{Ising}$}
As a final example, we briefly consider the $\text{Fib}\times\text{Ising}$ fusion category, dubbed in Anyonwiki as TriCritIsing \cite{gert_vercleyen_anyonwiki}. In this case there are six objects $\{1\times 1, 1\times \psi, 1 \times \sigma, \tau\times 1, \tau\times \psi, \tau\times\sigma\} \coloneqq \{\boldsymbol{1},\boldsymbol{2}, \boldsymbol{3},\boldsymbol{4},\boldsymbol{5},\boldsymbol{6}\}$. The fusion rules for this model are given in table \ref{tab:fibxising_fusion}, and the quantum dimensions of the objects are

\begin{equation}
(d_{\boldsymbol{1}}, d_{\boldsymbol{2}}, d_{\boldsymbol{3}}, d_{\boldsymbol{4}},d_{\boldsymbol{5}},d_{\boldsymbol{6}})  = (1, 1, \sqrt{2}, \varphi, \varphi, \varphi \sqrt{2}).
\end{equation}

\begin{table}[tbp]
\centering
\renewcommand{\arraystretch}{1.15}
\setlength{\tabcolsep}{4pt}
\[
\begin{array}{c|cccccc}
\otimes & \boldsymbol{1} & \boldsymbol{2} & \boldsymbol{3} & \boldsymbol{4} & \boldsymbol{5} & \boldsymbol{6} \\ \hline
\boldsymbol{1} & \boldsymbol{1} & \boldsymbol{2} & \boldsymbol{3} & \boldsymbol{4} & \boldsymbol{5} & \boldsymbol{6} \\
\boldsymbol{2} & \boldsymbol{2} & \boldsymbol{1} & \boldsymbol{3} & \boldsymbol{5} & \boldsymbol{4} & \boldsymbol{6} \\
\boldsymbol{3} & \boldsymbol{3} & \boldsymbol{3} & \boldsymbol{1} \oplus \boldsymbol{2} & \boldsymbol{6} & \boldsymbol{6} & \boldsymbol{4} \oplus \boldsymbol{5} \\
\boldsymbol{4} & \boldsymbol{4} & \boldsymbol{5} & \boldsymbol{6} & \boldsymbol{1} \oplus \boldsymbol{4} & \boldsymbol{2} \oplus \boldsymbol{5} & \boldsymbol{3} \oplus \boldsymbol{6} \\
\boldsymbol{5} & \boldsymbol{5} & \boldsymbol{4} & \boldsymbol{6} & \boldsymbol{2} \oplus \boldsymbol{5} & \boldsymbol{1} \oplus \boldsymbol{4} & \boldsymbol{3} \oplus \boldsymbol{6} \\
\boldsymbol{6} & \boldsymbol{6} & \boldsymbol{6} & \boldsymbol{4} \oplus \boldsymbol{5} & \boldsymbol{3} \oplus \boldsymbol{6} & \boldsymbol{3} \oplus \boldsymbol{6} & \boldsymbol{1} \oplus \boldsymbol{2} \oplus \boldsymbol{4} \oplus \boldsymbol{5}
\end{array}
\]
\caption{Fusion rules for $\mathrm{Fib}\times \mathrm{Ising}$.}
\label{tab:fibxising_fusion}
\end{table}

\par\medskip
\noindent\textbf{Integrable anyonic chains.}\quad
In this case the only choice of external object giving new non-trivial chains is $a = \tau\times\sigma \equiv \boldsymbol{6}$. There are three independent projectors contributing in this case
\begin{equation}
H^{\text{Fib}\times \text{Ising}} = \alpha_1 \mathrm{P}^{\boldsymbol{6}}_{\boldsymbol{2}} + \alpha_2 \mathrm{P}^{\boldsymbol{6}}_{\boldsymbol{4}} + \alpha_3  \mathrm{P}^{\boldsymbol{6}}_{\boldsymbol{5}}.  
\end{equation}
Using the constrained boost operator formalism, we find that in this case there are five independent integrable choices of parameters:
\begin{align}
& \alpha_1 = \alpha_2 = \alpha_3 \qquad \text{(TL)}, \\
&\alpha_3 = \alpha_1 + \alpha_2, \label{eq:FibIsingParam}\\
&\alpha_2 = \alpha_3 = 0  \qquad \text{(TL)}, \\
&\alpha_1 = \frac{\alpha_3}{2} \left(3+2 \sqrt{5}\pm\sqrt{25+2 \sqrt{5}}\right), \alpha_2 = \frac{\alpha_3}{2} \left(1\mp\sqrt{5-2 \sqrt{5}}\right) \label{eq:FibIsingIsolated} .
\end{align}
We note in this case there are two TL points, given that there are two projections onto invertible objects $\bs{6}\otimes \bs{6}\rightarrow \bs{1}$ and $\bs{6}\otimes \bs{6}\rightarrow \bs{2}$. The case \eqref{eq:FibIsingParam}, similarly to \eqref{eq:FibFibz}, corresponds to decoupled golden and Ising chains. The two isolated integrable points \eqref{eq:FibIsingIsolated} have a non-trivial coupling between the Fib and Ising chains, and we expect these to correspond to non-trivial CFTs.

\section{Conclusions and outlook}

In this paper we reviewed several aspects of integrable models on constrained Hilbert spaces, with a focus on anyonic chains. Using the constrained boost operator formalism we were able to reproduce several known constrained integrable models, as well as construct several new ones. These integrable models often correspond to critical spin chains in the $L\rightarrow \infty$ limit, highlighting the connection between fusion categories and CFTs.

There are several directions for further research. One avenue is to systematically study all of the fusion categories on Anyonwiki \cite{gert_vercleyen_anyonwiki}, and classify integrable/critical models therein. Several fusion categories there admit no integrable models beyond the TL case, such as `pseudo $\mathfrak{psu}(2)_6$'. However, we expect interesting models to crop up for more exotic fusion categories like Adj($\mathfrak{so}(16)_2$) and Rep$(\text{Dic}_{12})$. It would be interesting to study higher products of fusion categories such as $\text{Fib}\times\text{Fib}\times\text{Fib}$, and compare integrable points to critical points found in \cite{Antunes:2025huk}. Other interesting cases are Haagerup-Izumi models HI$(G)$ for which the $F$-symbols are not yet known, for example $G = \mathbb{Z}_2\times \mathbb{Z}_2$ or $G = \mathbb{Z}_4$. Finding the $F$-symbols and studying these cases in detail would help clarify the relationship between fusion categories and integrable/critical models.

The space of constrained spin chains is rich, and extends beyond the case of anyonic chains. Within the fusion category landscape, one idea is to study chains based on fusion categories of multiplicity higher than 1. A standard first example for this case would be the Rep$(A_4)$ fusion category \cite{Lootens:2024gfp, Perez-Lona:2024sds}. Beyond this, one could consider chains defined from the fusion of several non-invertible external objects $a_i$, rather than a single object $a$. For example, one could study a $\mathfrak{psu}(2)_5$ chain with alternating external objects $a_{2i} = \boldsymbol{1}$ and $a_{2i+1}=\boldsymbol{2}$. There is also relatively less work done on projectors which project beyond nearest-neighbor, as studied in \cite{Trebst2_2008}. Beyond the anyonic chain/fusion category case there are several interesting constrained Hilbert spaces to study. One such case arises in the dilatation operator of a $\mathcal{N}=2$ SCFT \cite{Gadde:2009dj, Bozkurt:2024tpz, Bozkurt:2025exl}. Despite some signs of integrability in this case, it does not appear to be `boost-integrable' in the way described in this paper. It would be interesting to study this case further and classify integrable models on this Hilbert space. There are also relatively-unstudied generalizations of the Rydberg-blockade constraint, which are likely to contain integrable models with exotic properties  \cite{Mukherjee:2021exb, Mohapatra:2023ive}.

In recent years, there has been the development of open-source numerical tools which allow for the study of one-dimensional quantum many-body systems directly in symmetry-adapted tensor-network language. In particular, TensorKit \cite{Devos:2025yoj} provides a flexible framework for tensor computations with internal symmetries, including abelian, non-abelian, and anyonic symmetries, while MPSKit \cite{2024zndo..10654900V} implements finite and infinite matrix-product-state algorithms for quantum lattice systems. Earlier versions of these softwares were already successfully applied to lattice models using data from the Haagerup fusion category \cite{Vanhove:2021zop}. One interesting direction of study is to apply these numerical tools to simulate anyonic chains of higher rank, beyond the usual DMRG approach which is best suited to Hilbert spaces of product form. This would allow for a deeper study of full phase portraits of anyonic systems. For example, the phase diagram of spin-1 $\mathfrak{su}(2)_k$ spin chains is still not fully understood \cite{Vernier_2017}, and the spin-$\frac32$ chains are fully open for study. Such tools may also allow for a deeper study of higher-rank cases, such as HI$(\mathbb{Z}_5)$.

The models we studied in this paper were mostly derived from unitary solutions of the pentagon equations. While there have been studies of non-unitary counterparts for the $\mathfrak{su}(2)_k$ case \cite{Ardonne_2011,Lootens:2019xjv}, the literature in this case is rather sparse, especially in light of recent developments in non-unitary CFTs \cite{Bianchini:2014uta, Bianchini:2015uea}. It would be interesting to reproduce the results of this paper for the effective central charge $c_{\text{eff}}$ in the case of non-unitary $\mathfrak{su}(2)_k$ chains. This would allow for a framework to study models in more complicated non-unitary fusion categories, for example finding a non-unitary HI$(\mathbb{Z}_3)$ CFT.

Finally, another promising direction is to extend the search for integrable structures beyond one-dimensional constrained Hilbert spaces. Recent work on higher-dimensional categorical lattice models, in particular fusion surface models, has shown that anyonic chain constructions admit natural $2+1$-dimensional analogs governed by fusion 2-categories and higher categorical symmetries \cite{Inamura:2023qzl, Eck:2024myo, Eck:2025ldx}. It would be interesting to ask whether analogs of the constrained boost formalism, commuting transfer matrices, or other integrability diagnostics can be formulated in this setting.

\vspace{1cm}
\noindent {\bf Acknowledgments.} 
We thank Balázs Pozsgay and Eric Vernier for collaboration on a related project. We thank Boris de Vos for bringing TensorKit and MPSKit to our attention. MdL was supported in part by SFI and the Royal Society for funding under
grants UF160578, RGF$\backslash$ R1$\backslash$ 181011, RGF$\backslash$8EA$\backslash$180167 and RF$\backslash$
ERE$\backslash$ 210373. MdL is also supported by ERC-2022-CoG - FAIM 101088193, which also supports LC. Research at Perimeter Institute is
supported in part by the Government of Canada through the Department of Innovation, Science and Economic Development and
by the Province of Ontario through the Ministry of Colleges and Universities.
\bigskip

\appendix

\addcontentsline{toc}{section}{References}
\bibliography{anyon_review}
\bibliographystyle{utphys}

\end{document}